\documentclass[preprint,prd,tightenlines,floatfix,
nofootinbib,eqsecnum,superscriptaddress]{revtex4-1}

\usepackage[T1]{fontenc}		% fnot encoding with polish letters !or OT4

\usepackage{amsmath,amsfonts,amssymb,amstext,mathrsfs}
\usepackage{amsthm}
\usepackage{mathpazo}

\usepackage[dvips]{graphicx}
\usepackage{epsf,float}
\usepackage{revsymb}

\usepackage{dcolumn}% Align table columns on decimal point
\usepackage{braket}
\usepackage{color,xcolor}
\usepackage{graphicx}
\usepackage{subfigure}
\usepackage{multirow}
\usepackage{tabularx}
\usepackage{pstricks}
\usepackage[section]{placeins}
\usepackage{booktabs}
\usepackage{array}

\usepackage{tablefootnote}

\usepackage{commath}

\usepackage{hyperref}
\usepackage{lineno}
\usepackage{hyphenat}

\newcommand{\Pom}{\mathbb{P}}
\newcommand{\Ode}{\mathbb{O}}
\newcommand{\Reg}{\mathbb{R}}

\newcommand{\bpa}{\mbox{\boldmath $p_{a}$}}

\newcommand{\bhkperp}{\mbox{\boldmath $\hat{k}_{\perp}$}}

\newcommand{\bpaap}{\mbox{\boldmath $p_{1}'$}}
\newcommand{\bpbbp}{\mbox{\boldmath $p_{2}'$}}
\newcommand{\bptpa}{\mbox{\boldmath $p_{t,1}'$}}
\newcommand{\bptpb}{\mbox{\boldmath $p_{t,2}'$}}

\renewcommand\slash[1]{\not \! #1}

\usepackage[normalem]{ulem}  % \sout{old text} for strikeout

\begin{document}

%------------------------
\nolinenumbers
%\linenumbers
%------------------------

\title{\boldmath 
Soft-photon radiation in high-energy proton-proton collisions\\
within the tensor-Pomeron approach:
Bremsstrahlung}

\vspace{0.6cm}

\author{Piotr Lebiedowicz}
%\orcid{0000-0003-1963-6263}
\email{Piotr.Lebiedowicz@ifj.edu.pl}
\affiliation{Institute of Nuclear Physics Polish Academy of Sciences, 
Radzikowskiego 152, PL-31342 Krak{\'o}w, Poland}

\author{Otto Nachtmann}
\email{O.Nachtmann@thphys.uni-heidelberg.de}
\affiliation{Institut f\"ur Theoretische Physik, Universit\"at Heidelberg,
Philosophenweg 16, D-69120 Heidelberg, Germany}

\author{Antoni Szczurek
%\orcid{0000-0001-5247-8442}
\footnote{Also at \textit{College of Natural Sciences, 
Institute of Physics, University of Rzesz{\'o}w, 
ul. Pigonia 1, PL-35310 Rzesz{\'o}w, Poland}.}}
\email{Antoni.Szczurek@ifj.edu.pl}
\affiliation{Institute of Nuclear Physics Polish Academy of Sciences, 
Radzikowskiego 152, PL-31342 Krak{\'o}w, Poland}

\begin{abstract}
We discuss diffractive processes
in proton-proton collisions
at small momentum transfers 
without and with photon radiation.
We consider the soft exclusive reactions 
$pp \to pp$, $p\bar{p} \to p\bar{p}$, 
and $pp \to pp \gamma$
within the tensor-pomeron and vector-odderon approach.
We compare our results with the data
for $pp$ and $p\bar{p}$ total cross sections,
for the ratio of real to imaginary part 
of the forward scattering amplitude, 
and for the elastic $pp$ cross sections,
especially those from TOTEM.
To describe the low-energy data more accurately 
the secondary reggeons must be included.
We write down the amplitudes 
for the photon bremsstrahlung
in high-energy proton-proton collisions
using the tensor-pomeron model.
These results are relevant for the c.m. energies
presently available at the Relativistic Heavy Ion Collider 
and at the LHC.
We present predictions for the proposed measurements
of soft photons with the planned future upgrade 
of the ALICE experiment at the LHC.
We investigate the limits of applicability of 
the soft-photon approximation (SPA) based on Low's theorem.
The corresponding SPA results are compared 
to those obtained from our complete model.
The regions of phase space are given quantitatively
where SPA and our complete tensor-pomeron results
are close to each other.
As an example, let $k_{\perp}$, $\rm y$, and $\omega$,
be the absolute value of the transverse momentum,
the rapidity, and the energy of the photon, respectively,
in the overall c.m. system.
For the region 
$1 \; {\rm MeV} < k_{\perp} < 100 \; {\rm MeV}$ and
$3.5  < |\rm y| < 5.0$, we find that the SPA Ansatz
with only the pole terms $\propto \omega^{-1}$ agrees
at the percent level with our complete model result up to
$\omega \cong 2$ GeV.
\end{abstract}

%\pacs{}

\maketitle

%----------------------------
\section{Introduction}
%----------------------------

With this article, we continue our investigations
of soft-photon radiation in diffractive hadronic high-energy reactions.
In Ref.~\cite{Lebiedowicz:2021byo}, we treat,
as a first example using our approach,
high-energy $\pi \pi$ scattering 
without and with photon radiation.
In the present paper, we extend these considerations
to $pp$ and $p \bar{p}$ scattering.

The emission of soft photons, that is, photons of
energy $\omega$ approaching zero, 
was treated in the seminal paper 
by Low~\cite{Low:1958sn}.
In that paper, it was shown that the term of order $\omega^{-1}$
in the amplitude for the emission reaction
can be obtained from the amplitude without photon emission.
To this order, the emission comes exclusively
from the external particles,
and this is a strict consequence of Quantum Field Theory (QFT).
Many soft-photon approximations (SPAs)
are based on this result.

Experimental studies trying to verify Low's theorem
\cite{Goshaw:1979kq,Chliapnikov:1984ed,Botterweck:1991wf,Banerjee:1992ut,Antos:1993wv,Tincknell:1996ks,Belogianni:1997rh,Belogianni:2002ib,Belogianni:2002ic,Abdallah:2005wn,Abdallah:2007aa,DELPHI:2010cit} 
have, in many cases,
found large deviations from the SPA calculations.
For a review of the experimental and theoretical
situations, see Ref.~\cite{Wong:2014pY}.
Clearly, more experimental and theoretical work
is needed in order to clarify this so-called
soft-photon problem.

From the experimental side there is, for instance,
the plan for a new multipurpose detector at the LHC, 
ALICE~3 \cite{Adamova:2019vkf}.
One physics aim for this new initiative is a measurement 
of ultrasoft photons at very low transverse momentum in 
$pp$, $pA$ and $AA$ collisions;
see, e.g., Refs.~\cite{QM2022_PBM,EMMI_RRTF}.
%In this context we would also like to mention 
%the EMMI workshop on
%``Real and virtual photon production 
%at ultra-low transverse momentum and low mass at LHC''.
%Corresponding link to the talks
%presented at this workshop can be found in~\cite{EMMI_RRTF}.

From the theoretical side, many authors have studied
soft-photon production following Ref.~\cite{Low:1958sn};
see, for instance, Refs.~\cite{Gribov:1966hs,Burnett:1967km,Bell:1969yw,Lipatov:1988ii,
DelDuca:1990gz,Gervais:2017yxv,Bern:2014vva,Lysov:2014csa,Bonocore:2021cbv}.
In our paper \cite{Lebiedowicz:2021byo}, we have presented
two types of soft-photon studies.
\begin{itemize}
\item[(1)] We have studied the amplitude for the reaction
$\pi \pi \to \pi \pi \gamma$ in the limit of the photon
c.m. energy $\omega$ going to zero. 
Using only rigorous QFT methods,
we have calculated the terms of order $\omega^{-1}$ and $\omega^{0}$.
We found agreement with the result of Low \cite{Low:1958sn}
for the $\omega^{-1}$ term, but we disagreed with
the $\omega^{0}$ term of Ref.~\cite{Low:1958sn}.
We have analyzed the origin of this disagreement,
and we give a critique of the corresponding results
of the papers \cite{Low:1958sn,Burnett:1967km,Lipatov:1988ii,
DelDuca:1990gz,Gervais:2017yxv,Bern:2014vva,Lysov:2014csa}
in Appendixes~A and B of Ref.~\cite{Lebiedowicz:2021byo}.
\item[(2)] We have discussed the reactions $\pi \pi \to \pi \pi$
and $\pi \pi \to \pi \pi \gamma$ at high energies
in a specific model, the tensor-pomeron model of Ref.~\cite{Ewerz:2013kda}.
The ``standard'' results obtained in this approach
were then compared to various soft-photon approximations.
\end{itemize}

With the present paper, we continue our line (2) of research.
Further results from the line (1) will be presented elsewhere.

One class of hadronic reactions one can study 
at the LHC is exclusive diffractive proton-proton collisions.
Examples are $pp \to pp$ elastic scattering
and central exclusive production of mesons,
for instance, $pp \to pp \pi^{+} \pi^{-}$. 
In this paper, we present the results of our investigations
of the following soft reactions at small momentum transfer:
\begin{eqnarray}
&&p + p \to p + p, \nonumber\\
&&p + \bar{p} \to p + \bar{p}, \nonumber\\
&&p + p \to p + p + \gamma\,.
\label{reactions}
\end{eqnarray}
We shall work within the tensor-pomeron model 
as proposed in Ref.~\cite{Ewerz:2013kda} 
for soft hadronic high-energy reactions.
There, the soft pomeron and the charge conjugation
$C = +1$ reggeons are described as 
effective rank-2 symmetric tensor exchanges,
and the odderon and the $C = -1$ reggeons are described
as effective vector exchanges.
Now, our task is to construct 
a soft-photon diffractive amplitude 
of the $pp \to pp \gamma$ reaction
which satisfies all theoretical constraints.
%We perform an analysis from a general QFT point of view.

Before coming to our present investigations, we make
remarks on some related works.
Exclusive diffractive photon bremsstrahlung
in $pp$ collisions was discussed earlier
in Refs.~\cite{Khoze:2010jv,Lebiedowicz:2013xlb,Khoze:2017igg}
within other approaches.
The bremsstrahlung-type emission of $\omega$ and $\pi^{0}$ mesons
was calculated in Refs.~\cite{Cisek:2011vt,Lebiedowicz:2013vya}.
It is also worth noting that the $pp \to pp \gamma$ reaction
has not yet been measured at high energies; however,
feasibility studies 
were performed for RHIC energies 
\cite{Chwastowski:2015mua} 
and for LHC energies 
\cite{Chwastowski:2016jkl,Chwastowski:2016zzl}.
%show a non-negligible potential of the measurement.
%Simulations of the bremsstrahlung processes considered
%were performed using GenEx generator \cite{Kycia:2014hea}
%based on calculations presented in \cite{Lebiedowicz:2013xlb}.

The theoretical methods which we shall develop in the present paper
for soft-photon production can also be used
in a completely different context:
for the production of ``dark photons``.
Indeed, an interesting proposal of new physics search
was discussed recently 
in Ref.~\cite{Foroughi-Abari:2021zbm}:
to study the forward production of 
dark vectors (photons)
and scalars via bremsstrahlung
in proton-proton collisions with the proposed Forward Physics Facility (FPF) 
at the High-Luminosity LHC \cite{Feng:2022inv}.
At the LHC, such weakly coupled long-lived particles,
with masses $m \sim 10$~MeV--1~GeV,
could be produced through 
light meson decays and bremsstrahlung
in the region that would be covered by the FPF. 

Our present paper is organized as follows.
In Sec.~\ref{sec:2}, we discuss the amplitudes
for the reactions listed in (\ref{reactions})
%$pp \to pp$, $p\bar{p} \to p\bar{p}$, and $pp \to pp \gamma$
within the tensor-pomeron approach.
In Sec.~\ref{sec:SPA}, we describe
two SPAs based on Low's theorem.
The results of our calculations are presented in Sec.~\ref{sec:3}.
Section~\ref{sec:3A} is devoted to a comparison of the model results 
to the available data 
on the total and elastic $pp$ and $p \bar{p}$ cross sections.
In Sec.~\ref{sec:3B}, we present our ``exact'' model 
or ``standard'' results 
for the $pp \to pp \gamma$ reaction
and a comparison to SPAs.
Subsection~\ref{sec:3C} contains
comments on the photon radiation in connection with 
diffractive excitation of the proton.
Section~\ref{sec:4} contains a summary and our conclusions.
Some details of the present model are given in 
Appendixes~\ref{sec:appendixA} and \ref{sec:appendixB}.

Throughout our paper, we use the metric and $\gamma$-matrix
conventions of Ref.~\cite{Bjorken:1965}. 
%with \mbox{$\sigma_{\mu \nu} = \frac{i}{2} (\gamma_{\mu} \gamma_{\nu} - \gamma_{\nu} \gamma_{\mu})$}.

%----------------------------
\section{Reactions $pp \to pp$, $p\bar{p} \to p\bar{p}$, and $pp \to pp \gamma$}
\label{sec:2}
%----------------------------

Here we discuss the reactions
\begin{eqnarray}
&&p (p_{a},\lambda_{a}) + p (p_{b},\lambda_{b}) \to p (p_{1},\lambda_{1}) + p (p_{2},\lambda_{2})\,,
\label{2.1}\\
&&p (p_{a},\lambda_{a}) + \bar{p} (p_{b},\lambda_{b}) \to p (p_{1},\lambda_{1}) + \bar{p} (p_{2},\lambda_{2})\,,
\label{2.1b}
\end{eqnarray}
and
\begin{eqnarray}
p (p_{a},\lambda_{a}) + p (p_{b},\lambda_{b}) \to p (p_{1}',\lambda_{1}) + p (p_{2}',\lambda_{2}) 
+ \gamma(k, \epsilon)\,.
\label{2.2}
\end{eqnarray}
The momenta are denoted by $p_{a}, \ldots, k$;
the helicities of the protons are denoted by 
$\lambda_{a}, \ldots, \lambda_{2}$;
and $\epsilon$ is the polarization vector of the photon.
The energy-momentum conservation in (\ref{2.1}),
(\ref{2.1b}), and (\ref{2.2}) requires
\begin{eqnarray}
&&p_{a} + p_{b} = p_{1} + p_{2}\,,
\label{2.3} \\
&&p_{a} + p_{b} = p_{1}' + p_{2}' + k\,.
\label{2.4a}
\end{eqnarray}

We consider soft hadronic high-energy reactions. 
We use standard formulas of the tensor-pomeron 
and vector-odderon model from Ref.~\cite{Ewerz:2013kda}.
In this model, the assumption is made
that the pomeron $\Pom$ 
and the charge-conjugation $C = +1$ reggeons
$f_{2 \Reg}$, $a_{2 \Reg}$ couple to hadrons like 
symmetric tensors of rank~2,
and the odderon $\Ode$ and the $C = -1$ reggeons 
$\omega_{\Reg}$, $\rho_{\Reg}$ couple to hadrons like vectors.
We do not treat $\gamma$ exchange in the following.

In Ref.~\cite{Ewerz:2016onn}, it is shown that 
the experimental results \cite{Adamczyk:2012kn} 
on the spin dependence of high-energy proton-proton
elastic scattering exclude 
a scalar character of the pomeron couplings but
are perfectly compatible with the tensor-pomeron model.
A vector coupling for the pomeron 
could definitely be ruled out 
as shown in Ref.~\cite{Britzger:2019lvc}.

\subsection{Reaction $pp \to pp$}
\label{sec:2A}

%-------------------------------------------------------------
\begin{figure}[!h]
\includegraphics[width=5.0cm]{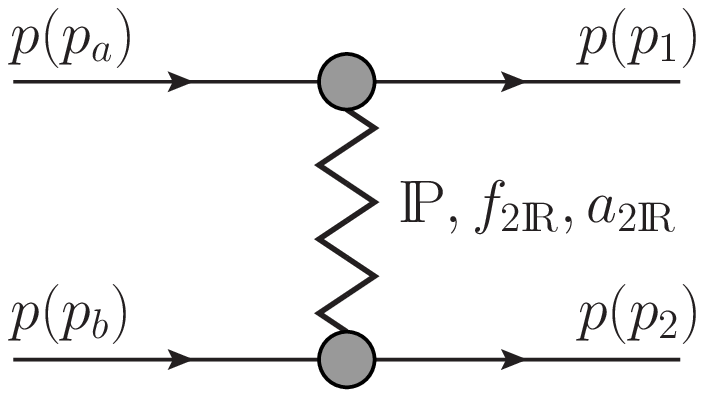}
\quad
\includegraphics[width=4.8cm]{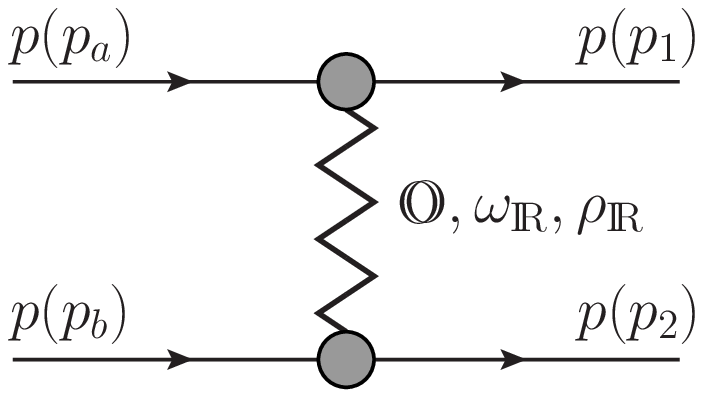}
\caption{The diagrams for $pp \to pp$ 
elastic scattering with $C = +1$ (left)
%($\Pom$, $f_{2 \Reg}$, $a_{2 \Reg}$) 
and $C = -1$ (right)
%($\Ode$, $\omega_{\Reg}$, $\rho_{\Reg}$) 
exchanges.
There are also the diagrams corresponding
to the exchange $p_{1} \leftrightarrow p_{2}$.}
\label{fig:pp_pp}
\end{figure}
%-------------------------------------------------------------
We consider first the reaction (\ref{2.1})
for off-shell protons.
Also, here, energy-momentum conservation (\ref{2.3}) holds.
We have the diagrams of Fig.~\ref{fig:pp_pp}
with $C = +1$ and $C = -1$ exchanges.
The kinematic variables are
\begin{eqnarray}
&& s = (p_{a} + p_{b})^{2} = (p_{1} + p_{2})^{2}\,, \nonumber \\
&& t = (p_{a} - p_{1})^{2} = (p_{b} - p_{2})^{2}\,, \nonumber \\
&& u = (p_{a} - p_{2})^{2} = (p_{b} - p_{1})^{2}\,, \nonumber \\
&& m_{a}^{2} = p_{a}^{2}\,,
m_{b}^{2} = p_{b}^{2}\,,
m_{1}^{2} = p_{1}^{2}\,,
m_{2}^{2} = p_{2}^{2}\,.
\label{2.4}
\end{eqnarray}
The interchange $p_{1} \leftrightarrow p_{2}$ implies 
$t \leftrightarrow u$,
where $u = -s -t + m_{a}^{2} + m_{b}^{2} + m_{1}^{2} + m_{2}^{2}$.
We are interested in the kinematic region
\begin{eqnarray}
\sqrt{s} \gg m_{p}\,, \quad \sqrt{|t|} \lesssim  m_{p}\,, \quad 
s \gg |m_{a}^{2}|, \,|m_{b}^{2}|,\, |m_{1}^{2}|,\, |m_{2}^{2}|. 
\label{2.5}
\end{eqnarray}
There we can neglect the diagrams 
with $p_{1} \leftrightarrow p_{2}$.

We denote the off-shell $pp$ scattering amplitude by
\begin{eqnarray}
{\cal M}^{(0)}(p_{a}, p_{b}, p_{1}, p_{2})\,.
\label{2.6}
\end{eqnarray}

In the tensor-pomeron model \cite{Ewerz:2013kda},
all exchanges at high energies are assumed to be describable
by effective single Regge poles.
Some standard references to Regge theory are
\cite{Collins:1977,Caneschi,Gribov:2009}.
For further literature, 
we refer to \cite{Ewerz:2013kda,Lebiedowicz:2021byo}.
The special feature of the tensor-pomeron model is,
as mentioned above,
that all $C = +1$ exchanges 
($\Pom$, $f_{2 \Reg}$, $a_{2 \Reg}$) 
are described as effective
rank-2 symmetric tensor exchanges
and that all $C = -1$ exchanges 
\mbox{($\Ode$, $\omega_{\Reg}$, $\rho_{\Reg}$)} are described
as effective vector exchanges.
To calculate the expressions for the diagrams of Fig.~\ref{fig:pp_pp},
we need the effective propagators for
the exchange objects and their proton-proton coupling vertices.
All these quantities are listed in Chap.~3 of Ref.~\cite{Ewerz:2013kda}.

To write down the amplitude (\ref{2.6}) in this
model in a convenient way we define
for the $C = +1$ exchanges the functions
\begin{eqnarray}
{\cal F}_{\Pom pp}(s,t) &=&
\big[ 3 \beta_{\Pom pp} F_{1}(t) \big]^{2}
\frac{1}{4s}(-is \alpha'_{\Pom})^{\alpha_{\Pom}(t)-1} \,,
\label{2.7}\\
{\cal F}_{f_{2 \Reg} pp}(s,t) &=&
\big[ \frac{g_{f_{2 \Reg} pp}}{M_{0}} F_{1}(t) \big]^{2}
\frac{1}{4s}(-is \alpha'_{f_{2 \Reg}})^{\alpha_{f_{2 \Reg}}(t)-1} \,,
\label{2.8}\\
{\cal F}_{a_{2 \Reg} pp}(s,t) &=&
\big[ \frac{g_{a_{2 \Reg} pp}}{M_{0}} F_{1}(t) \big]^{2}
\frac{1}{4s}(-is \alpha'_{a_{2 \Reg}})^{\alpha_{a_{2 \Reg}}(t)-1} \,;
\label{2.9}
\end{eqnarray}
and for the $C = -1$ exchanges the functions
\begin{eqnarray}
{\cal F}_{\Ode pp}(s,t) &=& - \eta_{\Ode}
\big[ 3 \beta_{\Ode pp} F_{1}(t) \big]^{2}
(-is \alpha'_{\Ode})^{\alpha_{\Ode}(t)-1} \,,
\label{2.10}\\
{\cal F}_{\omega_{\Reg} pp}(s,t) &=&
\big[ \frac{g_{\omega_{\Reg} pp}}{M_{-}} F_{1}(t) \big]^{2}
(-is \alpha'_{\omega_{\Reg}})^{\alpha_{\omega_{\Reg}}(t)-1} \,,
\label{2.11}\\
{\cal F}_{\rho_{\Reg} pp}(s,t) &=&
\big[ \frac{g_{\rho_{\Reg} pp}}{M_{-}} F_{1}(t) \big]^{2}
(-is \alpha'_{\rho_{\Reg}})^{\alpha_{\rho_{\Reg}}(t)-1} \,.
\label{2.12}
\end{eqnarray}
We define also the following quantities:
\begin{eqnarray}
&&{\cal F}_{T}(s,t) =
{\cal F}_{\Pom pp}(s,t) + {\cal F}_{f_{2 \Reg} pp}(s,t) 
+ {\cal F}_{a_{2 \Reg} pp}(s,t)\,,
\label{1aux}\\
&&{\cal F}_{V}(s,t) =
{\cal F}_{\Ode pp}(s,t) + {\cal F}_{\omega_{\Reg} pp}(s,t) 
+ {\cal F}_{\rho_{\Reg} pp}(s,t)\,.
\label{2aux}
\end{eqnarray}

All quantities occurring in
(\ref{2.7})--(\ref{2.12}) are as defined 
in Chaps. 3.1 and 3.2 of Ref.~\cite{Ewerz:2013kda}.
To make our article self-contained, we give in
Appendix~\ref{sec:appendixA} a list of the quantities
and their values which we use.
In considering the effective propagators
and vertices, some additional comments are in order:
%%%%%%%%%%%%%%%%
\begin{itemize}
%%%%%%%%%%%%%%%%
\item
%%%%%%%%%%%%%%%%
In Ref.~\cite{Ewerz:2013kda}, exchange degeneracy of the reggeons
with $C = +1$ ($\Reg_{+} = f_{2 \Reg}$, $a_{2 \Reg}$) and 
the reggeons with $C = -1$ 
($\Reg_{-} = \omega_{\Reg}$, $\rho_{\Reg}$)
and equality of the $\Reg_{+}$ and $\Reg_{-}$ trajectories 
was assumed; see also Ref.~\cite{Donnachie:2002en}.
This leads for the trajectories to $\alpha_{\Reg}(t) = 
\alpha_{\Reg_{+}}(t) = \alpha_{\Reg_{-}}(t)$.
We assume a standard linear form
$\alpha_{\Reg}(t) = \alpha_{\Reg}(0) + \alpha_{\Reg}'\, t$
with the intercept and the slope:
$\alpha_{\Reg}(0) = 0.5475$,
$\alpha_{\Reg}' = 0.9$~GeV$^{-2}$.
%%%%%%%%%%%%%%%%
\item
%%%%%%%%%%%%%%%%
The pomeron trajectory function in (\ref{2.7})
is also taken as linear in $t$,
$\alpha_{\Pom}(t) = 1 + \epsilon_{\Pom} + \alpha_{\Pom}' \,t$.
For the intercept parameter $\epsilon_{\Pom}$
and the slope parameter $\alpha_{\Pom}'$, we have
the default values from Ref.~\cite{Ewerz:2013kda}: 
$\epsilon_{\Pom} = 0.0808$,
$\alpha_{\Pom}' = 0.25$~GeV$^{-2}$.
The $\Pom$ and $\Reg$ intercept parameters
have been determined by Donnachie and Landshoff 
from a simultaneous fit to the $pp$ and $p \bar{p}$ scattering 
data for $\sqrt{s} > 10$~GeV \cite{Donnachie:1992ny}
and should be regarded rather as ``effective'' parameters.
The parameter $\alpha_{\Pom}' = 0.25$~GeV$^{-2}$
has been determined from the $pp$ elastic scattering data
\cite{Donnachie:1983hf} 
(see also Ref.~\cite{Donnachie:2002en}) and
is in good agreement with experimental findings
on the exclusive $\rho^{0}$ photoproduction 
\cite{ZEUS:1997rof,H1:2020lzc}.
The value for the coupling constant 
of the pomeron to protons is
$\beta_{\Pom pp} = 1.87\;{\rm GeV}^{-1}$;
see (3.44) and Sec.~6.3 of Ref.~\cite{Ewerz:2013kda}.
For simplicity,
in (\ref{2.7})--(\ref{2.12}),
the Dirac electromagnetic form factor of the proton is used. 
It is clear that this cannot
be strictly correct;
see the discussion in Chap.~3.2 of Ref.~\cite{Donnachie:2002en}
and in Ref.~\cite{Zhou:2006gj}.
It will be shown in Sec.~\ref{sec:3A} 
that to effectively describe 
the TOTEM data for the elastic $pp$ cross section 
($d\sigma/dt$) at $\sqrt{s} = 13$~TeV, 
taking into account only the leading $\Pom$ contribution,
we must change in Eq.~(\ref{2.7}) both
the $\epsilon_{\Pom}$ parameter 
and the pomeron-proton form factor $F_{1}(t)$.
%%%%%%%%%%%%%%%%
\item
%%%%%%%%%%%%%%%%
For the odderon exchange,
we will also consider an alternative Ansatz
corresponding to a double Regge pole (\ref{A9}).
Then, we have
\begin{eqnarray}
{\cal F}_{\Ode pp}(s,t) \to \widetilde{\cal F}_{\Ode pp}(s,t)
&=& {\cal F}_{\Ode pp}(s,t)
\left[
C_{1} + C_{2} \ln \left( -i s \alpha_{\Ode}' \right)
\right]\,,
\label{2.10_new}
\end{eqnarray}
where $C_{1}$ and $C_{2}$ are real constants.
In (\ref{2.10}) and (\ref{2.10_new}),
the factor $\eta_{\Ode} = \pm 1$ and 
the odderon trajectory function are unknown.
We assume
$\alpha_{\Ode}(t) = 1 + \epsilon_{\Ode} + \alpha_{\Ode}'\, t$
with the parameters $\epsilon_{\Ode}$ and $\alpha_{\Ode}'$
that should be determined by experiment.
Of course, the pomeron amplitude must dominate over the odderon
one for $s \to \infty$ in order to ensure positive
total $pp$ and $p \bar{p}$ cross sections.
Thus, we must require $\epsilon_{\Ode} \leqslant \epsilon_{\Pom}$.
For our study here, we assume
$\beta_{\Ode pp} = 0.1 \times \beta_{\Pom pp} \simeq 0.2\;{\rm GeV}^{-1}$,
$\alpha_{\Ode}' = \alpha_{\Pom}'$, $\epsilon_{\Ode} = 0.0800$,
$\eta_{\Ode} = -1$, and $(C_{1}, C_{2})$ as listed in (\ref{A10}).
How these parameters are determined from comparisons between
theory and experiment will be discussed in Sec.~\ref{sec:3A}.
%%%%%%%%%%%%%%%%
\end{itemize}
%%%%%%%%%%%%%%%%

In the following, we use tensor-product notation.
The first factors will always refer to the $p_{a}$-$p_{1}$ line,
and the second refer to the $p_{b}$-$p_{2}$ line of the diagrams
shown in Fig.~\ref{fig:pp_pp}.

In our model, the off-shell proton-proton 
scattering amplitude has the form
\begin{eqnarray}
&&{\cal M}^{(0)}(p_{a}, p_{b}, p_{1}, p_{2})
= {\cal M}^{(0)}_{\Pom} + {\cal M}^{(0)}_{f_{2 \Reg}} + {\cal M}^{(0)}_{a_{2 \Reg}}
+ {\cal M}^{(0)}_{\Ode} + {\cal M}^{(0)}_{\omega_{\Reg}} + {\cal M}^{(0)}_{\rho_{\Reg}}
\nonumber\\
&& = i {\cal F}_{T}(s,t)
\big[ \gamma^{\mu} \otimes \gamma_{\mu} (p_{a} + p_{1}, p_{b} + p_{2})
+ (\slash{p}_{b} + \slash{p}_{2})\otimes(\slash{p}_{a} + \slash{p}_{1})
- \frac{1}{2}(\slash{p}_{a} + \slash{p}_{1})\otimes(\slash{p}_{b} + \slash{p}_{2})
\big]
\nonumber\\
&& \quad  - 
{\cal F}_{V}(s,t)\,
\gamma^{\mu} \otimes \gamma_{\mu}\,.
\label{2.13}
\end{eqnarray}

With the helicities $\lambda_{a}$, $\lambda_{b}$, 
$\lambda_{1}$, $\lambda_{2} \in \lbrace -1/2, 1/2 \rbrace$,
we get for the on-shell matrix element
\begin{eqnarray}
&&\braket{p(p_{1},\lambda_{1}),p(p_{2},\lambda_{2})|{\cal T}|p(p_{a},\lambda_{a}),p(p_{b},\lambda_{b})}\nonumber\\
&& \equiv {\cal M}^{({\rm on\; shell}) \,pp}(s,t) 
\nonumber\\
&& = 
\bar{u}_{1} \otimes \bar{u}_{2}
{\cal M}^{(0)}(p_{a}, p_{b}, p_{1}, p_{2})
\left. u_{a} \otimes u_{b}\right|_{\rm on\; shell}
\nonumber\\
&& = 
i {\cal F}_{T}(s,t)
\big[ 
\bar{u}_{1} \gamma^{\mu} u_{a} \,
\bar{u}_{2} \gamma_{\mu} u_{b} \,
(p_{a} + p_{1}, p_{b} + p_{2})
+ \bar{u}_{1} \gamma^{\mu} u_{a} (p_{b}+p_{2})_{\mu}\,
  \bar{u}_{2} \gamma^{\nu} u_{b} (p_{a}+p_{1})_{\nu} 
\nonumber\\
&&
\quad - 2 m_{p}^{2} \,
\bar{u}_{1} u_{a} \, \bar{u}_{2} u_{b}
\big]
- {\cal F}_{V}(s,t)\,
\bar{u}_{1} \gamma^{\mu} u_{a} \,
\bar{u}_{2} \gamma_{\mu} u_{b}\,.
\label{2.14}
\end{eqnarray}
For brevity of notation, we define
here and in the following for the spinors 
$u_{a} = u(p_{a},\lambda_{a})$,
$\bar{u}_{1} = \bar{u}(p_{1},\lambda_{1})$, etc.
We also denote
${\cal M}^{({\rm on\; shell}) \,pp}(s,t)$ as the on-shell
$pp$ elastic scattering amplitude.

We consider now the high-energy 
small-angle limit where we have the simple relations
\begin{eqnarray}
&&\bar{u}(p_{1}, \lambda_{1}) \gamma^{\mu} u(p_{a}, \lambda_{a}) \cong  
(p_{a} + p_{1})^{\mu} \delta_{\lambda_{1} \lambda_{a}}\,, 
\nonumber \\
&& (p_{a} + p_{1}, p_{b} + p_{2}) \cong 2 s \,.
\label{HESA_limit}
\end{eqnarray}
From (\ref{2.14}) we get, for $s \to \infty$,
setting a possible odderon contribution to zero,
the pomeron contribution as leading term in the form
\begin{eqnarray}
{\cal M}^{({\rm on\; shell})\, pp}(s,t)
\to
i 8 s^{2} {\cal F}_{\Pom pp}(s,t) 
\delta_{\lambda_{1}\lambda_{a}} 
\delta_{\lambda_{2}\lambda_{b}}\,.
\label{2.14_high_energy}
\end{eqnarray}
In this high-energy small-angle limit the amplitude
calculated from the tensor-pomeron exchange
is the same as the standard Donnachie-Landshoff (DL) amplitude;
see the discussion in Chap.~6.1 of Ref.~\cite{Ewerz:2013kda}.

Coming back to the general case, we emphasize that in
the following we use the exact formulas
(\ref{2.13}) and (\ref{2.14}), without the approximations (\ref{HESA_limit}).

The total cross section for unpolarized protons, 
obtained from the forward-scattering amplitudes 
using the optical theorem, is
\begin{eqnarray}
&&\sigma_{\rm tot}(pp) =
\frac{1}{\sqrt{s(s-4 m_{p}^{2})}}\,
\frac{1}{4} \sum_{\lambda_{a}, \lambda_{b}} 
{\rm Im} \braket{p(p_{a},\lambda_{a}),p(p_{b},\lambda_{b})|{\cal T}|p(p_{a},\lambda_{a}),p(p_{b},\lambda_{b})}
\nonumber \\
&& \quad = \frac{1}{\sqrt{s(s-4 m_{p}^{2})}}
\big\lbrace 
{\rm Re}
{\cal F}_{T}(s,0)\,
8\big[ (s - 2 m_{p}^{2} )^{2} - m_{p}^{4} \big]
-
{\rm Im}
{\cal F}_{V}(s,0)\,
2 (s - 2 m_{p}^{2})
\big\rbrace \,. \quad
\label{2.15}
\end{eqnarray}
With (\ref{2.7})--(\ref{2.10_new}), this gives
\begin{eqnarray}
&&\sigma_{\rm tot}(pp) =
2 \big( 1-\frac{4m_{p}^{2}}{s} \big)^{-1/2}
\big\lbrace 
\big[
\big( 3 \beta_{\Pom pp} \big)^{2} 
(s \alpha'_{\Pom})^{\alpha_{\Pom}(0)-1}
\cos\big( \frac{\pi}{2} (\alpha_{\Pom}(0)-1) \big)
\nonumber \\
&& \qquad 
+
\big( \frac{g_{f_{2 \Reg} pp}}{M_{0}} \big)^{2} 
(s \alpha'_{f_{2 \Reg}})^{\alpha_{f_{2 \Reg}}(0)-1}
\cos\big( \frac{\pi}{2} (\alpha_{f_{2 \Reg}}(0)-1) \big) 
\nonumber \\
&& \qquad +
\big( \frac{g_{a_{2 \Reg} pp}}{M_{0}} \big)^{2} 
(s \alpha'_{a_{2 \Reg}})^{\alpha_{a_{2 \Reg}}(0)-1}
\cos\big( \frac{\pi}{2} (\alpha_{a_{2 \Reg}}(0)-1) \big)
\big]
\big( 1-\frac{4m_{p}^{2}}{s} +\frac{3m_{p}^{4}}{s^{2}} \big)
\nonumber \\
&& \qquad -
\big[
-\eta_{\Ode}
\big( 3 \beta_{\Ode pp} \big)^{2} 
(s \alpha'_{\Ode})^{\alpha_{\Ode}(0)-1}
\big[ \cos\big( \frac{\pi}{2} \alpha_{\Ode}(0) \big)
\big(C_{1} + C_{2} \ln \left(s \alpha_{\Ode}'\right) \big)
-C_{2} \frac{\pi}{2} \sin\big( \frac{\pi}{2} \alpha_{\Ode}(0) \big)
\big]
\nonumber \\
&& \qquad +
\big( \frac{g_{\omega_{\Reg} pp}}{ M_{-}} \big)^{2} 
(s \alpha'_{\omega_{\Reg}})^{\alpha_{\omega_{\Reg}}(0)-1}
\cos\big( \frac{\pi}{2} \alpha_{\omega_{\Reg}}(0) \big)
\nonumber \\
&& \qquad +
\big( \frac{g_{\rho_{\Reg} pp}}{ M_{-}} \big)^{2} 
(s \alpha'_{\rho_{\Reg}})^{\alpha_{\rho_{\Reg}}(0)-1}
\cos\big( \frac{\pi}{2} \alpha_{\rho_{\Reg}}(0) \big)
\big]
\big( 1-\frac{2m_{p}^{2}}{s}\big)
\big\rbrace
\,.
\label{2.16}
\end{eqnarray}
Here, we have used the expression (\ref{2.10_new})
inserted into (\ref{2aux}).
To get the expression with (\ref{2.10})
for the odderon exchange, we have to set $C_{1} = 1$
and $C_{2} = 0$.

\subsection{Reaction $p\bar{p} \to p\bar{p}$}
\label{sec:2B}

\begin{figure}[!h]
\includegraphics[width=6.5cm]{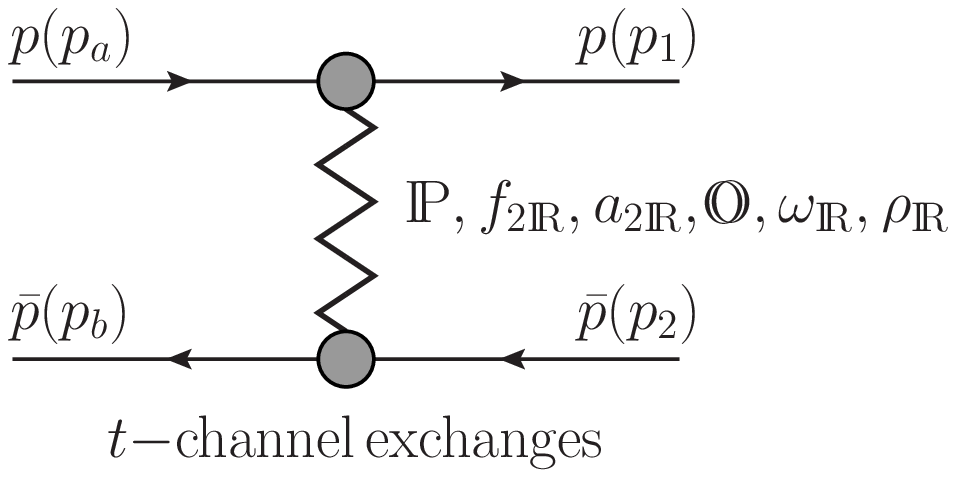}
\quad
\includegraphics[width=6.5cm]{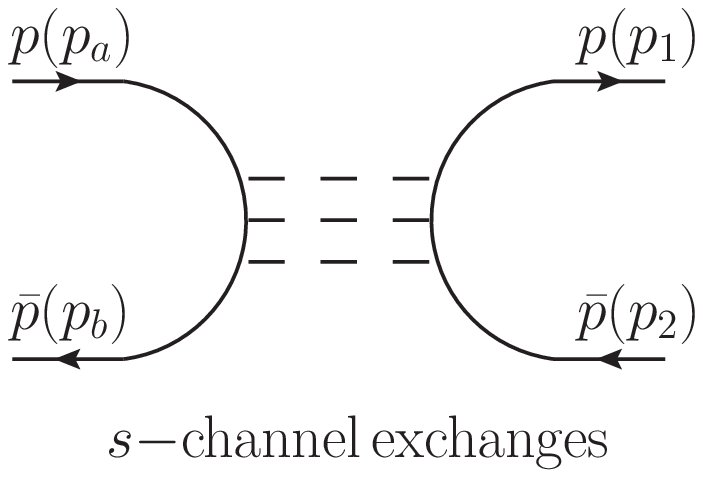}
\caption{The diagrams for $p \bar{p} \to p \bar{p}$ 
elastic scattering with the $t$- and $s$-channel exchanges.}
\label{fig:ppbar_ppbar}
\end{figure}

Here we study the reaction (\ref{2.1b}),
where $p$ and $\bar{p}$ can be on or off shell.
In our model, considering only hadronic exchanges, 
we have the diagrams 
shown in Fig.~\ref{fig:ppbar_ppbar}.
For high c.m. energies and small momentum transfers (\ref{2.5}), 
the $s$-channel exchanges should be negligible.
%-------------------------------------------------------------
%-------------------------------------------------------------

Now we get for the $p \bar{p}$ off-shell amplitude
\begin{eqnarray}
&&{\cal M}^{(0)}_{p\bar{p}}(p_{a}, p_{b}, p_{1}, p_{2})
%= {\cal M}^{(0)}_{\Pom} + {\cal M}^{(0)}_{f_{2 \Reg}} + {\cal M}^{(0)}_{a_{2 \Reg}}
%+ {\cal M}^{(0)}_{\Ode} + {\cal M}^{(0)}_{\omega_{\Reg}} + {\cal M}^{(0)}_{\rho_{\Reg}}
\nonumber\\
&& = i {\cal F}_{T}(s,t)
\big[ \gamma^{\mu} \otimes \gamma_{\mu} (p_{a} + p_{1}, p_{b} + p_{2})
+ (\slash{p}_{b} + \slash{p}_{2})\otimes(\slash{p}_{a} + \slash{p}_{1})
- \frac{1}{2}(\slash{p}_{a} + \slash{p}_{1})\otimes(\slash{p}_{b} + \slash{p}_{2})
\big]
\nonumber\\
&& \quad  + 
{\cal F}_{V}(s,t)\,
\gamma^{\mu} \otimes \gamma_{\mu}\,.
\label{16c}
\end{eqnarray}
For the on-shell amplitude, we get
\begin{eqnarray}
&&\braket{p(p_{1},\lambda_{1}),\bar{p}(p_{2},\lambda_{2})|{\cal T}|p(p_{a},\lambda_{a}),\bar{p}(p_{b},\lambda_{b})} =
\bar{u}_{1} \otimes \bar{v}_{b}
{\cal M}^{(0)}_{p\bar{p}}(p_{a}, p_{b}, p_{1}, p_{2})
u_{a} \otimes v_{2}
\nonumber\\
&& \quad =
i {\cal F}_{T}(s,t) 
\big[ 
\bar{u}_{1} \gamma^{\mu} u_{a} \,
\bar{v}_{b} \gamma_{\mu} v_{2} \,
(p_{a} + p_{1}, p_{b} + p_{2})
+ \bar{u}_{1} \gamma^{\mu} u_{a} (p_{b}+p_{2})_{\mu}\,
  \bar{v}_{b} \gamma^{\nu} v_{2} (p_{a}+p_{1})_{\nu} 
\nonumber\\
&&\quad
+ 2 m_{p}^{2} \,
\bar{u}_{1} u_{a} \, \bar{v}_{b} v_{2}
\big]
+ {\cal F}_{V}(s,t)\,
\bar{u}_{1} \gamma^{\mu} u_{a} \,
\bar{v}_{b} \gamma_{\mu} v_{2}\,.
\label{16d}
\end{eqnarray}
The total $p \bar{p}$ cross section is
\begin{eqnarray}
&&\sigma_{\rm tot}(p \bar{p}) =
\frac{1}{\sqrt{s(s-4 m_{p}^{2})}}\,
\frac{1}{4} \sum_{\lambda_{a}, \lambda_{b}} 
{\rm Im} \braket{p(p_{a},\lambda_{a}),\bar{p}(p_{b},\lambda_{b})|{\cal T}|p(p_{a},\lambda_{a}),\bar{p}(p_{b},\lambda_{b})}
\nonumber \\
&& \quad = \frac{1}{\sqrt{s(s-4 m_{p}^{2})}}
\big\lbrace 
{\rm Re}
{\cal F}_{T}(s,0)\,
8 \big[ (s - 2 m_{p}^{2} )^{2} - m_{p}^{4} \big]+
{\rm Im}
{\cal F}_{V}(s,0)
2 (s - 2 m_{p}^{2})
\big\rbrace \,. \quad
\label{16e}
\end{eqnarray}
With (\ref{2.7})--(\ref{2.10_new}), this gives 
\begin{eqnarray}
&&\sigma_{\rm tot}(p \bar{p}) =
2 \big( 1-\frac{4m_{p}^{2}}{s} \big)^{-1/2}
\big\lbrace 
\big[
\big( 3 \beta_{\Pom pp} \big)^{2} 
(s \alpha'_{\Pom})^{\alpha_{\Pom}(0)-1}
\cos\big( \frac{\pi}{2} (\alpha_{\Pom}(0)-1) \big)
\nonumber \\
&& \qquad 
+
\big( \frac{g_{f_{2 \Reg} pp}}{M_{0}} \big)^{2} 
(s \alpha'_{f_{2 \Reg}})^{\alpha_{f_{2 \Reg}}(0)-1}
\cos\big( \frac{\pi}{2} (\alpha_{f_{2 \Reg}}(0)-1) \big) 
\nonumber \\
&& \qquad +
\big( \frac{g_{a_{2 \Reg} pp}}{M_{0}} \big)^{2} 
(s \alpha'_{a_{2 \Reg}})^{\alpha_{a_{2 \Reg}}(0)-1}
\cos\big( \frac{\pi}{2} (\alpha_{a_{2 \Reg}}(0)-1) \big)
\big]
\big( 1-\frac{4m_{p}^{2}}{s} +\frac{3m_{p}^{4}}{s^{2}} \big)
\nonumber \\
&& \qquad +
\big[
-\eta_{\Ode}
\big( 3 \beta_{\Ode pp} \big)^{2} 
(s \alpha'_{\Ode})^{\alpha_{\Ode}(0)-1}
\big[ \cos\big( \frac{\pi}{2} \alpha_{\Ode}(0) \big)
\big(C_{1} + C_{2} \ln \left(s \alpha_{\Ode}'\right) \big)
-C_{2} \frac{\pi}{2} \sin\big( \frac{\pi}{2} \alpha_{\Ode}(0) \big)
\big]
\nonumber \\
&& \qquad +
\big( \frac{g_{\omega_{\Reg} pp}}{ M_{-}} \big)^{2} 
(s \alpha'_{\omega_{\Reg}})^{\alpha_{\omega_{\Reg}}(0)-1}
\cos\big( \frac{\pi}{2} \alpha_{\omega_{\Reg}}(0) \big)
\nonumber \\
&& \qquad +
\big( \frac{g_{\rho_{\Reg} pp}}{ M_{-}} \big)^{2} 
(s \alpha'_{\rho_{\Reg}})^{\alpha_{\rho_{\Reg}}(0)-1}
\cos\big( \frac{\pi}{2} \alpha_{\rho_{\Reg}}(0) \big)
\big]
\big( 1-\frac{2m_{p}^{2}}{s}\big)
\big\rbrace
\,.
\label{16f}
\end{eqnarray}

Comparing (\ref{2.13}) with (\ref{16c}),
(\ref{2.14}) with (\ref{16d}), and
(\ref{2.16}) with (\ref{16f}),
we see that for $pp$ and $p \bar{p}$ scattering
the $C = +1$ exchanges contribute with the same sign
and 
the $C = -1$ exchanges contribute with opposite sign,
as it should be.

\subsection{Reaction $pp \to pp \gamma$}
\label{sec:2C}

Now, we consider the reaction (\ref{2.2}).
Here, we have the energy-momentum relation (\ref{2.4a}).
The kinematic variables are
\begin{eqnarray}
&&s = (p_{a} + p_{b})^{2} = (p_{1}' + p_{2}' + k)^{2}\,, \nonumber \\
&&s' = (p_{a} + p_{b} - k)^{2} = (p_{1}' + p_{2}')^{2}\,, \nonumber \\
&&t_{1} = (p_{a} - p_{1}')^{2} = (p_{b} - p_{2}' - k)^{2}\,, \nonumber \\
&&t_{2} = (p_{b} - p_{2}')^{2} = (p_{a} - p_{1}' - k)^{2}\,.
\label{2.17}
\end{eqnarray}
%

%-------------------------------------------------------------
\begin{figure}[!h]
(a)\includegraphics[width=4.6cm]{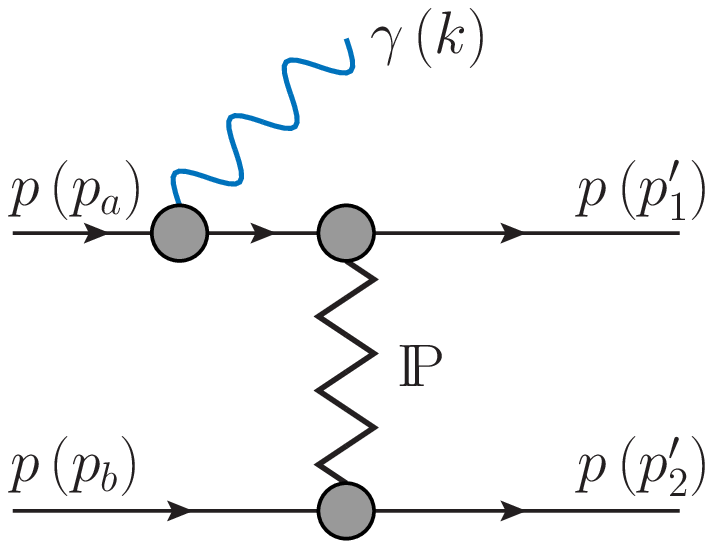}\quad
(b)\includegraphics[width=5.3cm]{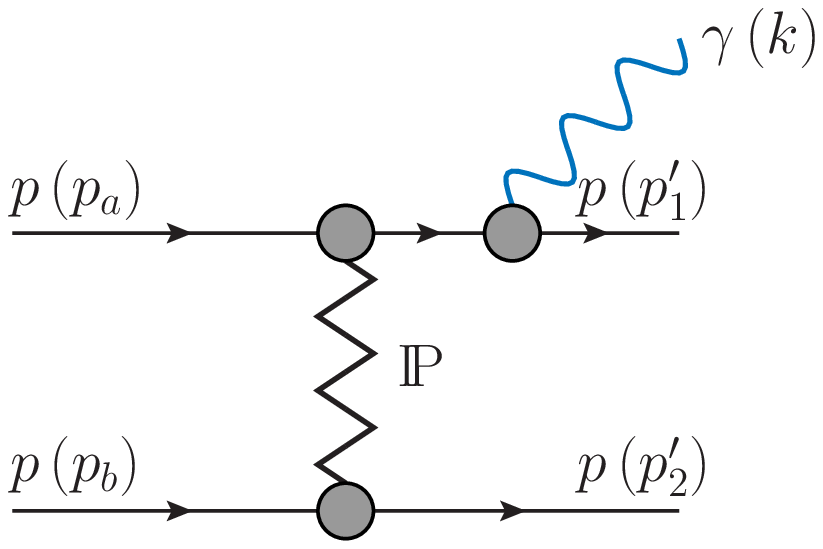}
(c)\includegraphics[width=4.5cm]{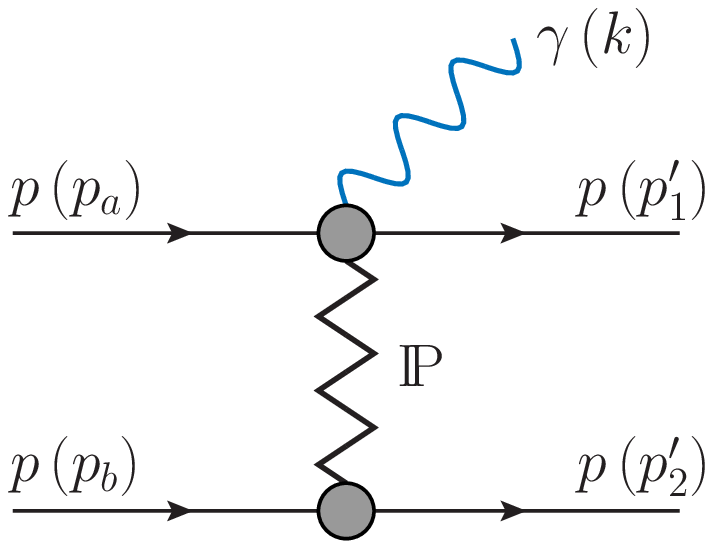}
(d)\includegraphics[width=4.6cm]{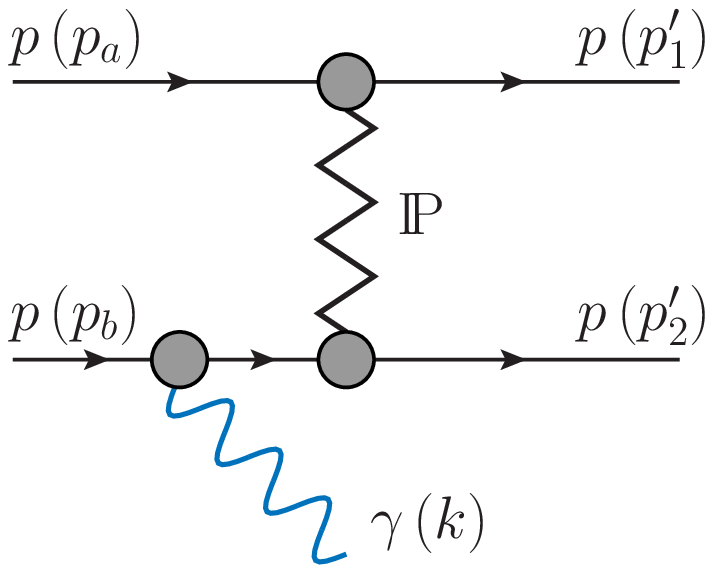}\quad
(e)\includegraphics[width=5.3cm]{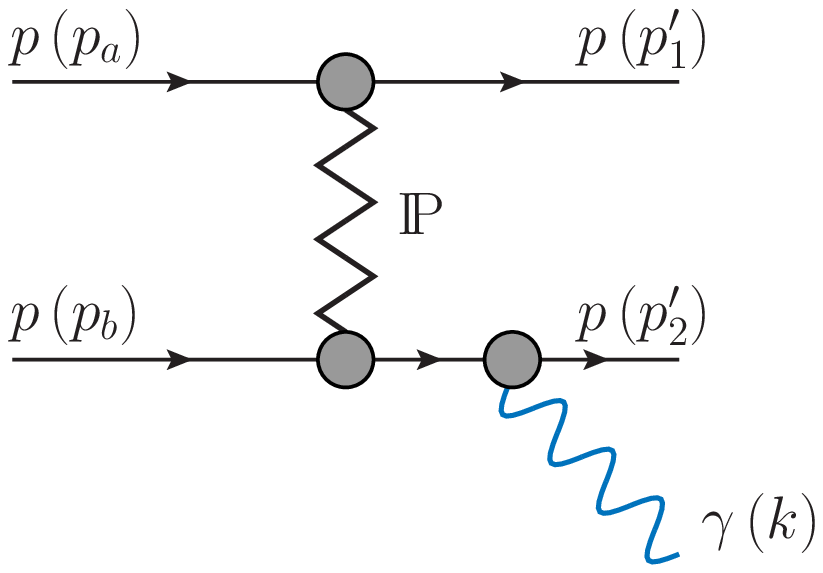}
(f)\includegraphics[width=4.5cm]{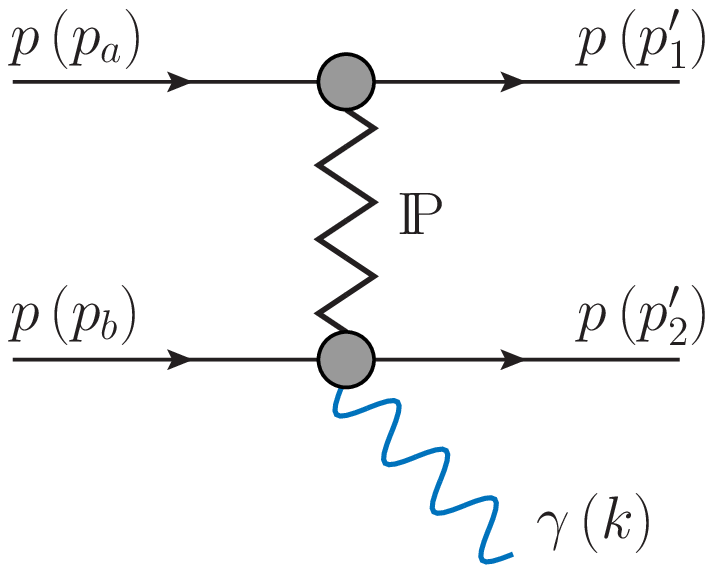}\\
(g)\includegraphics[width=5.8cm]{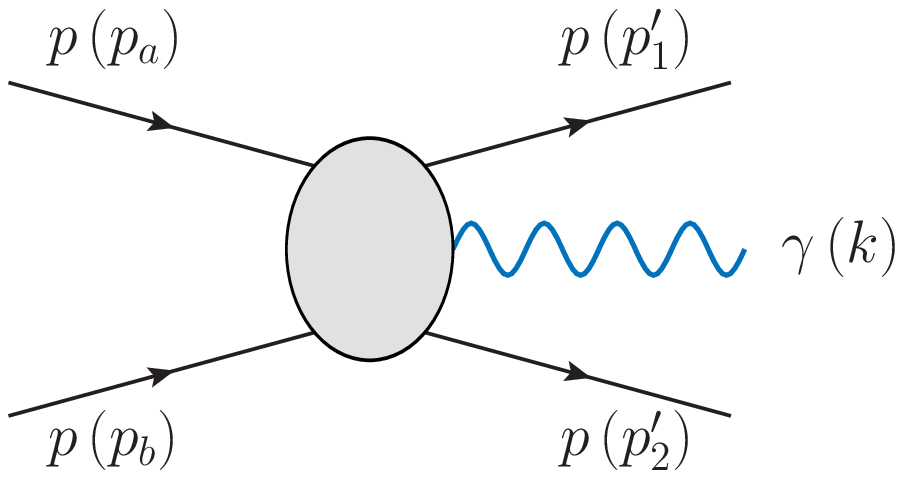}
\caption{Diagrams for the reaction $pp \to pp \gamma$
with exchange of the pomeron $\Pom$ (a--f)
and the ``structure'' term (g).
The diagrams for odderon and reggeon exchanges are 
as in (a--f)
with $\Pom$ replaced by $f_{2 \Reg}$, $a_{2 \Reg}$,
$\Ode$, $\omega_{\Reg}$, and $\rho_{\Reg}$.
In addition there are the diagrams 
corresponding to the interchange 
of the two final protons $p(p_{1}') \leftrightarrow p(p_{2}')$.
These are not shown here.}
\label{fig:pp_pp_gam}
\end{figure}
%-------------------------------------------------------------
For the photon emission process (\ref{2.2}),
we have seven types of diagrams shown in Fig.~\ref{fig:pp_pp_gam}.
In the diagrams (a), (b), (d), and (e),
the photon is emitted from the external proton lines.
The diagrams (c) and (f) correspond to contact terms.
The precise definition of the contact terms 
and of the ``structure'' term (g) will be given below.
For high c.m. energies $\sqrt{s}$ 
and small momentum transfers $|t_{1,2}|$
[see (\ref{2.17})],
the diagrams of Fig.~\ref{fig:pp_pp_gam} 
with $p_{1}' \leftrightarrow p_{2}'$
are expected to give negligible contributions.
We shall call the diagrams (a), (b), (d), and (e),
made gauge invariant by the addition of (c) and (f),
the bremsstrahlung diagrams.

The relevant ${\cal T}$-matrix element is
\begin{eqnarray}
&&\braket{p(p_{1}',\lambda_{1}),p(p_{2}',\lambda_{2}),
\gamma(k, \epsilon)|{\cal T}|p(p_{a},\lambda_{a}),p(p_{b},\lambda_{b})}  \nonumber\\
&&= 
(\epsilon^{\mu})^{*}
{\cal M}^{(\rm total)}_{\mu}(p_{a},\lambda_{a}; p_{b},\lambda_{b};
p_{1}',\lambda_{1};p_{2}',\lambda_{2};k)\,.
\label{2.18}
\end{eqnarray}
We consider ${\cal M}^{(\rm total)}_{\mu}$ for arbitrary $k$.
Gauge invariance requires
\begin{equation}
k^{\mu} {\cal M}^{(\rm total)}_{\mu} = 0\,.
\label{2.19}
\end{equation}
Let the sum of the diagrams of Fig.~\ref{fig:pp_pp_gam}
with $\Pom$ and all other exchanges be
\begin{equation}
{\cal M}_{\mu}(p_{1}',p_{2}') = {\cal M}_{\mu}^{(a)} + {\cal M}_{\mu}^{(b)}
+ {\cal M}_{\mu}^{(c)} + {\cal M}_{\mu}^{(d)} 
+ {\cal M}_{\mu}^{(e)} + {\cal M}_{\mu}^{(f)} + {\cal M}_{\mu}^{(g)}\,.
\label{2.20}
\end{equation}
The complete amplitude is then
\begin{equation}
{\cal M}^{(\rm total)}_{\mu} = {\cal M}_{\mu}(p_{1}',p_{2}')
- {\cal M}_{\mu}(p_{2}',p_{1}')\,,
\label{2.20a}
\end{equation}
where a corresponding exchange of helicities is understood.
The relative minus sign here
is due to the Fermi statistics, 
which requires the amplitude to be antisymmetric
under interchange of the two final protons.

The inclusive cross section for the real-photon yield of
the reaction (\ref{2.2}),
including a statistics factor 1/2
due to identical particles appearing in the final state,
is as follows
\begin{eqnarray}
&&d\sigma({pp \to pp \gamma}) =
\frac{1}{2}\frac{1}{2\sqrt{s(s-4 m_{p}^{2})}}\,
\frac{d^{3}k}{(2 \pi)^{3} \,2 k^{0}}
\int \frac{d^{3}p_{1}'}{(2 \pi)^{3} \,2 p_{1}'^{0}}
\frac{d^{3}p_{2}'}{(2 \pi)^{3} \,2 p_{2}'^{0}} 
\nonumber \\
&&\quad \times (2 \pi)^{4} \delta^{(4)}(p_{1}'+p_{2}'+k-p_{a}-p_{b})
\frac{1}{4}\sum_{p \, \rm spins}
{\cal M}_{\lambda}^{(\rm total)}
\big( {\cal M}_{\rho}^{(\rm total)} \big)^{*} 
(-g^{\lambda \rho})\,.
\label{incl_xs_2to3}
\end{eqnarray}

Let, in the c.m. system, $p_{1z}'$ and $p_{2z}'$
be the momentum components of $\bpaap$ and $\bpbbp$
in the direction of $\bpa$.
Then we can write (\ref{incl_xs_2to3}) with (\ref{2.20a}) as
\begin{eqnarray}
&&d\sigma({pp \to pp \gamma}) =
\frac{1}{2\sqrt{s(s-4 m_{p}^{2})}}\,
\frac{d^{3}k}{(2 \pi)^{3} \,2 k^{0}}
\int \frac{d^{3}p_{1}'}{(2 \pi)^{3} \,2 p_{1}'^{0}}
\frac{d^{3}p_{2}'}{(2 \pi)^{3} \,2 p_{2}'^{0}}\,\theta(p_{1z}'-p_{2z}')
\nonumber \\
&&\quad \times 
(2 \pi)^{4} \delta^{(4)}(p_{1}'+p_{2}'+k-p_{a}-p_{b})
\nonumber \\
&&\quad \times \frac{1}{4}
\sum_{p \, \rm spins}
\big( {\cal M}_{\lambda}(p_{1}',p_{2}')
- {\cal M}_{\lambda}(p_{2}',p_{1}') \big)
\big( {\cal M}_{\rho}(p_{1}',p_{2}')
- {\cal M}_{\rho}(p_{2}',p_{1}') \big)^{*} (-g^{\lambda \rho})\,. \qquad
\label{incl_xs_2to3_aux1}
\end{eqnarray}
We are interested in high c.m. energies $\sqrt{s}$ and 
small momentum transfers.
Then, for $p_{1z}' > p_{2z}'$, the amplitude 
${\cal M}_{\lambda}(p_{2}',p_{1}')$
is very small and can be neglected.
On the other hand, ${\cal M}_{\lambda}(p_{1}',p_{2}')$
is very small for $p_{1z}' < p_{2z}'$.
%For diffractive scattering the ${\cal M}_{\mu}(p_{2}',p_{1}')$
%(back scattering) is expected to correspond to very small cross
%section, and thus is of no practical interest
Therefore, from (\ref{incl_xs_2to3_aux1}),
we get with very high accuracy 
\begin{eqnarray}
&&d\sigma({pp \to pp \gamma}) =
\frac{1}{2\sqrt{s(s-4 m_{p}^{2})}}\,
\frac{d^{3}k}{(2 \pi)^{3} \,2 k^{0}}
\int \frac{d^{3}p_{1}'}{(2 \pi)^{3} \,2 p_{1}'^{0}}
\frac{d^{3}p_{2}'}{(2 \pi)^{3} \,2 p_{2}'^{0}}
\nonumber \\
&&\quad \times 
(2 \pi)^{4} \delta^{(4)}(p_{1}'+p_{2}'+k-p_{a}-p_{b})
\frac{1}{4}
\sum_{p \, \rm spins}
{\cal M}_{\lambda}(p_{1}',p_{2}')
\big( {\cal M}_{\rho}(p_{1}',p_{2}')\big)^{*} (-g^{\lambda \rho})\,. \qquad \;
\label{incl_xs_2to3_aux2}
\end{eqnarray}
In the following, we shall, for brevity of notation, set
${\cal M}_{\lambda} \equiv {\cal M}_{\lambda}(p_{1}',p_{2}')$.

For calculating ${\cal M}_{\lambda}$ 
from the diagrams of Fig.~\ref{fig:pp_pp_gam},
we use the following standard proton propagator 
and $\gamma pp$ vertex:\\
%\hspace*{2.0cm}
\newline
\includegraphics[width=90pt]{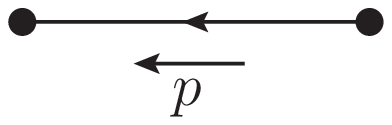} 
\vspace*{-1.9cm}
\begin{eqnarray}
iS_{F}(p) 
= \frac{i}{\slash{p} - m_{p} + i \epsilon}
=i \frac{\slash{p} + m_{p}}{p^{2} - m_{p}^{2} + i \epsilon}\,,
\label{2.21}
\end{eqnarray}
\includegraphics[width=150pt]{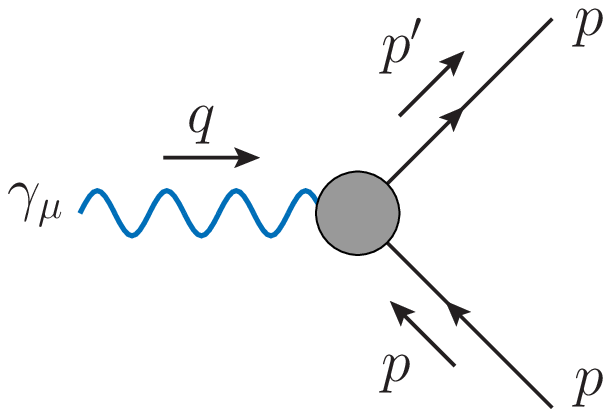} 
\vspace*{-3.5cm}
\begin{eqnarray} 
&&\qquad \qquad \qquad \qquad 
q = p' - p\,,
\nonumber \\
&&\qquad \qquad \qquad \qquad i \Gamma_{\mu}^{(\gamma pp)}(p',p) = -ie \big[ \gamma_{\mu} + 
\frac{i}{2 m_{p}} \sigma_{\mu \nu} q^{\nu} F_{2}(0)\big]\,, \nonumber \\
&&\qquad \qquad \qquad \qquad F_{2}(0) = \big( \frac{\mu_{p}}{\mu_{N}} - 1 \big),\;
\mu_{N} = \frac{e}{2 m_{p}},\; \frac{\mu_{p}}{\mu_{N}} = 2.7928\,,
\nonumber \\
&&\qquad \qquad \qquad \qquad 
e > 0,\; e = \sqrt{4 \pi \alpha_{\rm em}}\,.
\label{2.22}
\end{eqnarray}
%
%where $q = p' - p$.
We take the form factors in (\ref{2.22}) at $q^{2} = 0$
in order to be consistent with the Ward-Takahashi identity
\cite{Ward:1950xp,Takahashi:1957xn}:
\begin{equation}
(p'-p)^{\mu} \Gamma_{\mu}^{(\gamma pp)}(p',p) =
-e \big[ S_{F}^{-1}(p') - S_{F}^{-1}(p) \big]\,.
\label{2.25}
\end{equation}
In any case, we are finally interested in real photon emission where
\mbox{$k = -q$, $k^{2} = q^{2} = 0$}.

Now, we list our results for the photon-bremsstrahlung amplitudes
${\cal M}_{\mu}^{(a)}, \ldots, {\cal M}_{\mu}^{(f)}$
corresponding to the pomeron-exchange diagrams (a)--(f) 
from Fig.~\ref{fig:pp_pp_gam}
including the other exchanges
$f_{2 \Reg}$, $a_{2 \Reg}$, 
$\omega_{\Reg}$, $\rho_{\Reg}$, $\Ode$.

With the off-shell $pp$ scattering amplitude (\ref{2.13}),
the standard proton propagator (\ref{2.21}), 
and the $\gamma pp$ vertex function (\ref{2.22}),
we get the following radiative amplitudes:
\begin{eqnarray}
{\cal M}_{\mu}^{(a)}&=&
-
\bar{u}_{1'} \otimes \bar{u}_{2'} \,
{\cal M}^{(0)}(p_{a}-k,p_{b},p_{1}',p_{2}')\,
\big(
S_{F}(p_{a}-k)
\Gamma_{\mu}^{(\gamma pp)}(p_{a}-k,p_{a})\,
u_{a} \big) \otimes u_{b} \nonumber\\
&=&
e \bar{u}_{1'} \otimes \bar{u}_{2'}
\big\lbrace
i {\cal F}_{T}(s',t_{2}) 
\big[ 
\gamma^{\alpha} \otimes \gamma_{\alpha}
(p_{a} - k + p_{1}', p_{b} + p_{2}')
+ (\slash{p}_{b}+\slash{p}_{2}') \otimes (\slash{p}_{a} - \slash{k} +\slash{p}_{1}')
\nonumber\\
&&
- \frac{1}{2} (\slash{p}_{a} - \slash{k} +\slash{p}_{1}') \otimes (\slash{p}_{b}+\slash{p}_{2}')
\big] 
- {\cal F}_{V}(s',t_{2})\,
\gamma^{\alpha} \otimes \gamma_{\alpha} \big\rbrace
\nonumber\\
&& \times  
\big[
\frac{\slash{p}_{a} - \slash{k} + m_{p}}{(p_{a}-k)^{2}-m_{p}^{2}+i \varepsilon} 
\big(\gamma_{\mu} - \frac{i}{2 m_{p}} \sigma_{\mu \nu} k^{\nu} F_{2}(0) \big)u_{a} 
\big] 
\otimes
u_{b}
\,,
\label{2.26} \\
{\cal M}_{\mu}^{(b)} &=&
-
\big( \bar{u}_{1'} 
\Gamma_{\mu}^{(\gamma pp)}(p_{1}',p_{1}'+k)
S_{F}(p_{1}'+k) \big)
\otimes \bar{u}_{2'} \,
{\cal M}^{(0)}(p_{a},p_{b},p_{1}'+k,p_{2}')\,
u_{a} \otimes u_{b} \nonumber\\
&=&
e \big[\bar{u}_{1'} 
\big(\gamma_{\mu} - \frac{i}{2 m_{p}} \sigma_{\mu \nu} k^{\nu} F_{2}(0) \big)
\frac{\slash{p}_{1}' + \slash{k} + m_{p}}{(p_{1}'+k)^{2}-m_{p}^{2}+i \varepsilon} 
\big] \otimes \bar{u}_{2'}\nonumber\\
&& \times  
\big\lbrace
i {\cal F}_{T}(s,t_{2}) 
\big[ 
\gamma^{\alpha} \otimes \gamma_{\alpha}
(p_{a} + p_{1}'+ k, p_{b} + p_{2}')
+ (\slash{p}_{b} + \slash{p}_{2}') \otimes (\slash{p}_{a} + \slash{p}_{1}' + \slash{k})
\nonumber\\
&& 
- \frac{1}{2} (\slash{p}_{a} + \slash{p}_{1}' 
+ \slash{k}) \otimes (\slash{p}_{b}+\slash{p}_{2}')
\big]
- {\cal F}_{V}(s,t_{2})\,
\gamma^{\alpha} \otimes \gamma_{\alpha} \big\rbrace
u_{a} \otimes u_{b}
\,.
\label{2.27}
\end{eqnarray}
Here, the functions ${\cal F}_{T}$ and ${\cal F}_{V}$
are defined in (\ref{1aux}) and (\ref{2aux}), respectively.

Using (\ref{2.25}), we find
\begin{eqnarray}
k^{\mu}{\cal M}_{\mu}^{(a)}&=&
-e \bar{u}_{1'} \otimes \bar{u}_{2'} \,
{\cal M}^{(0)}(p_{a}-k,p_{b},p_{1}',p_{2}')\,
u_{a} \otimes u_{b}\,,
\label{2.28}\\
k^{\mu}{\cal M}_{\mu}^{(b)}&=&
e \bar{u}_{1'} \otimes \bar{u}_{2'} \,
{\cal M}^{(0)}(p_{a},p_{b},p_{1}'+k,p_{2}')\,
u_{a} \otimes u_{b}\,.
\label{2.29}
\end{eqnarray}

Now, we impose the gauge-invariance condition,
which must hold also for the photon emission
from the $p_{a}$-$p_{1}'$ lines 
in Fig.~\ref{fig:pp_pp_gam} alone:
\begin{equation}
k^{\mu} \big({\cal M}_{\mu}^{(a)}+{\cal M}_{\mu}^{(b)}+{\cal M}_{\mu}^{(c)} \big)= 0\,.
\label{2.30}
\end{equation}
We obtain then
\begin{eqnarray}
&&k^{\mu} {\cal M}_{\mu}^{(c)} = -k^{\mu} {\cal M}_{\mu}^{(a)} - k^{\mu} {\cal M}_{\mu}^{(b)}
\nonumber\\
&& \quad =
e \bar{u}_{1'} \otimes \bar{u}_{2'}
\big[ {\cal M}^{(0)}(p_{a}-k,p_{b},p_{1}',p_{2}') 
     - {\cal M}^{(0)}(p_{a},p_{b},p_{1}'+k,p_{2}')\big] u_{a} \otimes u_{b}
\nonumber\\
&& \quad =
e \bar{u}_{1'} \otimes \bar{u}_{2'}
\big\lbrace
i {\cal F}_{T}(s',t_{2})
\big[ 
\gamma^{\alpha} \otimes \gamma_{\alpha}
(p_{a} - k + p_{1}', p_{b} + p_{2}')
+ (\slash{p}_{b}+\slash{p}_{2}') \otimes (\slash{p}_{a} - \slash{k} +\slash{p}_{1}')
\nonumber\\
&& \quad 
- \frac{1}{2} (\slash{p}_{a} - \slash{k} +\slash{p}_{1}') \otimes (\slash{p}_{b}+\slash{p}_{2}')
\big]
- {\cal F}_{V}(s',t_{2})\,
\gamma^{\alpha} \otimes \gamma_{\alpha} 
\nonumber\\
&& \quad - 
i {\cal F}_{T}(s,t_{2})
\big[ 
\gamma^{\alpha} \otimes \gamma_{\alpha}
(p_{a} + p_{1}'+ k, p_{b} + p_{2}')
+ (\slash{p}_{b} + \slash{p}_{2}') \otimes (\slash{p}_{a} + \slash{p}_{1}' + \slash{k})
\nonumber\\
&& \quad  
- \frac{1}{2} (\slash{p}_{a} + \slash{p}_{1}' + \slash{k}) \otimes (\slash{p}_{b}+\slash{p}_{2}')
\big]
+ {\cal F}_{V}(s,t_{2})\,
\gamma^{\alpha} \otimes \gamma_{\alpha} \big\rbrace
u_{a} \otimes u_{b} \,.
\label{2.31}
\end{eqnarray}

We have from (\ref{2.17})
\begin{eqnarray}
s'&=& (p_{a} + p_{b}-k)^{2} \nonumber\\
  &=& (p_{a} + p_{b})^{2} - 2(k, p_{a} + p_{b}) + k^{2} \nonumber\\
  &=& s - (k, 2p_{a} + 2p_{b} - k)
\label{2.33}
\end{eqnarray}
and define
\begin{eqnarray}
\varkappa = \frac{(k, 2p_{a} + 2p_{b}-k)}{s}\,.
\label{2.33_kappa}
\end{eqnarray}
Now, using the expressions given by
Eqs.~(\ref{2.7})--(\ref{2.12}),
(\ref{2.33}), and (\ref{2.33_kappa}), we get:
\begin{eqnarray}
&&{\cal F}_{\Pom pp}(s',t_{2}) = {\cal F}_{\Pom pp}(s,t_{2})
\big[ 1 + \varkappa \big(2 - \alpha_{\Pom}(t_{2}) \big) g_{\Pom}(\varkappa, t_{2}) \big]\,,
\label{2.34}\\
&&g_{\Pom}(\varkappa, t_{2}) = \frac{1}{\big(2 - \alpha_{\Pom}(t_2)\big)\varkappa}
\big[ (1-\varkappa)^{\alpha_{\Pom}(t_2)-2}-1 \big]\,,
\label{2.35}
\end{eqnarray}
and analogously for $f_{2 \Reg}$ and $a_{2 \Reg}$, and
\begin{eqnarray}
&&{\cal F}_{\Ode pp}(s',t_{2}) = {\cal F}_{\Ode pp}(s,t_{2})
\big[ 1 + \varkappa \big(1 - \alpha_{\Ode}(t_{2}) \big) g_{\Ode}(\varkappa, t_{2}) \big]\,,
\label{2.36}\\
&&g_{\Ode}(\varkappa, t_{2}) = \frac{1}{\big(1 - \alpha_{\Ode}(t_2)\big)\varkappa}
\big[ (1-\varkappa)^{\alpha_{\Ode}(t_2)-1}-1 \big]\,,
\label{2.37}
\end{eqnarray}
and analogously for $\omega_{\Reg}$ and $\rho_{\Reg}$.

From (\ref{2.34})--(\ref{2.37}) and supplementing 
with the reggeons we get
\begin{eqnarray}
&&{\cal F}_{T}(s',t) = {\cal F}_{T}(s,t) +
\varkappa \, \Delta{\cal F}_{T}(s,t,\varkappa)\,, 
\label{2.39}\\
&&{\cal F}_{V}(s',t) = {\cal F}_{V}(s,t) +
\varkappa \, \Delta{\cal F}_{V}(s,t,\varkappa)\,,
\label{2.39_aux}
\end{eqnarray}
where
\begin{eqnarray}
\Delta{\cal F}_{T}(s,t,\varkappa) &=&
  (2 - \alpha_{\Pom}(t)) \,g_{\Pom}(\varkappa ,t)\,
  {\cal F}_{\Pom pp}(s,t)+ (2 - \alpha_{f_{2\Reg}}(t)) \,g_{f_{2\Reg}}(\varkappa ,t)\,
{\cal F}_{f_{2\Reg} pp}(s,t)\nonumber\\ 
&&+ (2 - \alpha_{a_{2\Reg}}(t)) \,g_{a_{2\Reg}}(\varkappa ,t)\,
{\cal F}_{a_{2\Reg} pp}(s,t)\,,
\label{2.40} \\
\Delta{\cal F}_{V}(s,t,\varkappa) &=&
  (1 - \alpha_{\Ode}(t)) \,g_{\Ode}(\varkappa ,t)\,
  {\cal F}_{\Ode pp}(s,t) + (1 - \alpha_{\omega_{\Reg}}(t)) \,g_{\omega_{\Reg}}(\varkappa ,t)\,
{\cal F}_{\omega_{\Reg} pp}(s,t)\nonumber\\ 
&&+ (1 - \alpha_{\rho_{\Reg}}(t)) \,g_{\rho_{\Reg}}(\varkappa ,t)\,
{\cal F}_{\rho_{\Reg} pp}(s,t)\,.
\label{2.41}
\end{eqnarray}

The formulas for the odderon in (\ref{2.36}), (\ref{2.37}),
and (\ref{2.41}), apply for a single-pole odderon
(\ref{2.10}), (\ref{A7}).
The corresponding formulas for a double-pole odderon,
(\ref{2.10_new}), (\ref{A9}),
are given in Appendix~\ref{sec:appendixA};
see (\ref{A24})--(\ref{A26}).

Inserting all this in (\ref{2.31}), we get
\begin{eqnarray}
&&k^{\mu}{\cal M}_{\mu}^{(c)} =
e \bar{u}_{1'} \otimes \bar{u}_{2'}
\big\lbrace
-i {\cal F}_{T}(s,t_{2})
\big[ 
2\gamma^{\alpha} \otimes \gamma_{\alpha}
(k,p_{b} + p_{2}')
+ 2(\slash{p}_{b}+\slash{p}_{2}') \otimes \slash{k}
- \slash{k} \otimes (\slash{p}_{b}+\slash{p}_{2}')
\big]
\nonumber\\
&& \quad +
i \frac{(k,2 p_{a} + 2p_{b}-k)}{s}
\Delta{\cal F}_{T}(s,t_{2},\varkappa)
\nonumber\\
&& \quad \times  
\big[ 
\gamma^{\alpha} \otimes \gamma_{\alpha}
(p_{a} + p_{1}'- k, p_{b} + p_{2}')
+ (\slash{p}_{b} + \slash{p}_{2}') \otimes (\slash{p}_{a} + \slash{p}_{1}' - \slash{k})
- \frac{1}{2} (\slash{p}_{a} + \slash{p}_{1}' - \slash{k}) \otimes (\slash{p}_{b}+\slash{p}_{2}')
\big]
\nonumber\\
&& \quad - \frac{(k,2 p_{a} + 2p_{b}-k)}{s}
\Delta{\cal F}_{V}(s,t_{2},\varkappa)\,
\gamma^{\alpha} \otimes \gamma_{\alpha} \big\rbrace
u_{a} \otimes u_{b} \,.\nonumber \\
\label{2.42}
\end{eqnarray}
Hence, the simplest solution of (\ref{2.42})
for ${\cal M}_{\mu}^{(c)}$ has the form
\begin{eqnarray}
&&{\cal M}_{\mu}^{(c)} =
e \bar{u}_{1'} \otimes \bar{u}_{2'}
\big\lbrace
-i {\cal F}_{T}(s,t_{2})
\big[ 
2\gamma^{\alpha} \otimes \gamma_{\alpha}
(p_{b} + p_{2}')_{\mu}
+ 2(\slash{p}_{b}+\slash{p}_{2}') \otimes \gamma_{\mu}
- \gamma_{\mu} \otimes (\slash{p}_{b}+\slash{p}_{2}')
\big]
\nonumber\\
&& \quad +
i \frac{(2 p_{a} + 2p_{b}-k)_{\mu}}{s}
\Delta{\cal F}_{T}(s,t_{2},\varkappa)
\nonumber\\
&& \quad \times  
\big[ 
\gamma^{\alpha} \otimes \gamma_{\alpha}
(p_{a} + p_{1}'- k, p_{b} + p_{2}')
+ (\slash{p}_{b} + \slash{p}_{2}') \otimes (\slash{p}_{a} + \slash{p}_{1}' - \slash{k})
- \frac{1}{2} (\slash{p}_{a} + \slash{p}_{1}' - \slash{k}) \otimes (\slash{p}_{b}+\slash{p}_{2}')
\big]
\nonumber\\
&& \quad - \frac{(2 p_{a} + 2p_{b}-k)_{\mu}}{s}
\Delta{\cal F}_{V}(s,t_{2},\varkappa)\,
\gamma^{\alpha} \otimes \gamma_{\alpha} \big\rbrace
u_{a} \otimes u_{b} \,.\nonumber \\
\label{2.43}
\end{eqnarray}
We \underline{define} (\ref{2.43}) 
as ${\cal M}_{\mu}^{(c)}$.
Possible additions to this solution of the form
\begin{eqnarray}
\widetilde{\cal M}_{\mu}\,, \quad k^{\mu} \widetilde{\cal M}_{\mu} = 0
\label{2.44}
\end{eqnarray}
can and will be considered as a part of 
${\cal M}_{\mu}^{(g)}$ for which we require
\begin{equation}
k^{\mu} {\cal M}_{\mu}^{(g)} = 0\,.
\label{2.45}
\end{equation}

For the diagrams (d), (e), and (f), we get
\begin{eqnarray}
&&{\cal M}_{\mu}^{(d)} = 
\left.{\cal M}_{\mu}^{(a)}
\right|_{\mathop{^{(p_{a}, \,\lambda_{a}) \leftrightarrow (p_{b},\, \lambda_{b})}_{(p_{1}',\, \lambda_{1}) \leftrightarrow (p_{2}', \,\lambda_{2})}}}\,,
\label{2.46d}\\
&&{\cal M}_{\mu}^{(e)}=
\left.{\cal M}_{\mu}^{(b)}
\right|_{\mathop{^{(p_{a}, \,\lambda_{a}) \leftrightarrow (p_{b},\, \lambda_{b})}_{(p_{1}',\, \lambda_{1}) \leftrightarrow (p_{2}', \,\lambda_{2})}}}\,,
\label{2.46e}\\
&&{\cal M}_{\mu}^{(f)}=
\left.{\cal M}_{\mu}^{(c)}
\right|_{\mathop{^{(p_{a}, \,\lambda_{a}) \leftrightarrow (p_{b},\, \lambda_{b})}_{(p_{1}',\, \lambda_{1}) \leftrightarrow (p_{2}', \,\lambda_{2})}}} \,.
\label{2.46f}
\end{eqnarray}
With the exchanges 
$(p_{a}, \lambda_{a}) \leftrightarrow (p_{b}, \lambda_{b})$
and
$(p_{1}', \lambda_{1}) \leftrightarrow (p_{2}', \lambda_{2})$,
we shall also exchange the order of the factors 
in the tensor products.
In this way, the first factors in the tensor products
always refer to the $p_{a}$-$p_{1}'$ line,
and the second factors refer to the $p_{b}$-$p_{2}'$ line,
respectively, in the diagrams of Fig.~\ref{fig:pp_pp_gam}.

We shall call
\begin{eqnarray}
{\cal M}_{\mu}^{(\rm standard)}&=&
{\cal M}_{\mu}^{(a)} + {\cal M}_{\mu}^{(b)}
+ {\cal M}_{\mu}^{(c)} + {\cal M}_{\mu}^{(d)} 
+ {\cal M}_{\mu}^{(e)} + {\cal M}_{\mu}^{(f)}
\label{2.50}
\end{eqnarray}
our standard, or ``conventional'', amplitude.
We have by construction
\begin{equation}
k^{\mu} {\cal M}_{\mu}^{(\rm standard)} = 0\,.
\label{2.51}
\end{equation}
All ``anomalous'' terms are then subsumed 
in ${\cal M}_{\mu}^{(g)}$, which satisfies (\ref{2.45})
and has no singularity for $k_{\mu} \to 0$.

The explicit forms of the amplitudes
${\cal M}_{\mu}^{(a)}, \ldots, {\cal M}_{\mu}^{(f)}$
in (\ref{2.26}), (\ref{2.27}), (\ref{2.43}),
and (\ref{2.46d})--(\ref{2.46f}),
calculated in the tensor-pomeron approach,
are a main result of our present paper.
But it turns out that these forms are not very convenient
for numerical computations.
Therefore, in Appendix~\ref{sec:appendixB},
we rewrite these amplitudes in a different form which
is more easy to handle numerically.

The intermediate protons in the diagrams of 
Fig.~\ref{fig:pp_pp_gam}~\mbox{(a, b, d, e)}
are off shell, and in principle
we should take care of that fact in our model.
But at present, we set possible cutoff form factors 
for off-shell protons
in the pomeron-proton and photon-proton vertices 
and in the proton propagator to 1.
Note that a naive usage of form factors 
can violate the Ward-Takahashi identity (\ref{2.25}).
We expect, however, that in our regions of interest,
in the small $k_{\perp}$ and $\omega$ ranges
and especially in the soft-photon limit,
the off-shell effects should be small.

%----------------------------
\section{Soft-photon approximation approach}
\label{sec:SPA}
%----------------------------

In the following, we shall compare our exact model results,
which we shall call standard results,
for the amplitude (\ref{2.50}), using (\ref{B3})--(\ref{B16}),
to two soft-photon approximations: SPA1 and SPA2.
In these SPAs, we consider only the pomeron-exchange terms 
for the radiative amplitudes
${\cal M}_{\mu}^{(a)}, \ldots, {\cal M}_{\mu}^{(f)}$ in (\ref{2.50}).
%and we use (\ref{B3}) with only the $j=1$ terms.}
Below, we list the explicit expressions for real photon
emission in $pp$ scattering.
\begin{enumerate}
\item[SPA1:] 
Here we keep only the pole terms
$\propto \omega^{-1}$ 
for ${\cal M}_{\mu}^{(a)}, \ldots, {\cal M}_{\mu}^{(f)}$ 
in (\ref{2.50}).
For real photons ($k^{2} = 0$), 
neglecting gauge terms $\propto k_{\mu}$,
and with
$p_{1}' \to p_{1}$, $p_{2}' \to p_{2}$,
we get
\begin{eqnarray}
{\cal M}_{\mu}
&\to& {\cal M}_{\mu, \;{\rm SPA1}}
\nonumber \\
&&= e{\cal M}^{({\rm on\; shell})\,pp}(s,t)
\Big[ 
-\frac{p_{a \mu}}{(p_{a} \cdot k)}
+\frac{p_{1 \mu}}{(p_{1} \cdot k)}
-\frac{p_{b \mu}}{(p_{b} \cdot k)}
+\frac{p_{2 \mu}}{(p_{2} \cdot k)} \Big]. \quad
\label{4.39}
\end{eqnarray}
Here, ${\cal M}^{({\rm on\; shell})\,pp}(s,t)$ 
is the on-shell $pp$-scattering amplitude 
given by (\ref{2.14}).
The result (\ref{4.39}) is easily obtained
from (\ref{B3}), (\ref{B4}), and (\ref{B15}).
Inserting (\ref{4.39}) into (\ref{incl_xs_2to3_aux2}),
we get the following SPA1 result 
for the inclusive photon cross section where,
for consistency, we neglect the photon momentum $k$
in the energy-momentum conserving $\delta^{(4)}(.)$ function:
\begin{eqnarray}
d\sigma({pp \to pp \gamma})_{\,\rm SPA1} &=&
\frac{d^{3}k}{(2 \pi)^{3} \,2 k^{0}}
\int d^{3}p_{1} \,d^{3}p_{2}\,e^{2}\;
\frac{d\sigma(pp \to pp)}{d^{3}p_{1}d^{3}p_{2}}\nonumber \\
&&\times 
\Big[ 
-\frac{p_{a \mu}}{(p_{a} \cdot k)}
+\frac{p_{1 \mu}}{(p_{1} \cdot k)}
-\frac{p_{b \mu}}{(p_{b} \cdot k)}
+\frac{p_{2 \mu}}{(p_{2} \cdot k)} \Big]\nonumber \\
&&\times 
\Big[ 
-\frac{p_{a \nu}}{(p_{a} \cdot k)}
+\frac{p_{1 \nu}}{(p_{1} \cdot k)}
-\frac{p_{b \nu}}{(p_{b} \cdot k)}
+\frac{p_{2 \nu}}{(p_{2} \cdot k)} \Big] (-g^{\mu \nu})\,.
\quad
\label{4.40}
\end{eqnarray}
Here,
\begin{eqnarray}
\frac{d\sigma(pp \to pp)}{d^{3}p_{1}d^{3}p_{2}} &=&
\frac{1}{2\sqrt{s(s-4 m_{p}^{2})}}\,
\frac{1}{(2 \pi)^{3} \,2 p_{1}^{0}\,(2 \pi)^{3} \,2 p_{2}^{0}} 
\nonumber\\
&&\times (2 \pi)^{4} \delta^{(4)}(p_{1}+p_{2}-p_{a}-p_{b})\,
\frac{1}{4}\sum_{p \, \rm spins}
|{\cal M}^{({\rm on\; shell})\,pp}(s,t)|^{2}\,.\qquad \;
\label{4.41}
\end{eqnarray}
In (\ref{4.40}) and (\ref{4.41}), we have a frequently used SPA.
One takes the momentum distribution of the particles 
without radiation [see (\ref{4.41})] and multiplies it
with the square of the emission
factor in the square brackets in (\ref{4.39}).

%%%%%%%%%%%%%%%%%%%%%%%%%%%%%%%%%%%%%%%%%%%%%%%
\item[SPA2:] 
%%%%%%%%%%%%%%%%%%%%%%%%%%%%%%%%%%%%%%%%%%%%%%%
Here we keep the energy-momentum relation (\ref{2.4a}).
We consider again real photon emission and
make in (\ref{incl_xs_2to3_aux2}) the replacement
\begin{eqnarray}
{\cal M}_{\mu}
&\to & {\cal M}_{\mu, \;{\rm SPA2}}
= {\cal M}_{{\Pom},\mu}^{(a + b + c)\,1}
+
{\cal M}_{{\Pom},\mu}^{(d + e + f)\,1}\,.
\label{4.42}
\end{eqnarray}
The amplitude (\ref{4.42})
corresponds to that given in (\ref{B4}) 
plus (\ref{B15}),
taking only the pomeron exchange into account.
Here, the squared momentum
transfer is $t_{2}$ for the diagrams of 
Fig.~\ref{fig:pp_pp_gam}~(a--c)
and $t_{1}$ for those of Fig.~\ref{fig:pp_pp_gam}~(d--f),
where $t_{1}$ and $t_{2}$ are defined in (\ref{2.17}).
Also, in the calculation of the photon distributions,
we keep the correct energy-momentum conserving $\delta^{(4)}(.)$ function
in (\ref{incl_xs_2to3_aux2}).
\end{enumerate}

We also consider a slightly modified
SPA1 amplitude where we use
the high-energy small-angle limit relations (\ref{HESA_limit}).
We get then $\widehat{{\cal M}}_{\mu, \;{\rm SPA1}}$ as in
(\ref{4.39}) but with (\ref{2.14_high_energy})
inserted for ${\cal M}^{({\rm on\; shell})\, pp}$:
\begin{eqnarray}
\widehat{{\cal M}}_{\mu, \;{\rm SPA1}}
= i e 8 s^{2} {\cal F}_{\Pom pp}(s,t) 
\delta_{\lambda_{1}\lambda_{a}} 
\delta_{\lambda_{2}\lambda_{b}}
\Big[ 
-\frac{p_{a \mu}}{(p_{a} \cdot k)}
+\frac{p_{1 \mu}}{(p_{1} \cdot k)}
-\frac{p_{b \mu}}{(p_{b} \cdot k)}
+\frac{p_{2 \mu}}{(p_{2} \cdot k)} \Big]. \quad
\label{4.39_aux}
\end{eqnarray}
For the SPA2, we examine, furthermore, the approximation
\begin{eqnarray}
\widehat{{\cal M}}_{\mu, \;{\rm SPA2}}
&=& 
%e {\cal M}^{({\rm on\; shell})\,pp}(s,t_{2}) 
i e 8 s^{2} {\cal F}_{\Pom pp}(s,t_{2}) 
\delta_{\lambda_{1}\lambda_{a}} 
\delta_{\lambda_{2}\lambda_{b}}
\Big[ 
-\frac{p_{a \mu}}{(p_{a} \cdot k)}
+\frac{p_{1 \mu}'}{(p_{1}' \cdot k)} \Big]
\nonumber\\
&&+ 
%e {\cal M}^{({\rm on\; shell})\,pp}(s,t_{1})
i e 8 s^{2} {\cal F}_{\Pom pp}(s,t_{1}) 
\delta_{\lambda_{1}\lambda_{a}} 
\delta_{\lambda_{2}\lambda_{b}} 
\Big[
-\frac{p_{b \mu}}{(p_{b} \cdot k)}
+\frac{p_{2 \mu}'}{(p_{2}' \cdot k)} \Big]
\,.
\label{SPA2_approx}
\end{eqnarray}
How these approximations
(\ref{4.39_aux}) and (\ref{SPA2_approx})
agree with
${\cal M}_{\mu, \;{\rm SPA1}}$ from the formula~(\ref{4.39})
and
${\cal M}_{\mu, \;{\rm SPA2}}$ from the formula~(\ref{4.42})
will be discussed at the end of Sec.~\ref{sec:3B}.

%-----------------------------------
\section{Results}
\label{sec:3}
%-----------------------------------

Below, we show our results 
for total and elastic $pp$ and $p \bar{p}$ scattering 
(Sec.~\ref{sec:3A})
and
results for the $pp \to pp \gamma$ reaction 
from the bremsstrahlung mechanism
(Sec.~\ref{sec:3B}).

%-----------------------------------
\subsection{Comparison of the model results with the total and elastic $pp$ and $p \bar{p}$ cross section data}
\label{sec:3A}
%-----------------------------------

Here, we compare our results
obtained from Eqs.~(\ref{2.14}) and (\ref{16d}) 
to the experimental data 
for $\sigma_{\rm tot}$, $d\sigma/dt$, 
and $\rho$, that is,
the ratio of the real part to imaginary part of the forward 
scattering amplitude,
\begin{eqnarray}
\rho = \frac{{\rm Re}{\cal T}(s,0)}{{\rm Im}{\cal T}(s,0)}\,.
\label{rho_ratio}
\end{eqnarray}
We emphasize that in the present paper it is \underline{not} our purpose
to perform a precision fit to the available experimental data
on these quantities, say, from ISR to LHC energies.
Here ISR means the CERN Intersecting Storage Rings which
operated with a maximum c.m. energy of 63 GeV.
Such fits have been, for instance, 
given in Refs.~\cite{Donnachie:2013xia,Jenkovszky:2017efs,Jenkovszky:2018itd,
Khoze:2018kna,Martynov:2018sga,
Broilo:2020kqg,Bence:2021uji,Godizov:2021ksd}.
Our aim here is to obtain a reasonable description
of the data mainly at $\sqrt{s} = 13$~TeV.
We want to fix the parameters of our model in this way.
The \underline{main} aim of our paper is to give
then predictions for soft-photon radiation 
in proton-proton elastic collisions for the LHC energy range; 
see Sec.~\ref{sec:3B}.

%The differential cross section $d\sigma/dt$
%has been measured by the TOTEM and ATLAS experiments 
%at $\sqrt{s} = 2.76$~TeV \cite{TOTEM:2018psk},
%7~TeV \cite{TOTEM:2011vxg,TOTEM:2013lle,ATLAS:2014vxr}, 
%8~TeV \cite{TOTEM:2016lxj,ATLAS:2016ikn},
%and 13~TeV \cite{TOTEM:2017asr,TOTEM:2017sdy,TOTEM:2018hki}.
%There is also measurement at $\sqrt{s} = 200$~GeV
%by the STAR Collaboration \cite{STAR:2020phn}.
%from extrapolation
%of the $d\sigma(pp)/dt$ to the optical point at $-t = 0$.
%We also compare with the ``favoured'' model 
%for the $pp$ amplitude
%used by the experimentalists (TOTEM).
%for 0.04~GeV$^{2} < |t| < 4$~GeV$^{2}$
%and in \cite{TOTEM:2018psk}
%at $\sqrt{s} = 2.76$~TeV for 0.36~GeV$^{2} < |t| < 0.74$~GeV$^{2}$. 

\begin{figure}[!ht]
\includegraphics[width=0.55\textwidth]{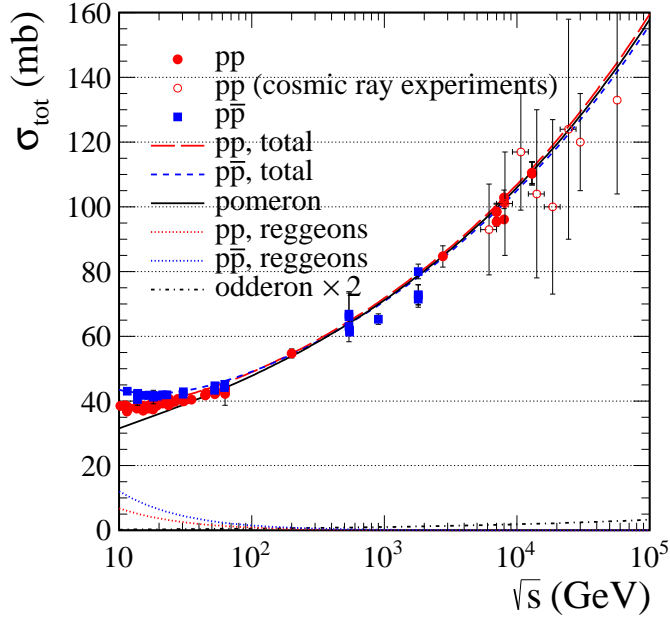}
\caption{\label{fig:sigma}
\small
The total cross sections 
for $pp$ and $p \bar{p}$ scattering
as a function of the center-of-mass energy $\sqrt{s}$.
The red long-dashed line and the blue dashed line
correspond to our complete results
for $pp$ and $p \bar{p}$ scattering, respectively.
Here, the pomeron parameters are as in (\ref{A1}),
(\ref{A2}), (\ref{A11}), and (\ref{A12})
with $\epsilon_{\Pom} = 0.0865$.
The reggeon parameters are given 
in (\ref{A3})--(\ref{A6}) and we take here 
for the odderon the parameters (\ref{A9}), (\ref{A10}), and
(\ref{A22}), with $(C_{1}, C_{2}) = (-2.0, 0.3)$.
The black solid line corresponds to the pomeron exchange alone,
the dotted lines to the reggeon exchanges
(lower for $pp$, upper for $p \bar{p}$),
and the dot-dashed line corresponds to the odderon contribution
increased by a factor of 2 for visualization.
The experimental data are from Ref.~\cite{TOTEM:2016lxj},
from Eq.~(8) and Figs.~4 and 5 of Ref.~\cite{TOTEM:2017asr},
from Ref.~\cite{TOTEM:2017sdy},
and from the Particle Data Group \cite{Zyla:2020zbs} 
(see Ref.~\cite{PDG_CrossSections}).}
\end{figure}

In Fig.~\ref{fig:sigma},
we present the total $pp$ (see the red points and lines) 
and $p \bar{p}$ (see the blue points and lines)
cross sections for $\sqrt{s} > 10$~GeV.
%In this experiment the elastic differential cross section
%has been measured in the squared four-momentum transfer range
%$0.045 \leqslant -t \leqslant 0.135$~GeV$^{2}$.
We show our complete theoretical results including 
the pomeron, the reggeon, and the odderon exchanges
and their separate contributions.
First, we notice that, of course,
the high-energy cross sections
are dominated by the pomeron exchange 
(see the black solid line).
The cross sections
with the pomeron contribution alone are
$\sigma_{\rm tot}(pp) = 102.0$~mb for $\sqrt{s} = 8$~TeV 
and
$\sigma_{\rm tot}(pp) = 110.9$~mb for $\sqrt{s} = 13$~TeV.
For the odderon exchange,
we use the double-pole Ansatz (\ref{A9})
with the parameters (\ref{A10})
and $(C_{1}, C_{2}) = (-2.0, 0.3)$
adjusted to describe $\rho = 0.1$
found by TOTEM at $\sqrt{s} = 13$~TeV.
See the discussion below in connection with Fig.~\ref{fig:rho}
which presents $\rho$ as function of $\sqrt{s}$.
There we also show the results for two more values
$(C_{1}, C_{2}) = (-1.0, 0.1)$ and $(-1.5, 0.2)$.
These lead to values of $\rho > 0.1$ at $\sqrt{s} = 13$~TeV.
Here we are interested in the total cross sections where
the odderon effects are very small.
We get, for instance, for the two cases
$(C_{1}, C_{2}) = (-1.0, 0.1)$ and $(-2.0, 0.3)$ 
the total cross sections,
$\sigma_{\rm tot}(pp) = 102.2$ and 102.9~mb for $\sqrt{s} = 8$~TeV
and
$\sigma_{\rm tot}(pp) = 111.2$ and 111.9~mb for $\sqrt{s} = 13$~TeV,
respectively.
Our results are in good agreement
with the TOTEM results:
$\sigma_{\rm tot}(pp)|_{\rm exp} = (102.9 \pm 2.3)$~mb
and
$\sigma_{\rm tot}(pp)|_{\rm exp} = (103.0 \pm 2.3)$~mb
for $\sqrt{s} = 8$~TeV \cite{TOTEM:2016lxj},
$\sigma_{\rm tot}(pp)|_{\rm exp} = (110.6 \pm 3.4)$~mb \cite{TOTEM:2017asr}
and
$\sigma_{\rm tot}(pp)|_{\rm exp} = (110.3 \pm 3.5)$~mb \cite{TOTEM:2017sdy}
for $\sqrt{s} = 13$~TeV.
\footnote{
The cross sections $\sigma_{\rm tot}(pp)|_{\rm exp}$
from Refs.~\cite{TOTEM:2017asr,TOTEM:2017sdy} 
for $\sqrt{s} = 13$~TeV 
were determined in a completely independent way
and, therefore, were combined for uncertainty reduction.
The result is then
$\sigma_{\rm tot}(pp)|_{\rm exp} = (110.5 \pm 2.4)$~mb;
see Eq.~(21) of Ref.~\cite{TOTEM:2017sdy}.}
Note that we get for large energies a total cross section 
for $pp$ exceeding that for $p\bar{p}$ collisions,
$\sigma_{\rm tot}(pp) > \sigma_{\rm tot}(p\bar{p})$.
%--------------------------------------------------------
%--------------------------------------------------------

\begin{figure}[!ht]
\includegraphics[width=0.49\textwidth]{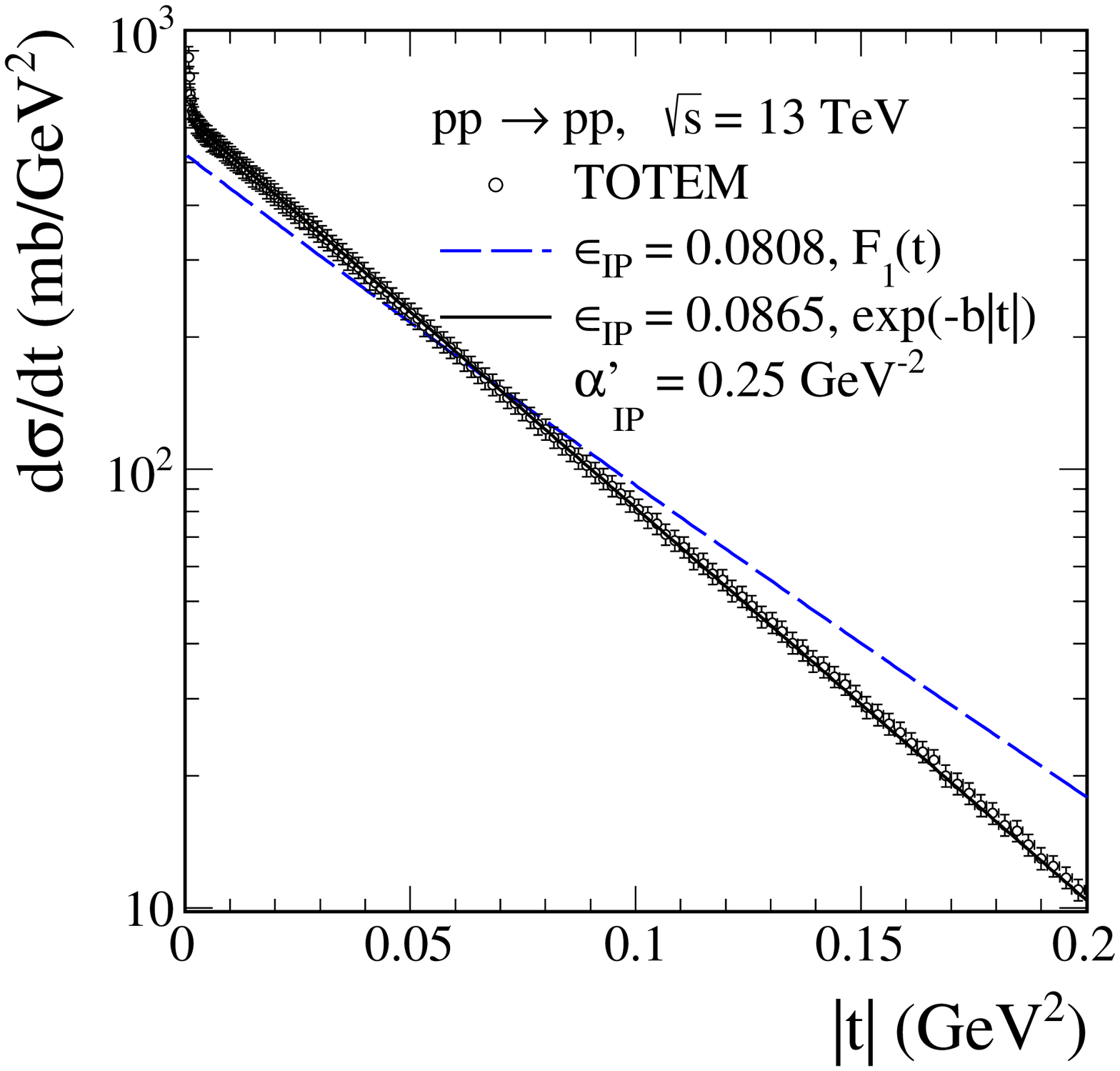}
\includegraphics[width=0.49\textwidth]{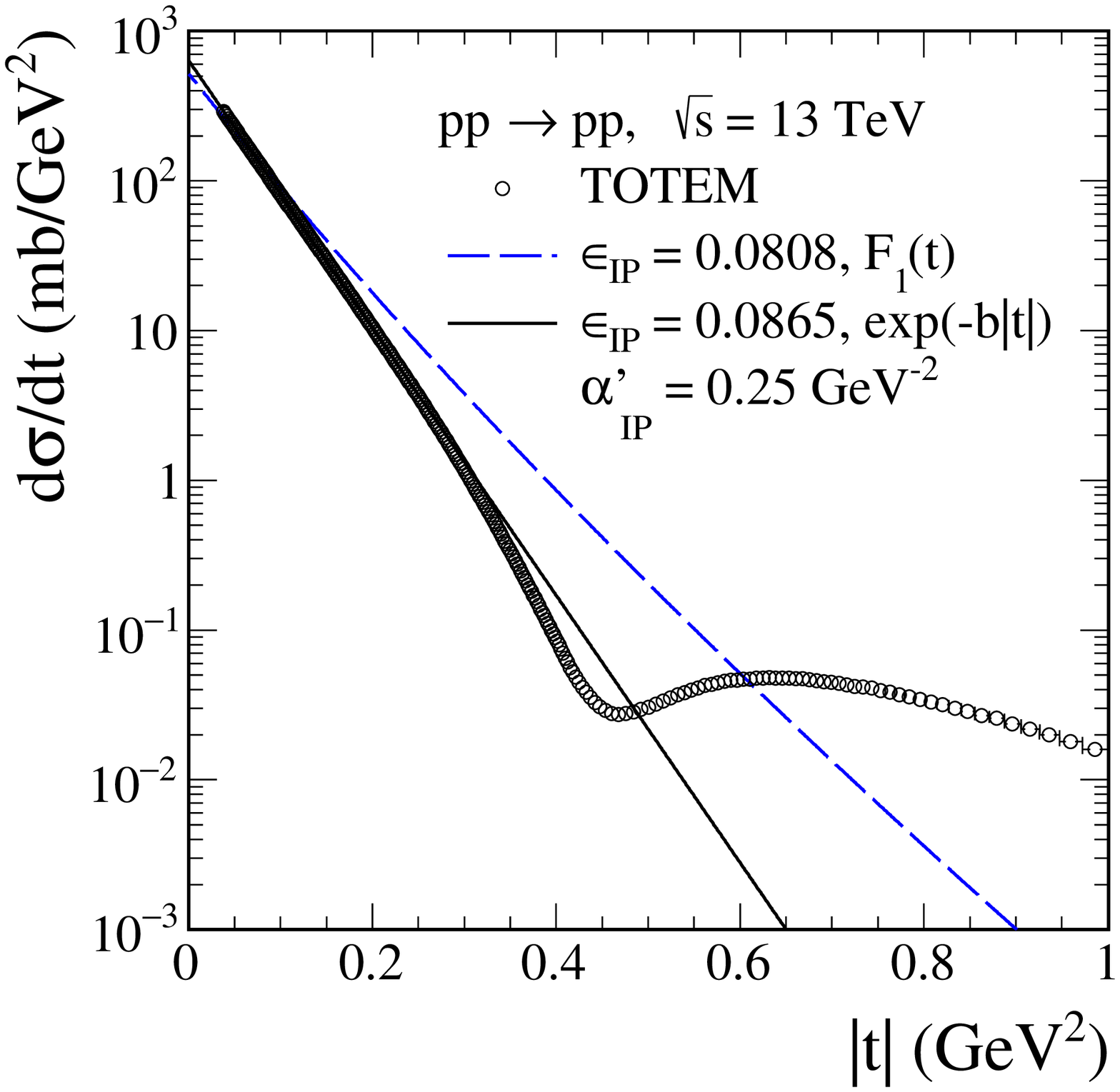}
\caption{\label{fig:dsig_dt}
\small
The differential $pp \to pp$ cross sections
measured by the TOTEM experiments
at $\sqrt{s} = 13$~TeV \cite{TOTEM:2017sdy,TOTEM:2018hki}.
We show our theoretical results obtained with two
different pomeron parameter sets: 
1) $\epsilon_{\Pom} = 0.0808$ and 
the Dirac form factor $F_{1}(t)$
(see the blue dashed lines),
and 2) $\epsilon_{\Pom} = 0.0865$ and
the exponential form factor $F(t) = \exp(-b\,|t|)$ 
with $b = 2.95$~GeV$^{-2}$ (see the black solid lines).
In both cases, we take 
$\alpha_{\Pom}' = 0.25$~GeV$^{-2}$ from (\ref{A2}).
Here, reggeon and odderon contributions are set to zero.}
\end{figure}

\begin{figure}[!ht]
\includegraphics[width=0.49\textwidth]{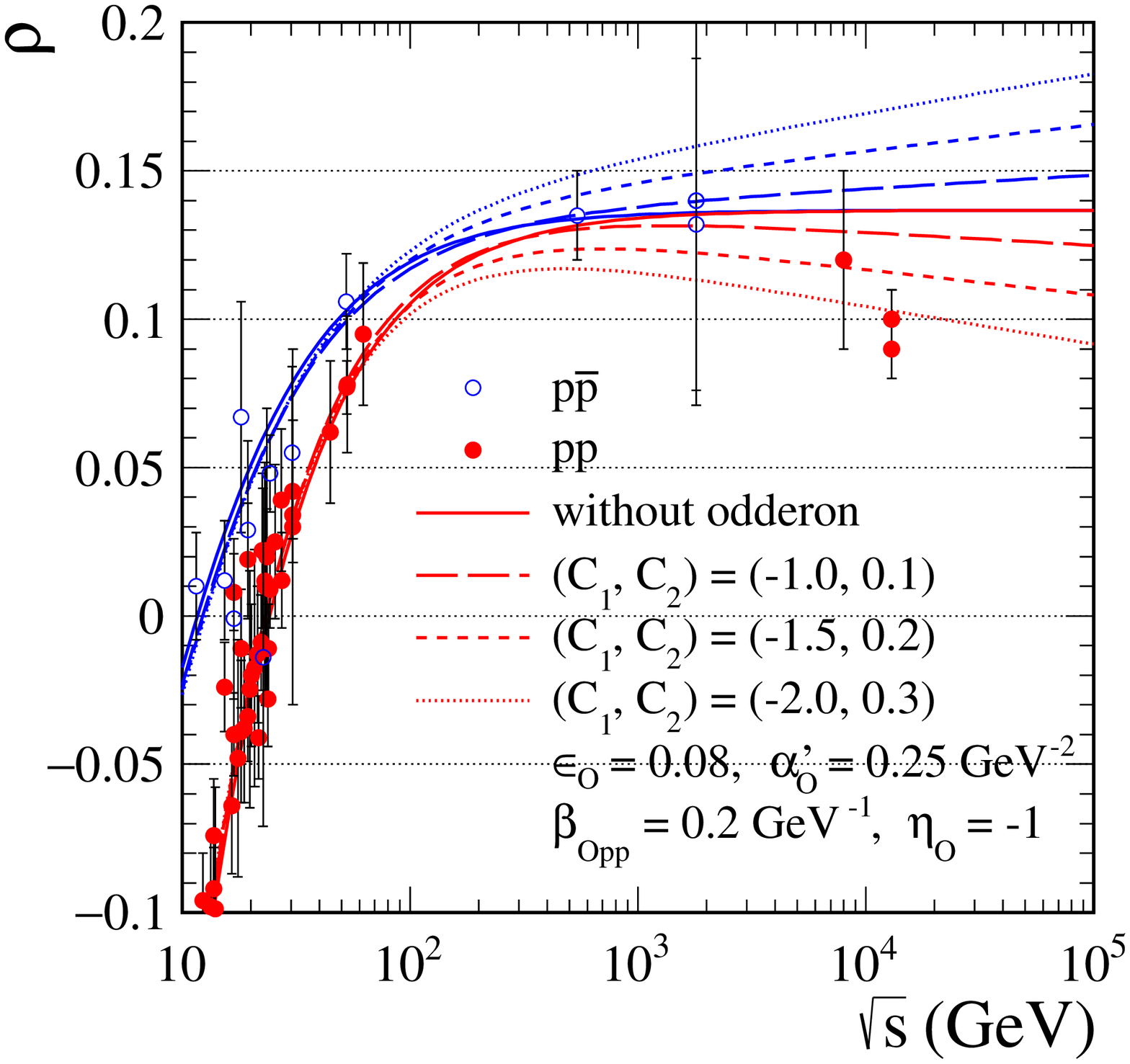}
\includegraphics[width=0.49\textwidth]{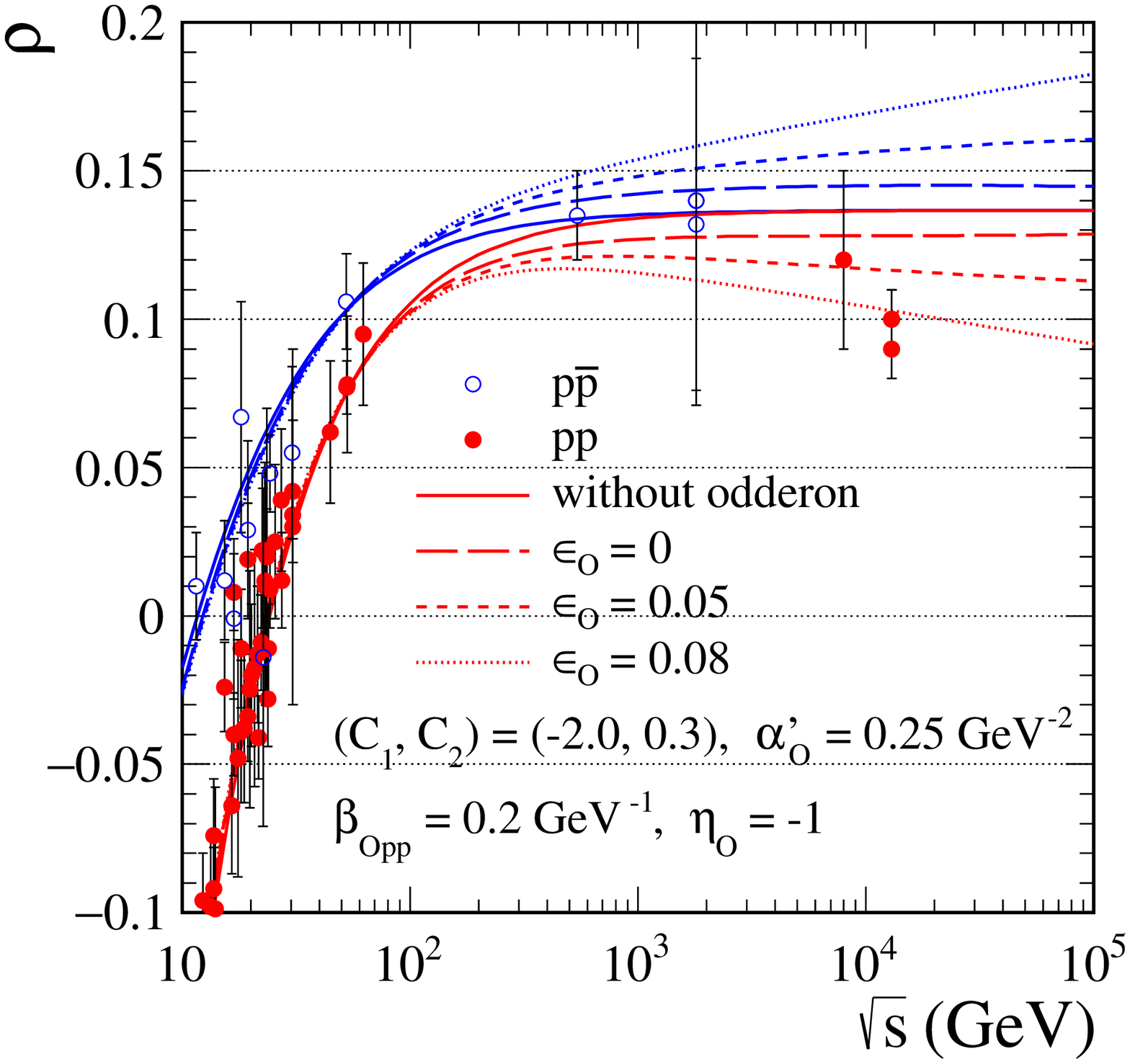}
\caption{\label{fig:rho}
\small
Our results for the $\rho$ parameter as function 
of the c.m. energy $\sqrt{s}$
for $p \bar{p}$ (the blue upper lines) 
and $pp$ (the red lower lines) scattering
together with the experimental data
from Refs.~\cite{TOTEM:2016lxj,TOTEM:2017sdy}
and from Ref.~\cite{PDG_CrossSections}.
Our complete results include
$\Pom$, $f_{2 \Reg}$, $a_{2 \Reg}$, $\omega_{\Reg}$, 
$\rho_{\Reg}$, and $\Ode$ exchanges.
The solid lines are without the odderon contribution,
and the dashed and dotted lines
are with the odderon for different values of the parameters 
as described in the figure legends.}
\end{figure}

In Fig.~\ref{fig:dsig_dt}, we show
the $pp$ elastic differential cross sections
for our model and the TOTEM data 
\cite{TOTEM:2017sdy,TOTEM:2018hki}
measured at $\sqrt{s} = 13$~TeV.
We find a quite good description of the data in the region 
0.003~GeV$^{2}$ $\leqslant -t \leqslant 0.26$~GeV$^{2}$
taking $\epsilon_{\Pom} = 0.0865$
and assuming the form factor $F(t) = \exp(-b\,|t|)$ 
with only one parameter, $b = 2.95$~GeV$^{-2}$.
For~comparison, we show also the results
for $\epsilon_{\Pom} = 0.0808$ and
the Dirac form factor $F_{1}(t)$.
We~should~not expect our single-pomeron exchange model 
to give a precision fit for $d\sigma/dt$ 
in the diffractive dip region and beyond.
The common lore is that in order to produce
the dip one needs the interference of
various terms in the amplitude, at least three terms:
$\Pom + \Pom \Pom + ggg$; see, e.g.,
Refs.~\cite{Donnachie:1983hf,Donnachie:2013xia}.
In Refs.~\cite{Jenkovszky:2017efs,Jenkovszky:2018itd}, 
the authors discussed
the so-called break effect, which leads to a smooth deflection 
of the linear exponential behavior
of the diffraction cone observed near $t = -0.1$~GeV$^{2}$.
The effect is related to the basic analytic properties of the theory 
($t$-channel unitarity) and reflects the presence of the pion cloud 
around the nucleon.
This small, but interesting, effect could easily be included
in our parametrization of the $t$ dependence of the $pp$
elastic cross section. But, as mentioned above,
for our purposes, we only need a reasonable description
of the $pp$ scattering data for $\sqrt{s} = 13$~TeV.
And, as we show in Fig.~\ref{fig:dsig_dt},
we obtain such a description in the $t$ range of interest for us
with our single-exponential form factor $F(t)$.
%The dip (diffraction minimum), on the other hand, 
%is generally accepted as a consequence
%of $s$-channel unitarity or
%absorption corrections to the scattering amplitude;
%see the discussion in \cite{Jenkovszky:2017efs}.}
%see, e.g.,
%\cite{Martynov:2018sga,Khoze:2018kna,
%Bence:2021uji,Godizov:2021ksd}.
%%\cite{Szanyi:2019kkn,Csorgo:2019ewn}
%as in diffractive electromagnetic processes 
%\cite{Donnachie:1998gm,Britzger:2019lvc,Dosch:2022mop}.
%double-pomeron ($\Pom \Pom$) exchange,
%or assumption that there are not two pomerons, 
%but only one, where the values
%of its intercept (and presumably of its slope) 
%are modified by pQCD interactions \cite{Dosch:2022mop}.
%--------------------------------------------------------
%--------------------------------------------------------

In Fig.~\ref{fig:rho}, we show our results for $\rho$
for the two elastic scattering processes 
$pp$ and $p\bar{p}$
together with the experimental data.
The solid lines represent results of our model without 
the odderon contribution.
The pomeron and reggeon parameters are the same as in
Fig.~\ref{fig:sigma}.
In the left panel,
the dashed and dotted lines show the results with the odderon exchange
for different values of $(C_{1}, C_{2})$ in (\ref{2.10_new})
and with the parameters (\ref{A10}), (\ref{A22}).
The $C = -1$ reggeons lead to a splitting 
of the $pp$ and $p \bar{p}$ results at smaller energies,
while the odderon mainly affects the higher energies.
The odderon parameters $(C_{1}, C_{2})$ are in essence
determined by the $\rho$ value at $\sqrt{s} = 13$~TeV.
The values $\rho = 0.103$, 0.116, and 0.129 lead to
%The values $\rho = 0.10$, 0.115 and 0.13 lead to
$(C_{1}, C_{2}) = (-2.0, 0.3)$, $(-1.5, 0.2)$, and $(-1.0, 0.1)$,
respectively.
In the right panel of Fig.~\ref{fig:rho},
we show the results for $(C_{1}, C_{2}) = (-2.0, 0.3)$ and
for $\epsilon_{\Ode} = 0$ and 0.05 to compare with
$\epsilon_{\Ode} = 0.08$.
We can see that the $\rho$ values mentioned 
can also be obtained by changing~$\epsilon_{\Ode}$.
%The odderon trajectory function is taken as linear in $t$,
%$\alpha_{\Ode}(t) = 1 + \epsilon_{\Ode} + \alpha_{\Ode}'\, t$.
%For our study here we assumed
%$\alpha_{\Ode}' = \alpha_{\Pom}' = 0.25$~GeV$^{-2}$.
%--------------------------------------------------------
%--------------------------------------------------------

The TOTEM Collaboration has reported direct measurements
of the $\rho$ parameter (\ref{rho_ratio})
through the Coulomb-nuclear interference
in the very small $|t|$ region.
The TOTEM results for $\rho$ extracted from 
the data on the proton-proton differential cross section are:
\begin{itemize}
\item 
$\rho = 0.12 \pm 0.03$ 
at $\sqrt{s} = 8$~TeV \cite{TOTEM:2016lxj};
\item 
$\rho = 0.09 \pm 0.01$ or $\rho = 0.10 \pm 0.01$
at $\sqrt{s} = 13$~TeV \cite{TOTEM:2017sdy},
depending on different physics assumptions and mathematical modelings.
\end{itemize}

We can make the following statement.
If we take the above $\rho$ values from TOTEM as representing the truth,
then, in the framework of our model,
we need an odderon contribution at $t = 0$.
Moreover, we could not find a reasonable description
of the low- and high-energy data for $\rho$ with a single-pole odderon.
For these reasons, we consider here a double-pole odderon.
To fit the TOTEM results for $\rho$, 
we have to choose the odderon parameters 
$(C_{1}, C_{2}) = (-2.0, 0.3)$.
These findings are consistent 
with the results of Refs.~\cite{Martynov:2017zjz,Martynov:2018pye}.
These authors claim that a small value of $\rho$
can be considered as the experimental
discovery of the odderon for small $|t|$, 
namely, in its maximal form.
The odderon was first introduced
%in 1973
on theoretical grounds in
%on the theoretical basis of asymptotic theorems 
Ref.~\cite{Lukaszuk:1973nt}.
For a review of odderon physics we refer 
the reader to Ref.~\cite{Ewerz:2003xi}.
However, recent reanalyses of the TOTEM results 
\cite{Pancheri:2018yhd,Ezhela:2020hws}
have shown that the values of $\rho$ may be larger than those
reported by the TOTEM Collaboration.
For illustration,
in Ref.~\cite{Pancheri:2018yhd}, 
one finds $\rho \simeq 0.135$ at $\sqrt{s} = 8$~TeV 
and $\rho \simeq 0.133$ at $\sqrt{s} = 13$~TeV.
Reference~\cite{Ezhela:2020hws} gives
$\rho = 0.123 \pm 0.010$ at $\sqrt{s} = 13$~TeV.
To be consistent with these values,
we should take in our model $(C_{1}, C_{2}) = (-1.0, 0.1)$.

In view of these ongoing discussions in the literature,
we think that the low value $\rho = 0.1$ 
at $\sqrt{s} = 13$~TeV reported by TOTEM probably
does not prove that there is
an odderon contribution at $t = 0$;
see, for instance, Refs.~\cite{Donnachie:2019ciz,Donnachie:2022aiq}.
However, there are good reasons to believe that 
there is an odderon effect at larger $|t|$ 
(in the dip-bump region)
that leads to a very significant difference 
between the differential cross sections
of elastic $pp$ and $p \bar{p}$ scattering,
as was first seen at the CERN ISR at $\sqrt{s} = 53$~GeV
\cite{Breakstone:1985pe}.
%ISR data at low $|t|$ region \cite{ABCDHW:1984yby}
%more precise data on pp elastic scattering at 53 GeV:
%deKerret:1977gvr
%more precise data on pp elastic scattering at 53 GeV:
%\cite{deKerret:1977gvr}
Remarkably, such a difference had been predicted 
theoretically in Ref.~\cite{Donnachie:1983hf}.
The advantage of corresponding
measurements at higher energies,
e.g., at the LHC,
is that subleading reggeon exchanges are 
very small there,
and thus an observation of differences 
between $pp$ and $p \bar{p}$
would be a clear signal of the odderon.
Indeed, strong evidence for the odderon
has been given in Ref.~\cite{TOTEM:2020zzr}
based on a model-independent analysis
of the combined TOTEM and D0 results
in the dip-bump region.

%-----------------------------------
\subsection{Our standard results for the $pp \to pp \gamma$ reaction and comparison with soft-photon approximations}
\label{sec:3B}
%-----------------------------------

First, we present our exact model 
or standard bremsstrahlung results
for the $pp \to pp \gamma$ reaction (see Sec.~\ref{sec:2C})
for the proton-proton collision energy $\sqrt{s} = 13$~TeV.
Below, $k_{\perp}$ is the absolute value 
of the photon transverse momentum,
$\omega = k^{0}$ is the center-of-mass photon energy,
and $\rm{y}$ is the rapidity of the photon.
Then, we will compare the resulting distributions
with those obtained via variants of the SPA 
discussed in Sec.~\ref{sec:SPA}.

We see from Fig.~\ref{fig:sigma} that at $\sqrt{s} = 13$~TeV
the $pp$-scattering cross section is completely
dominated by the pomeron exchange.
Therefore, in the following, we shall take into account
only the pomeron-exchange contribution
as shown in the diagrams of 
Fig.~\ref{fig:pp_pp_gam}~\mbox{(a)--(f)}.

\begin{figure}[!ht]
\includegraphics[width=0.49\textwidth]{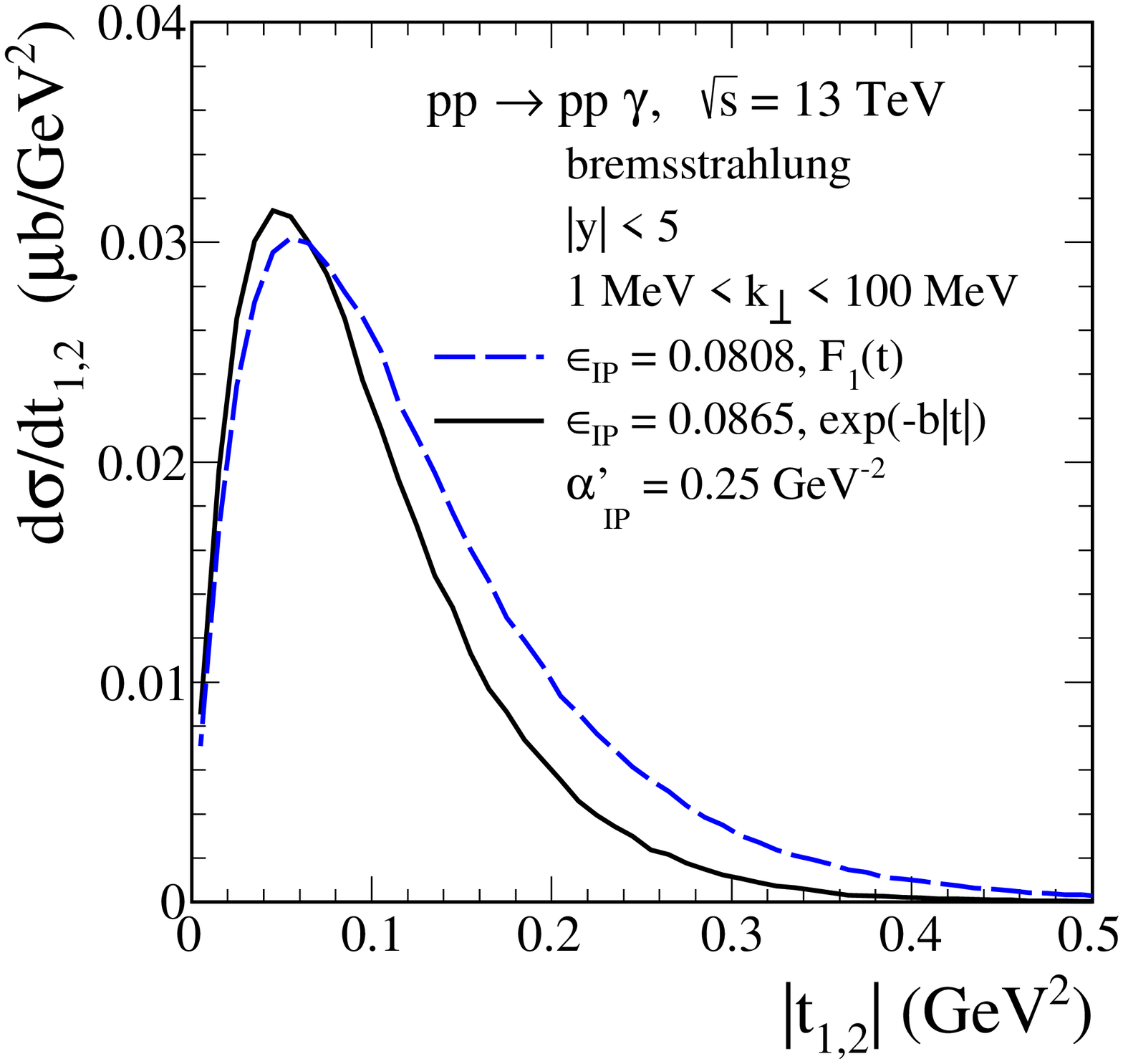}
\includegraphics[width=0.49\textwidth]{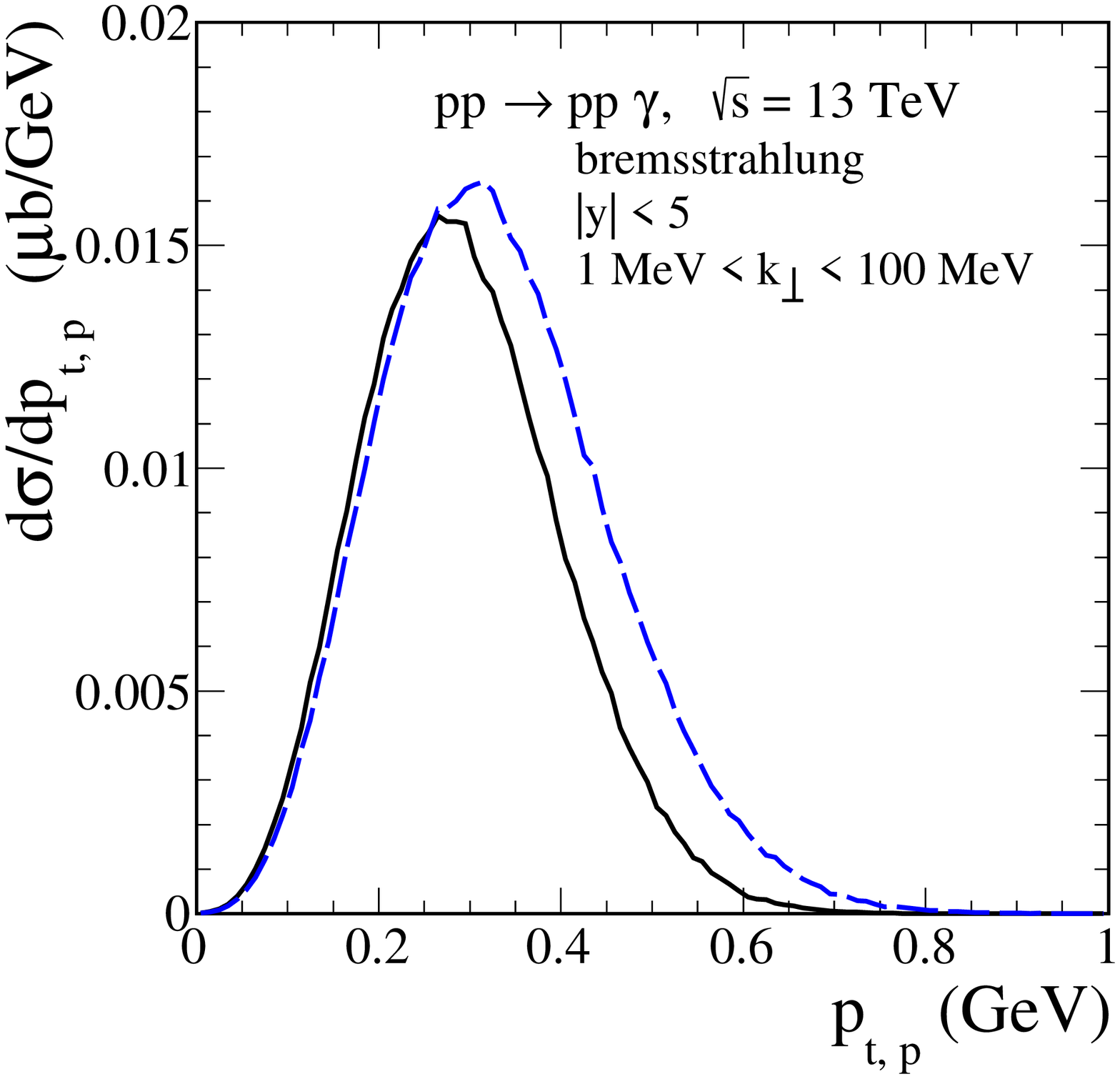}
\caption{\label{fig:ff}
\small
The distributions in 4-momentum transfer squared
$|t_{1,2}|$ and in the transverse momentum 
of the outgoing protons for the $pp \to pp \gamma$ reaction
calculated for $\sqrt{s} = 13$~TeV, $|\rm y| < 5$,
and $1\, {\rm MeV} < k_{\perp} < 100\, {\rm MeV}$.
The meaning of the lines is the same as in Fig.~\ref{fig:dsig_dt}.}
\end{figure}

In Fig.~\ref{fig:ff}, we show
the distributions in $|t_{1,2}|$,
where $t_{1,2}$ is either $t_{1}$ or $t_{2}$ 
defined in (\ref{2.17}),
and in transverse momentum of the proton
(here, $p_{t,p} = |\bptpa|$ or $|\bptpb|$).
At $t_{1,2} = 0$ and $p_{t,p} = 0$, 
all contributions vanish.
We can see how the $t_{1,2}$ dependence
is sensitive to the form factor in the $\Pom pp$ vertex (\ref{A11})
and $\epsilon_{\Pom}$ the pomeron intercept parameter; 
see the discussion after Eq.~(\ref{A2}).
In the following, we take in our calculations
$\epsilon_{\Pom} = 0.0865$ and 
the exponential pomeron-proton form factor
$F(t) = \exp(-b\,|t|)$ with $b = 2.95$~GeV$^{-2}$
adjusted to the TOTEM data (see Fig.~\ref{fig:dsig_dt}).
We see from Fig.~\ref{fig:ff} that photons with
$1\, {\rm MeV} < k_{\perp} < 100\, {\rm MeV}$
and $|\rm y| < 5$ come predominantly from $pp$ collisions
with momentum transfers between the protons
of order $p_{t,p} \sim \sqrt{|t_{1,2}|} \sim 0.3$~GeV.
Very small values of $p_{t,p}$ and/or 
$|t_{1,2}|$ hardly contribute.
%--------------------------------------------------------
%--------------------------------------------------------

\begin{figure}[!ht]
\includegraphics[width=0.49\textwidth]{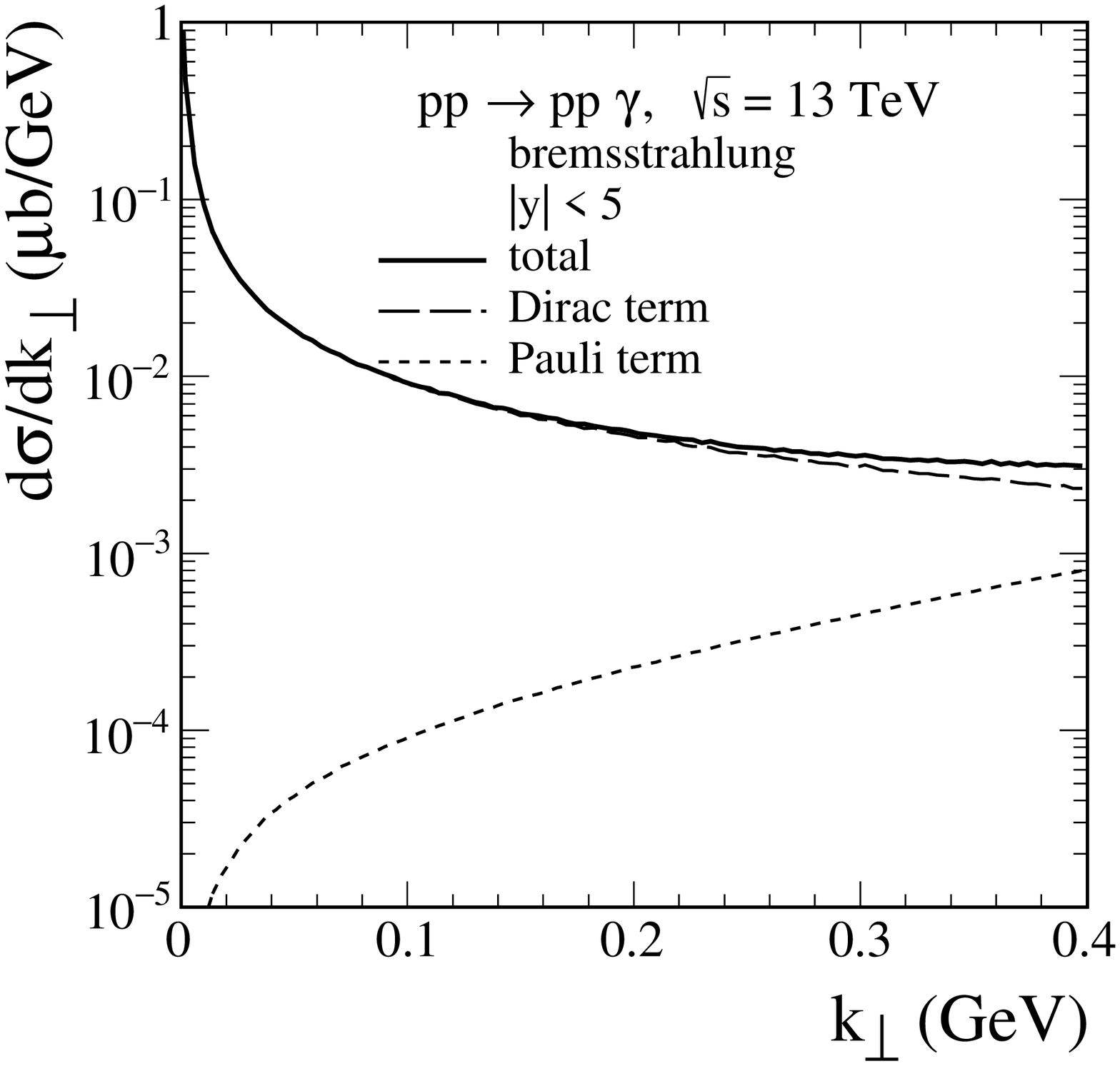}
\includegraphics[width=0.49\textwidth]{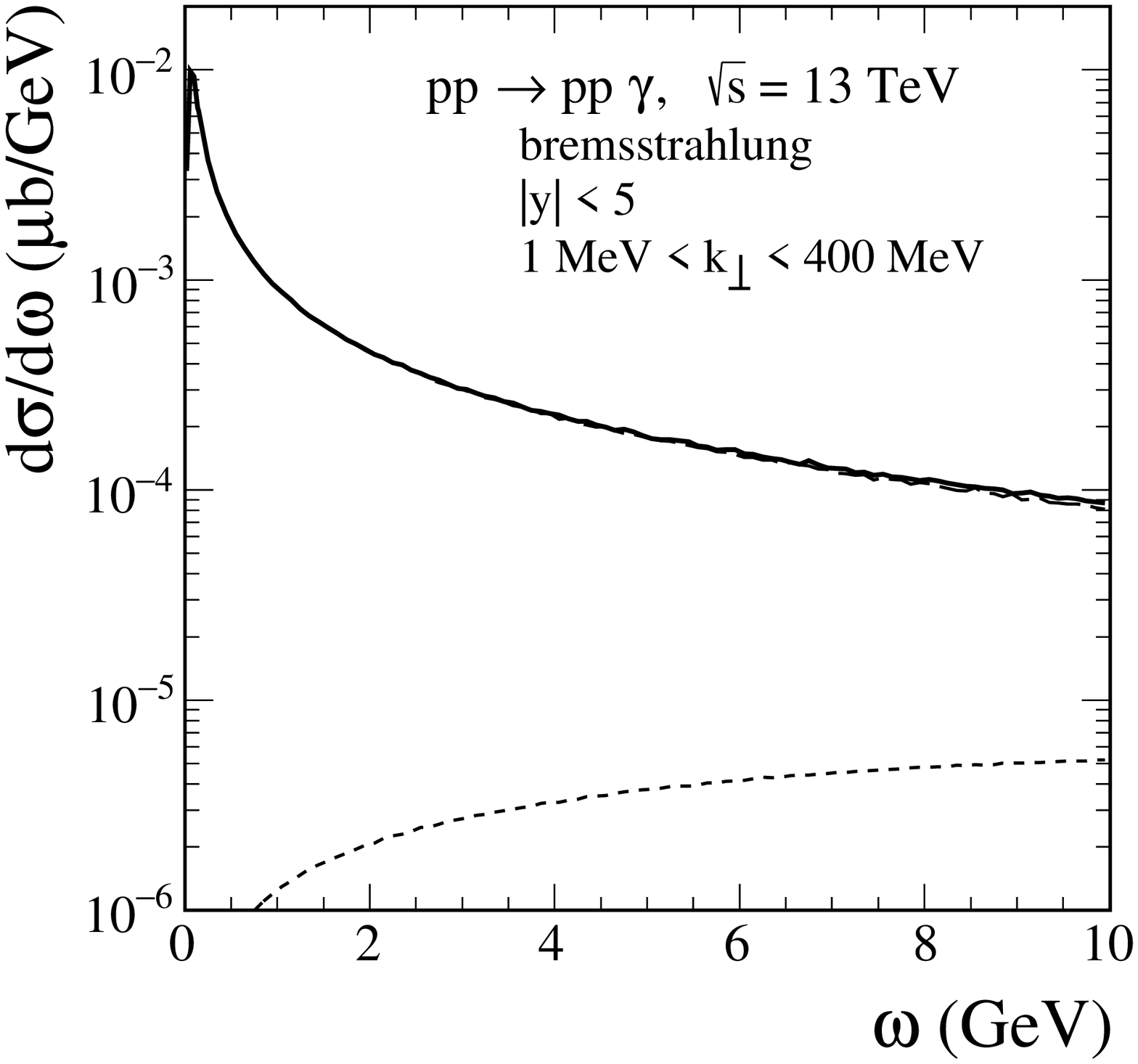}
\includegraphics[width=0.49\textwidth]{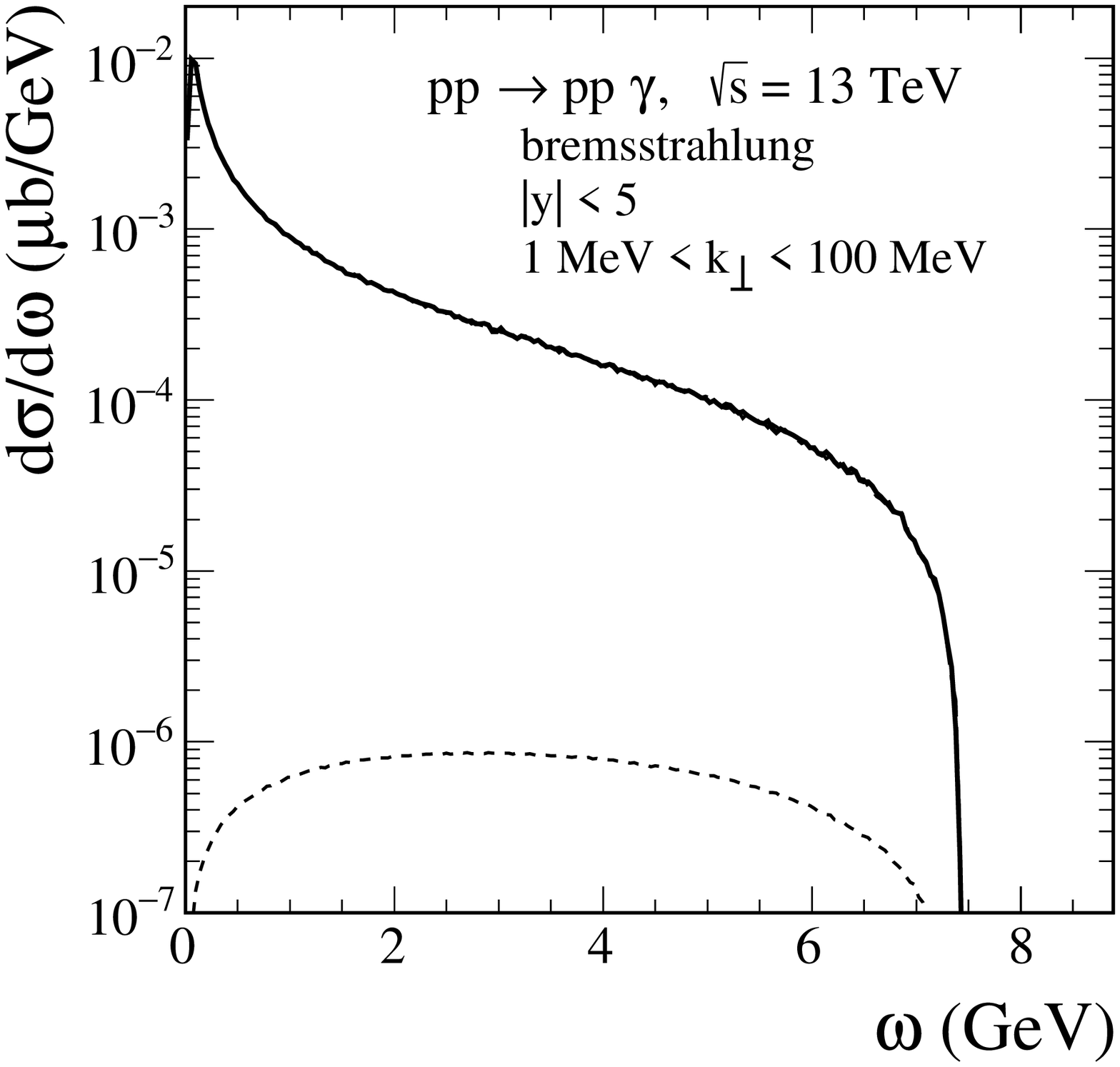}
\includegraphics[width=0.49\textwidth]{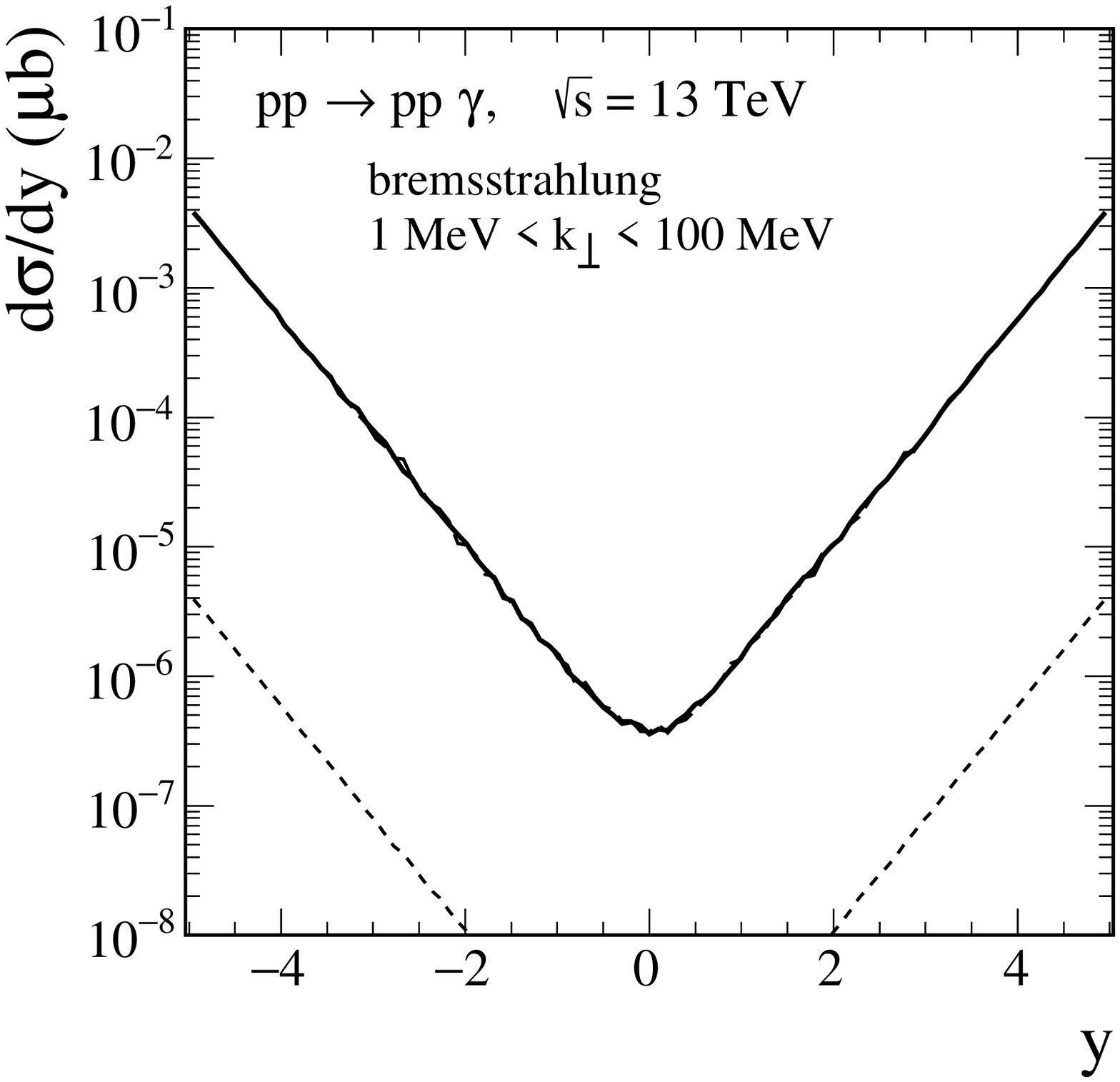}
\caption{\label{fig:deco}
\small
The differential cross sections for the $pp \to pp \gamma$ reaction
calculated for $\sqrt{s} = 13$~TeV, $|\rm y| < 5$,
and for the $k_{\perp}$ intervals as
specified in the figure legends.
Shown are our standard results (total)
and the results for the Dirac and Pauli terms alone.}
\end{figure}

In Fig.~\ref{fig:deco}, we show the results obtained for
$\sqrt{s} = 13$~TeV, $|\rm y| < 5$, and 
for two $k_{\perp}$ intervals:
$1 \, {\rm MeV} < k_{\perp} < 400 \, {\rm MeV}$ (top panels)
and $1 \, {\rm MeV} < k_{\perp} < 100 \, {\rm MeV}$ (bottom panels).
We see that in the small $k_{\perp}$ and $\omega$ regions 
the Dirac term from the $\gamma pp$ vertex function~(\ref{2.22})
dominates, while for larger values,
the anomalous magnetic moment of the proton (Pauli term) 
plays an important role.
Of course, for the complete result, all contributions
to ${\cal M}_{\mu}^{(\rm standard)}$
from the diagrams of Fig.~\ref{fig:pp_pp_gam}
with Dirac and Pauli terms have to be added coherently.
For more details on the size of various contributions to
${\cal M}_{\mu}^{(\rm standard)}$, we refer
to the discussions in Appendix~\ref{sec:appendixB}.
We get the integrated cross sections
for $\sqrt{s} = 13$~TeV and
in the $k_{\perp}$ range 
$1 \, {\rm MeV} < k_{\perp} < 100 \, {\rm MeV}$:
$\sigma(pp \to pp \gamma) = 0.21$~nb for $|\rm y| < 3.5$
and $\sigma(pp \to pp \gamma) = 4.01$~nb 
for~$3.5 < |\rm y| < 5$.
%--------------------------------------------------------
%--------------------------------------------------------

\begin{figure}[!ht]
\includegraphics[width=0.50\textwidth]{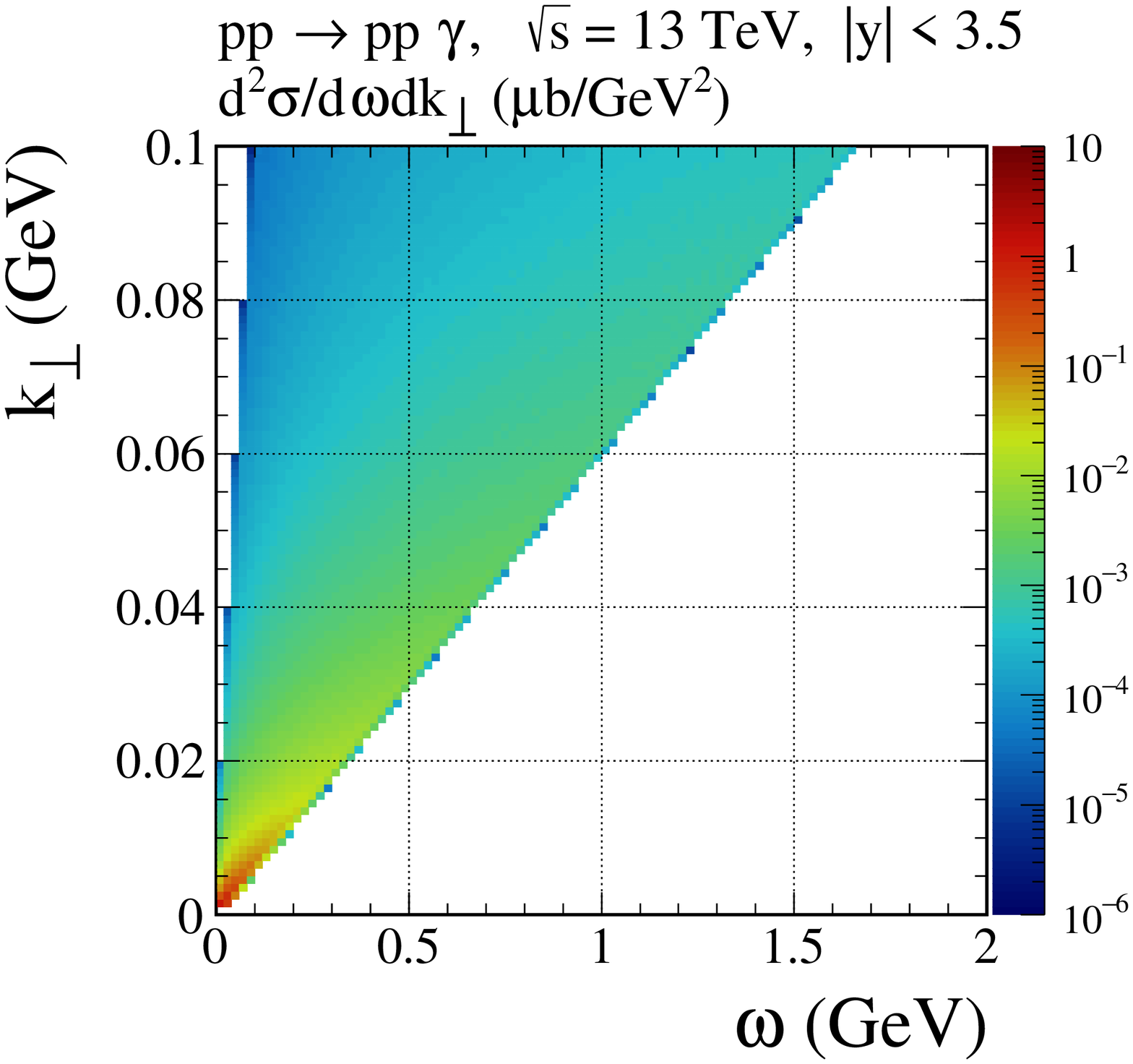}
\includegraphics[width=0.49\textwidth]{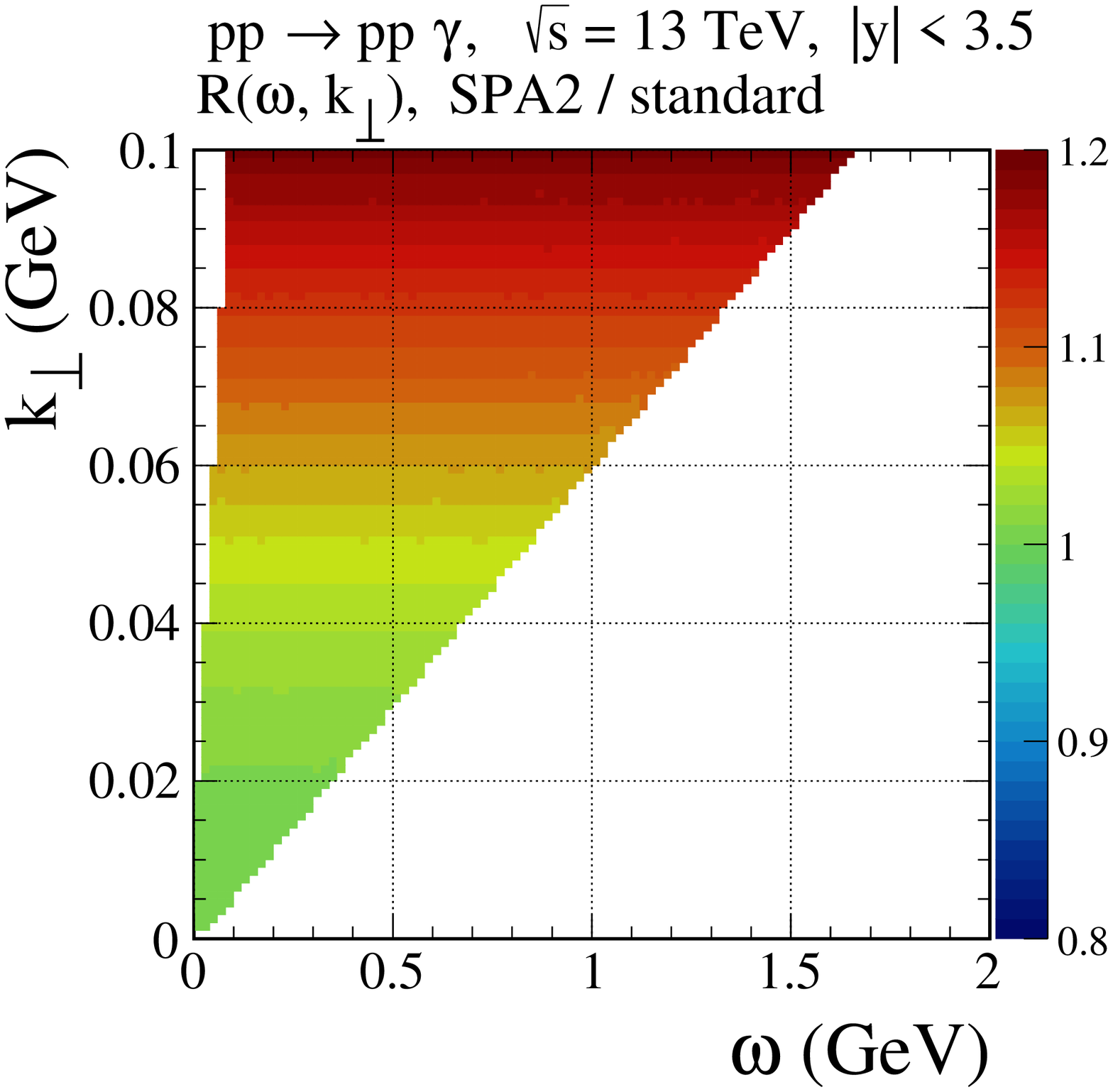}\\
\includegraphics[width=0.50\textwidth]{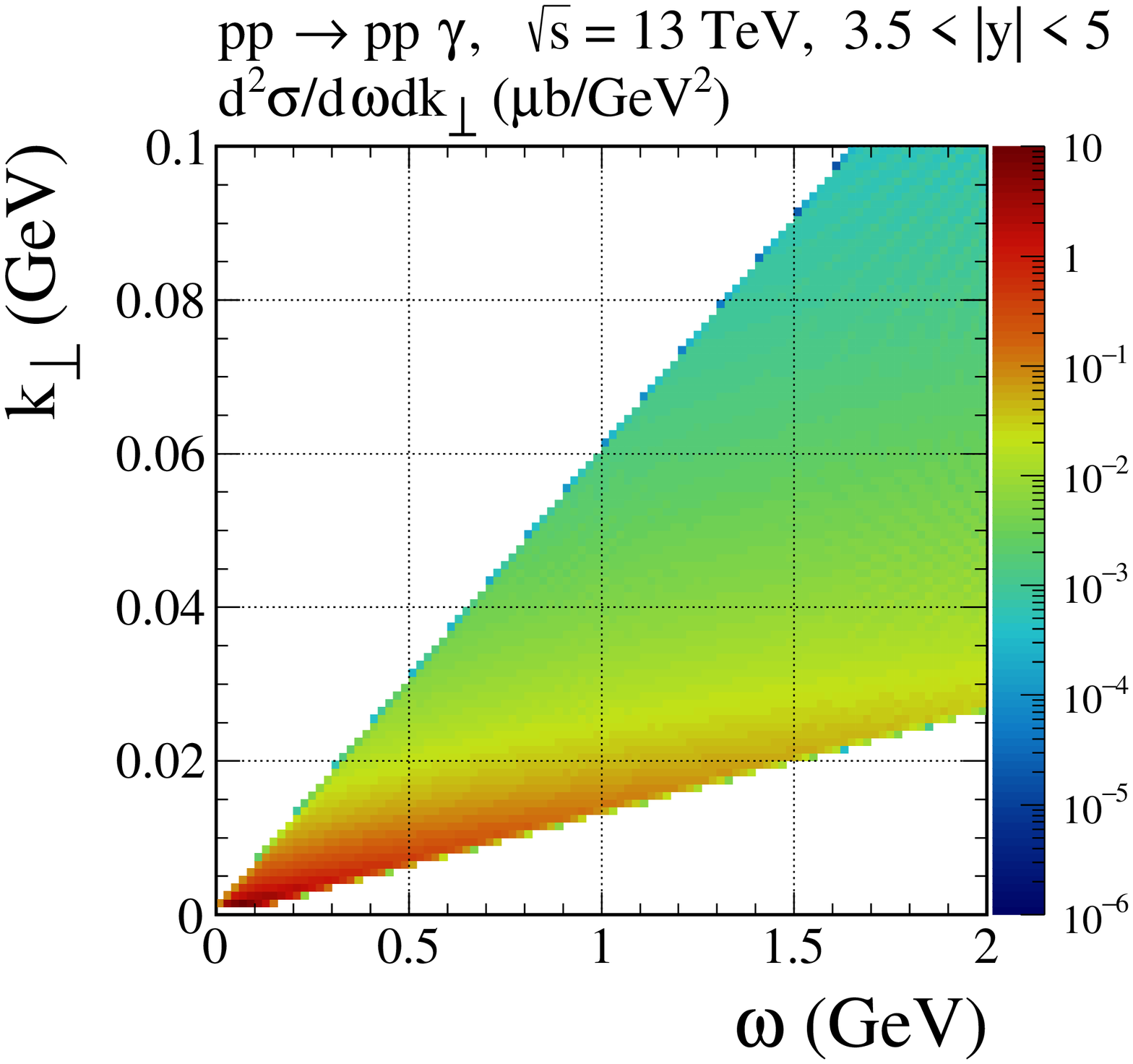}
\includegraphics[width=0.49\textwidth]{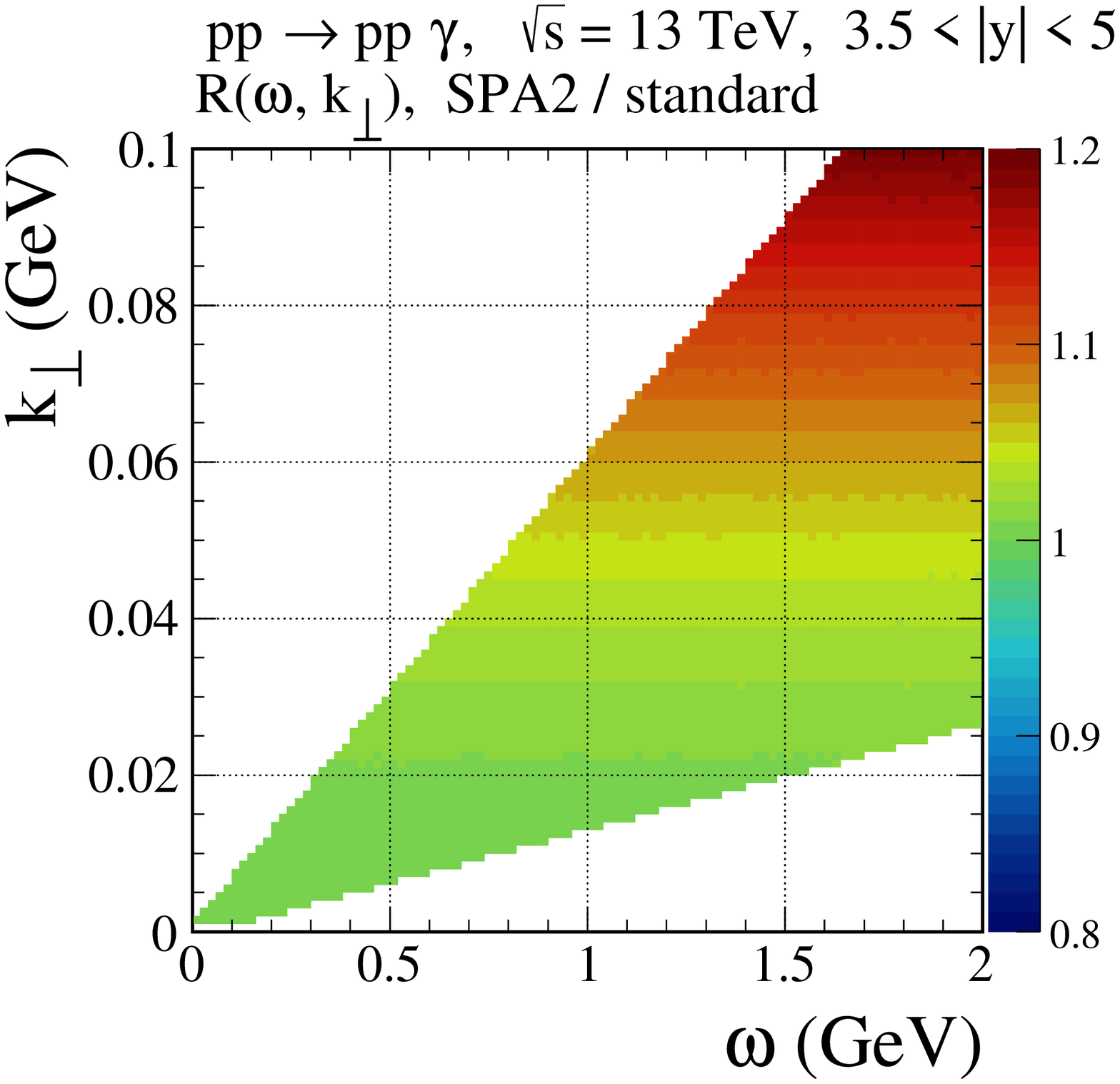}
\caption{\label{fig:map_y24}
\small
The left panels show the cross section
$d^{2}\sigma / d\omega dk_{\perp}$
as a function of ($\omega$, $k_{\perp}$)
calculated for $\sqrt{s} = 13$~TeV.
The top panels are for
$1 \, {\rm MeV} < k_{\perp} < 100 \, {\rm MeV}$,
$|\rm y| < 3.5$, and
the bottom panels are for $3.5 < |\rm y| < 5$.
The right panels show the ratio 
${\rm R}(\omega, k_{\perp})$
as defined in (\ref{ratio}).}
\end{figure}

In Fig.~\ref{fig:map_y24} (left panels),
we show the two-dimensional differential cross sections 
in the $\omega$-$k_{\perp}$ plane
calculated for $\sqrt{s} = 13$~TeV,
$1 \, {\rm MeV} < k_{\perp} < 100 \, {\rm MeV}$,
$|\rm y| < 3.5$ (the top panel) and 
$3.5 < |\rm y| < 5$ (the bottom panel).
The latter rapidity range is relevant
for the planned ALICE~3 measurements 
\cite{Adamova:2019vkf,QM2022_PBM,EMMI_RRTF} 
of ultrasoft photons.
Large ${\rm y}$ is near the $\omega$ axis,
and ${\rm y} = 0$ corresponds to the line $\omega = k_{\perp}$,
both in accordance with $\omega = k_{\perp} \cosh{\rm y}$.
%An upper cut on ${\rm y}$ is effecting 
%the upper limit of the allowed $\omega$,
%since $\omega = k_{\perp} \cosh{\rm y}$.
The phase-space region where
$\omega < k_{\perp}$ is forbidden.
We see that, due to our cuts on ${\rm y}$ and $k_{\perp}$
applied in phase space,
there are additional regions that are not accessible kinematically.

In the right panels of Fig.~\ref{fig:map_y24},
we show the ratio
\begin{eqnarray}
{\rm R}(\omega, k_{\perp}) =
\frac{d^{2}\sigma_{\rm SPA2} / d\omega dk_{\perp}}
     {d^{2}\sigma_{\rm standard} / d\omega dk_{\perp}}\,.
\label{ratio}
\end{eqnarray}
One can see that the SPA2 given by (\ref{4.42})
stays within 1$\, \%$ accuracy
for $k_{\perp} \lesssim 22$~MeV
and $\omega \lesssim 0.35$~GeV 
considering $|\rm y| < 3.5$
and up to $\omega \cong 1.7$~GeV 
for $3.5 < |\rm y| < 5.0$.
It is difficult to draw the ratio 
${\rm R}(\omega, k_{\perp})$ for SPA1/standard
due to different integration procedures in two different codes:
one for the exact three-body phase space (standard approach) 
and one for the two-body phase space supplemented 
by additional integration over photon 3-momentum (SPA1).
%--------------------------------------------------------
%--------------------------------------------------------

\begin{figure}[!ht]
\includegraphics[width=0.49\textwidth]{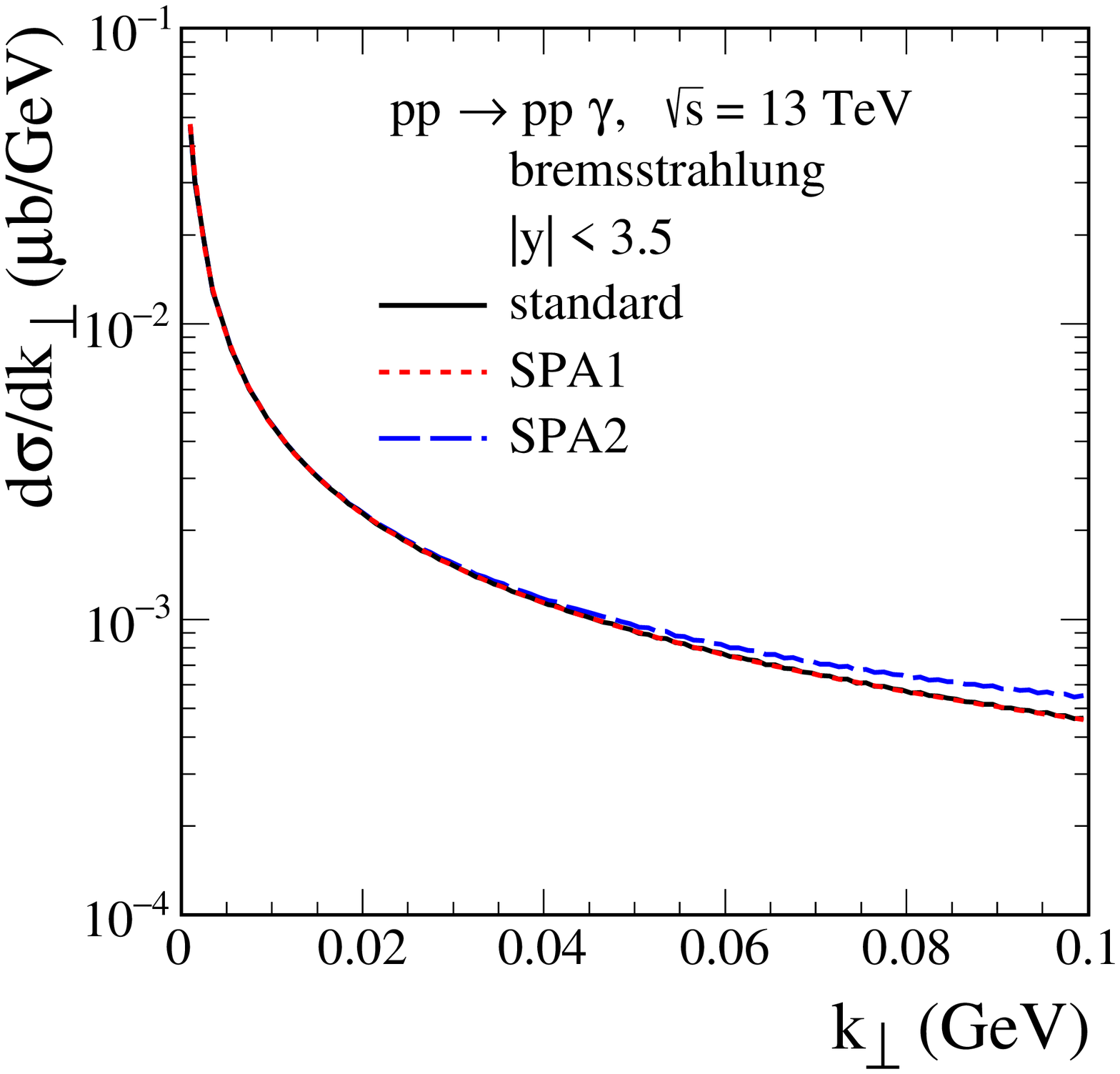}
\includegraphics[width=0.49\textwidth]{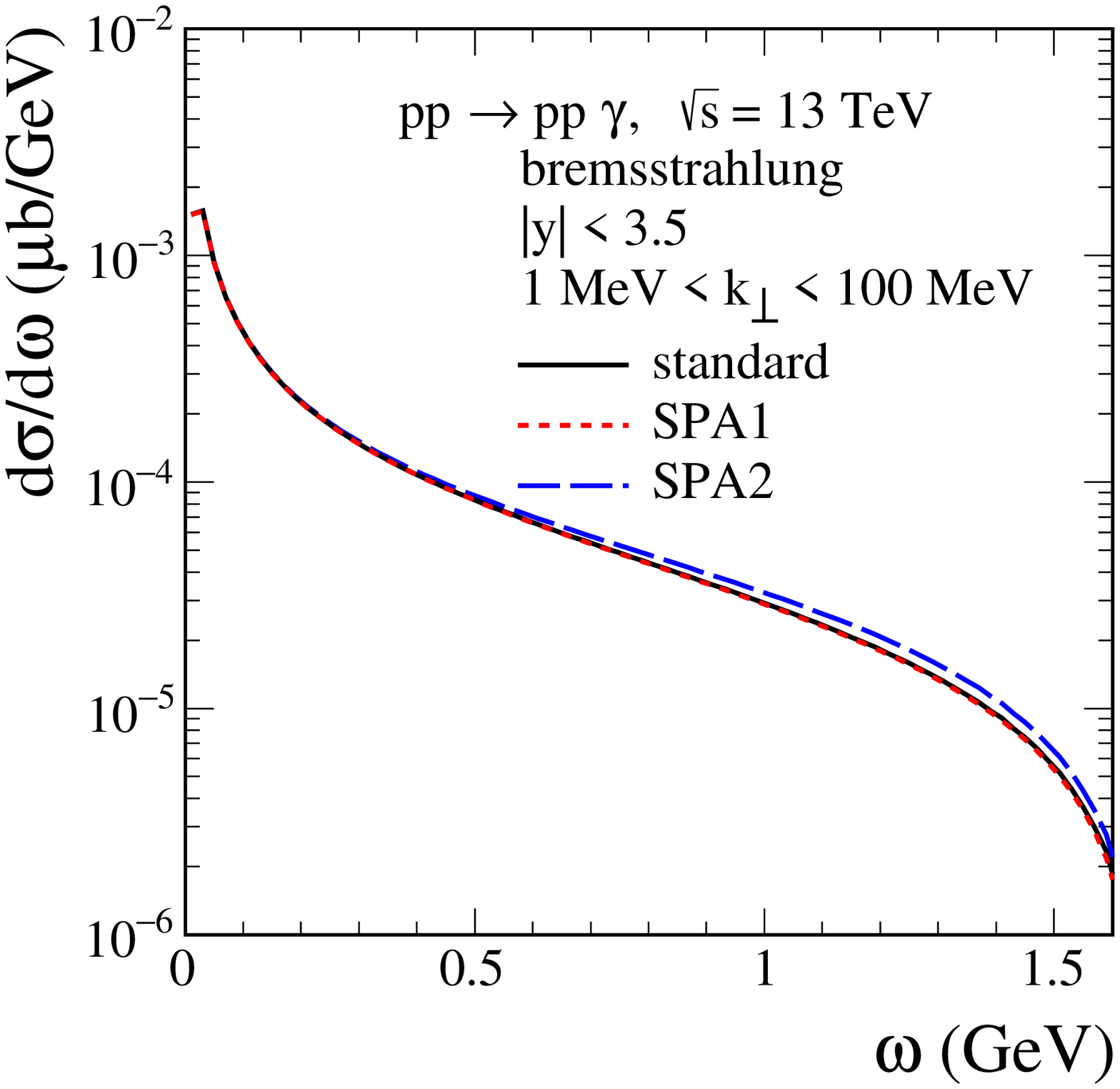}\\
\includegraphics[width=0.49\textwidth]{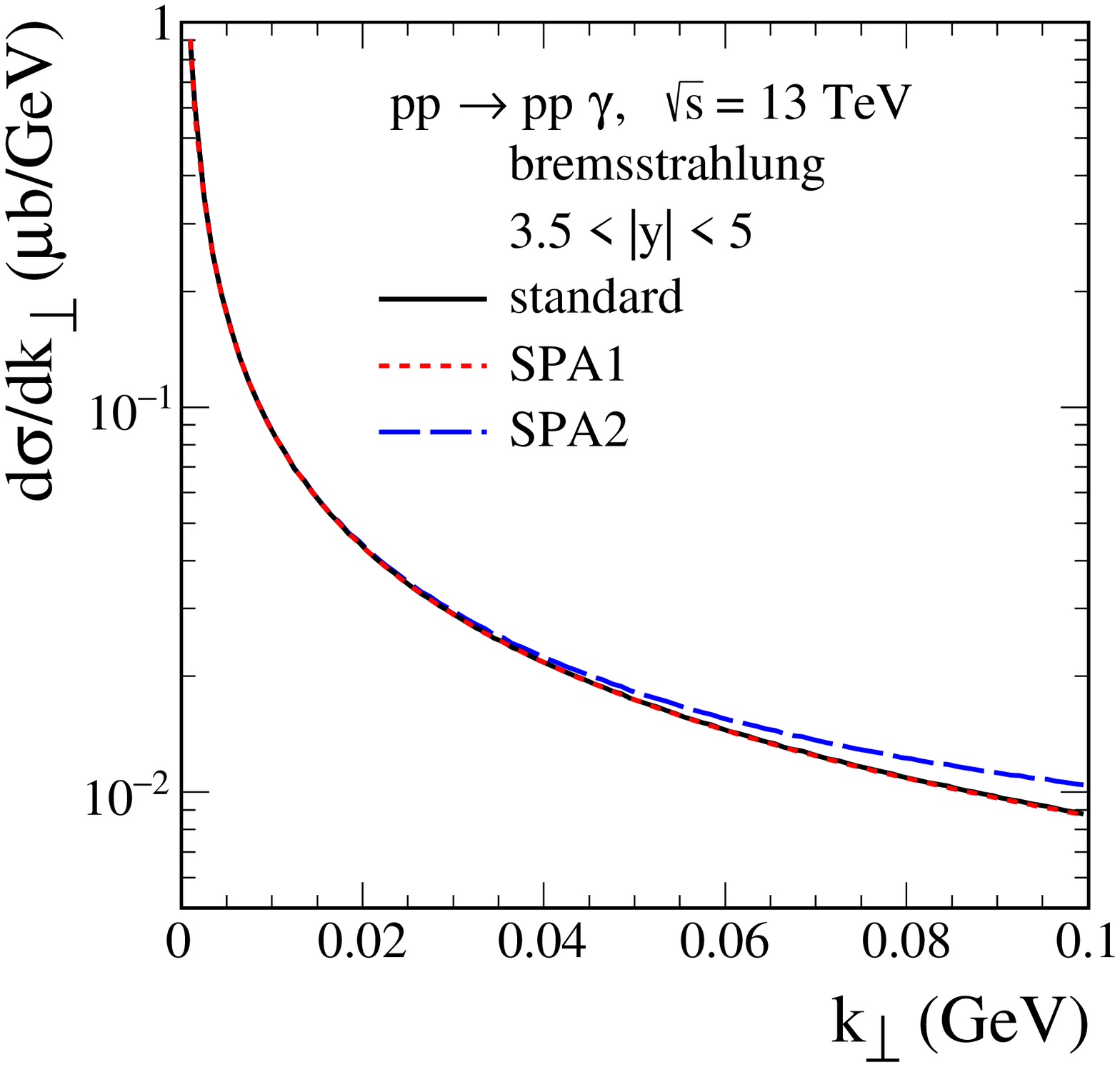}
\includegraphics[width=0.49\textwidth]{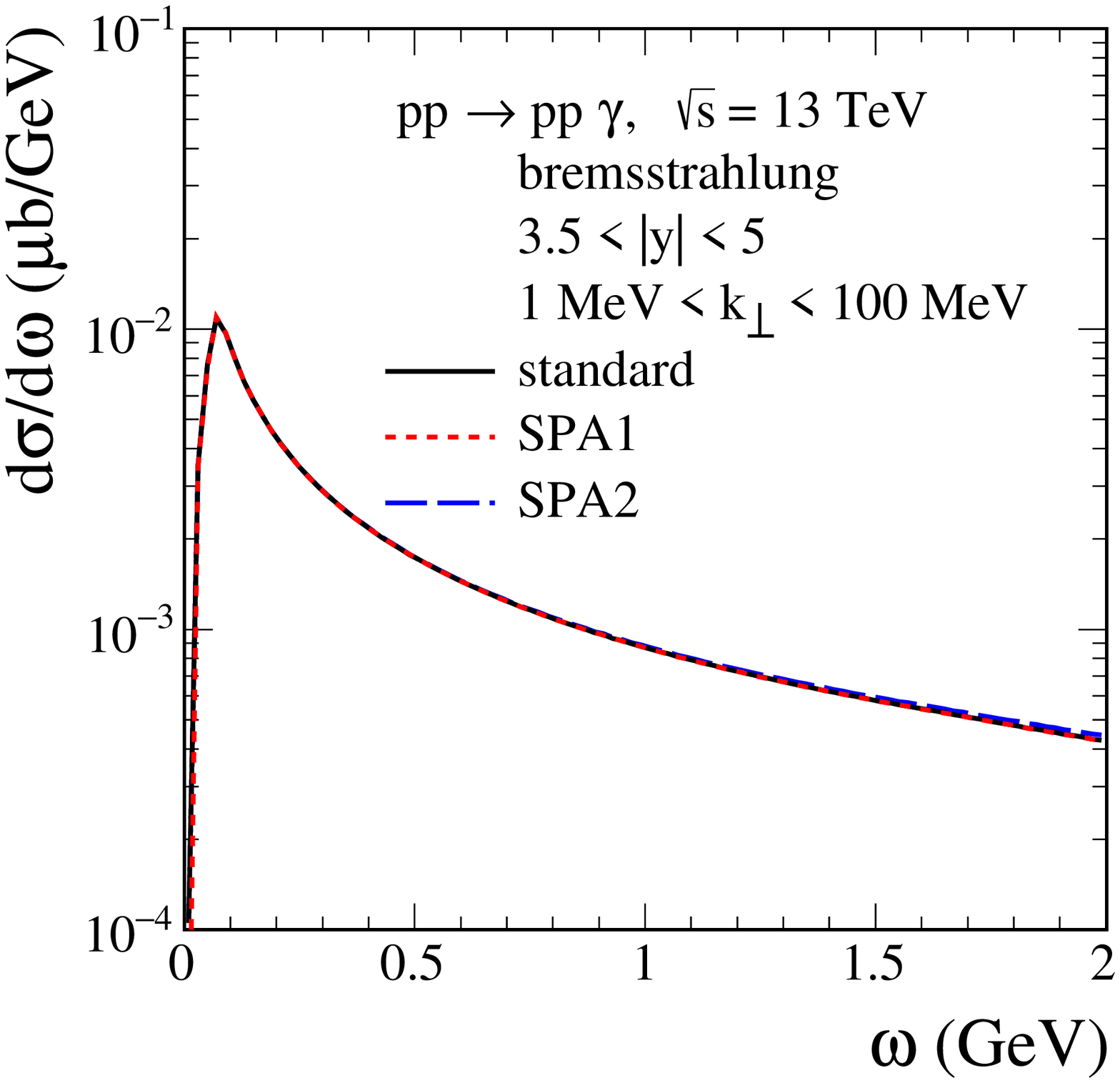}
\caption{\label{fig:y24}
\small
The differential distributions 
in transverse momentum of the photon
and in the energy of the photon
for the $pp \to pp \gamma$ reaction.
The calculations were done for $\sqrt{s} = 13$~TeV,
$1 \, {\rm MeV} < k_{\perp} < 100 \, {\rm MeV}$,
$|\rm y| < 3.5$ (the top panels)
and $3.5 < |\rm y| < 5$ (the bottom panels).
The black solid line corresponds to the standard result
(\ref{2.50}),
the red dotted line corresponds to SPA1~(\ref{4.39}), and
the blue long-dashed line corresponds to SPA2~(\ref{4.42}).}
\end{figure}

In Fig.~\ref{fig:y24}, we compare our standard result
to various SPAs on a semilogarithmic scale.
We see that both, the SPA1 (\ref{4.39}) and SPA2 (\ref{4.42}), 
follow the standard results very well.
Surprisingly, the SPA1, which does not have the correct
energy-momentum relations, fares somewhat better than SPA2.
But let us now have a closer look at these kinematic regions
on a linear scale.
%--------------------------------------------------------
%--------------------------------------------------------

\begin{figure}[!ht]
\includegraphics[width=0.49\textwidth]{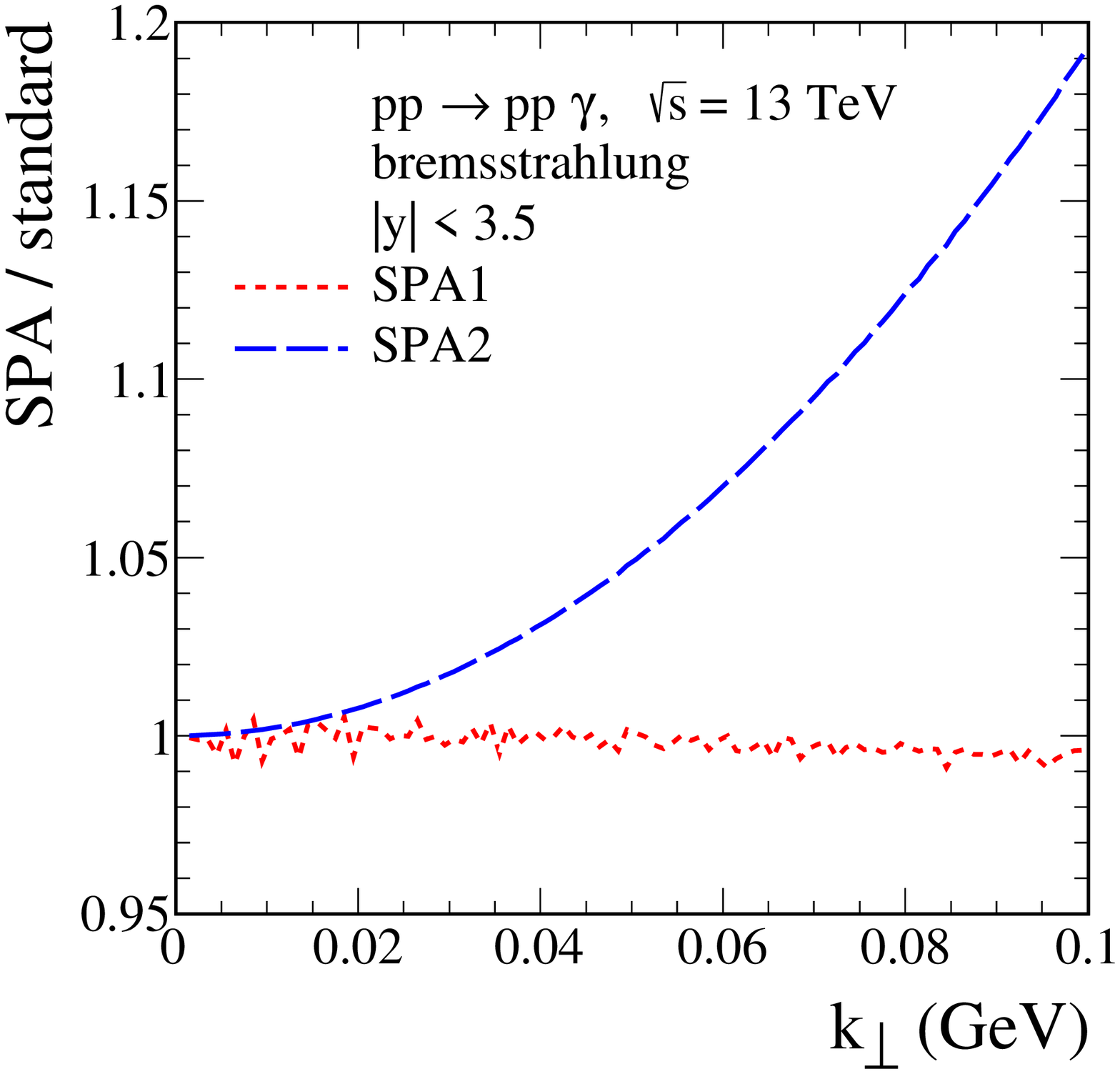}
\includegraphics[width=0.49\textwidth]{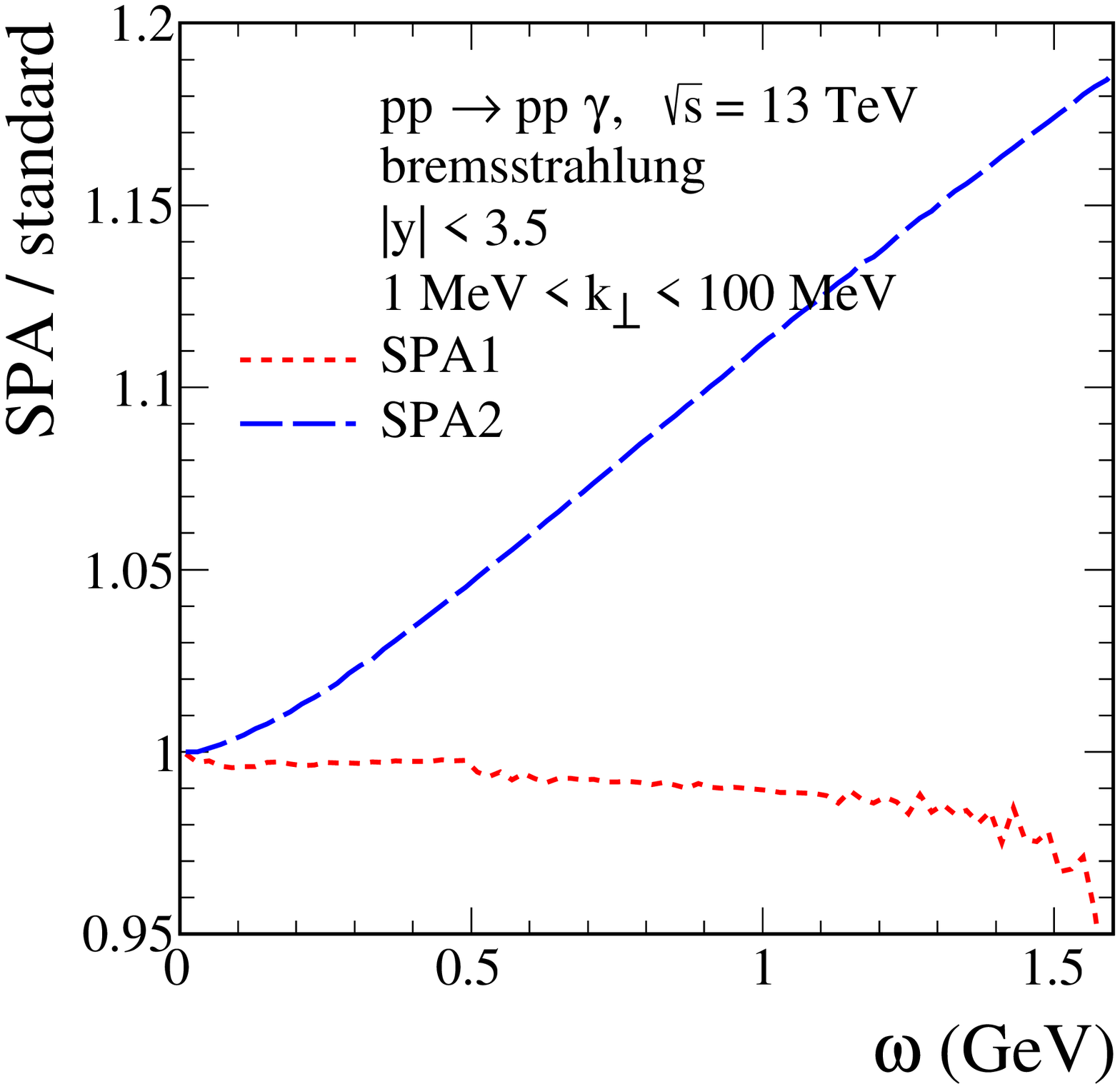}\\
\includegraphics[width=0.49\textwidth]{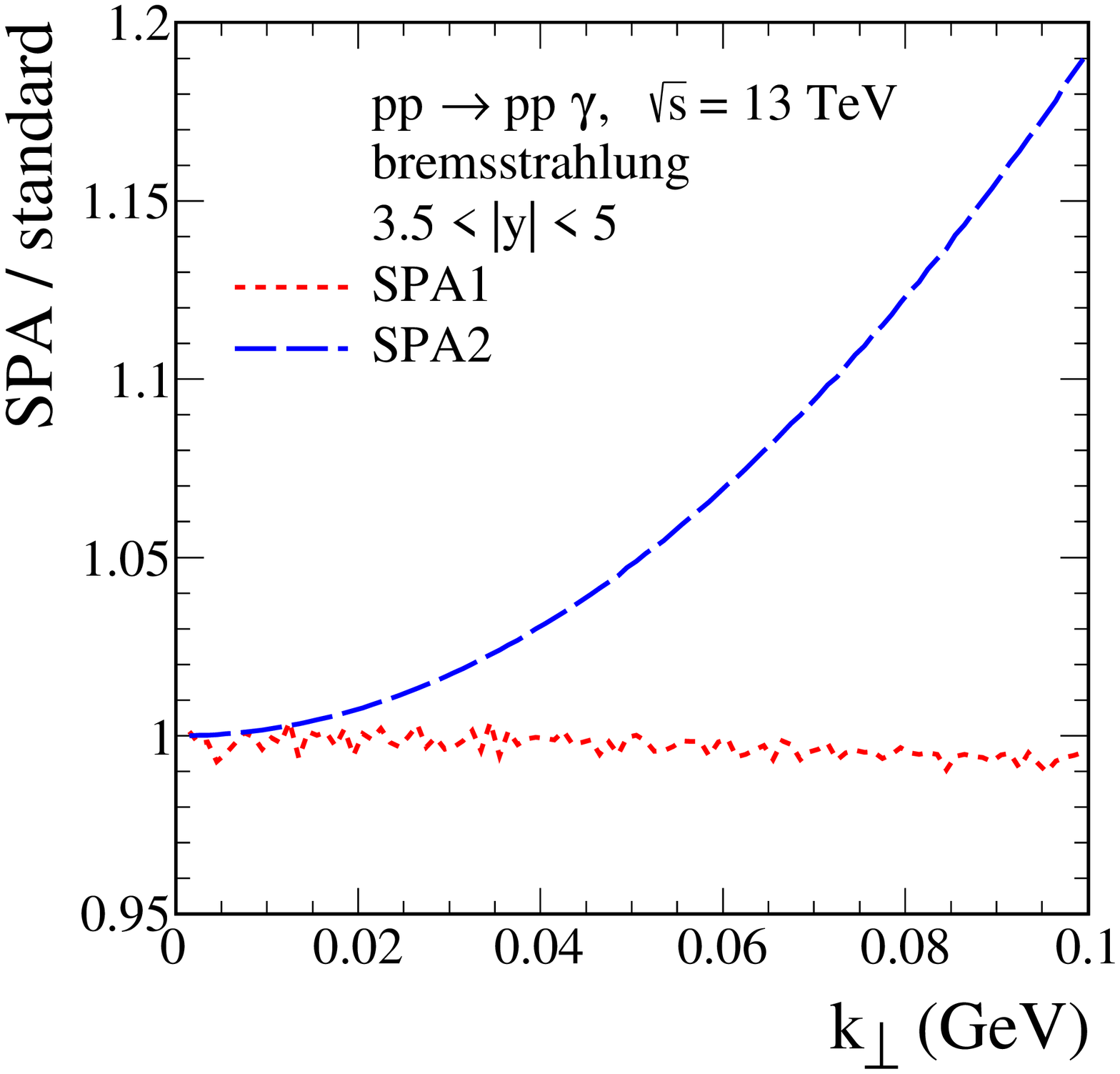}
\includegraphics[width=0.49\textwidth]{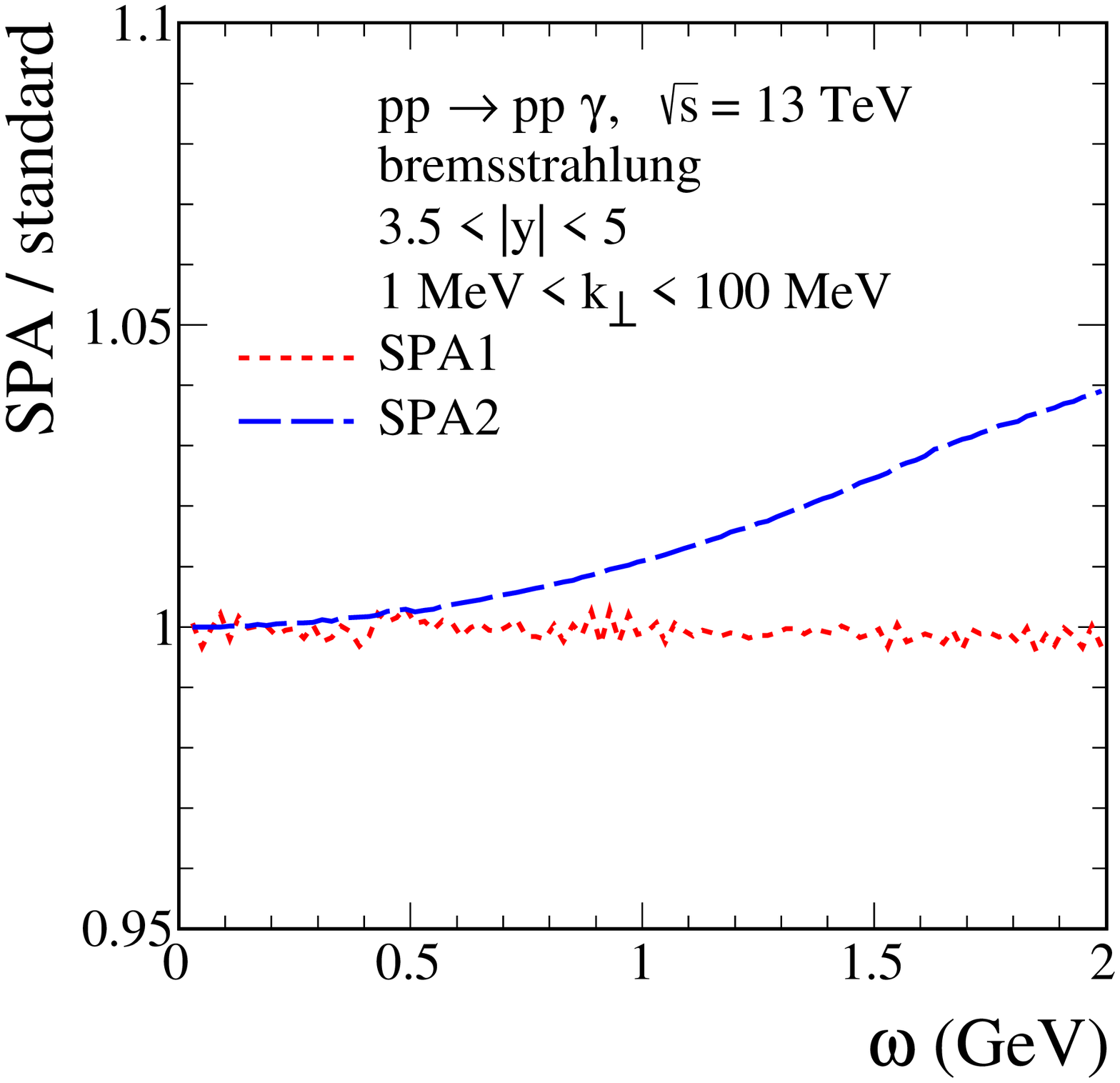}
\caption{\label{fig:ratios_y24}
\small
The ratios $\sigma_{\rm SPA}/\sigma_{\rm standard}$
given by (\ref{ratio_kt}) and (\ref{ratio_omega}), respectively.
The red dotted line corresponds to SPA1~(\ref{4.39}), 
and the blue long-dashed line corresponds to SPA2~(\ref{4.42}).
The oscillations in the SPA1 results 
are of numerical origin.}
\end{figure}

Figure~\ref{fig:ratios_y24} shows the ratios 
of the SPAs to the standard cross sections,
\begin{eqnarray}
&&\frac{d\sigma_{\rm SPA}/dk_{\perp}}{d\sigma_{\rm standard}/dk_{\perp}}\,
\label{ratio_kt} \,, \\
&&\frac{d\sigma_{\rm SPA}/d\omega}{d\sigma_{\rm standard}/d\omega}\,,
\label{ratio_omega}
\end{eqnarray}
as functions of $k_{\perp}$ and $\omega$, respectively.
The fluctuations of the ratio SPA1/standard 
(see the red dashed lines)
are due to a different organization of integration
in the two codes.
One can see that the deviations of the SPA1 
from the standard results
are up to around 1$\,\%$ for $|\rm y| < 3.5$,
$1 \, {\rm MeV} < k_{\perp} < 100 \, {\rm MeV}$, and
$\omega \lesssim 0.7$~GeV.
For the forward region, $3.5 < |\rm y| < 5$,
this accuracy occurs even up to larger $\omega \simeq 2$~GeV.
For the SPA2, the deviations increase rapidly 
with growing $k_{\perp}$ and $\omega$
(see also the right panels of Fig.~\ref{fig:map_y24}).
%--------------------------------------------------------
%--------------------------------------------------------

\begin{figure}[!ht]
\includegraphics[width=0.48\textwidth]{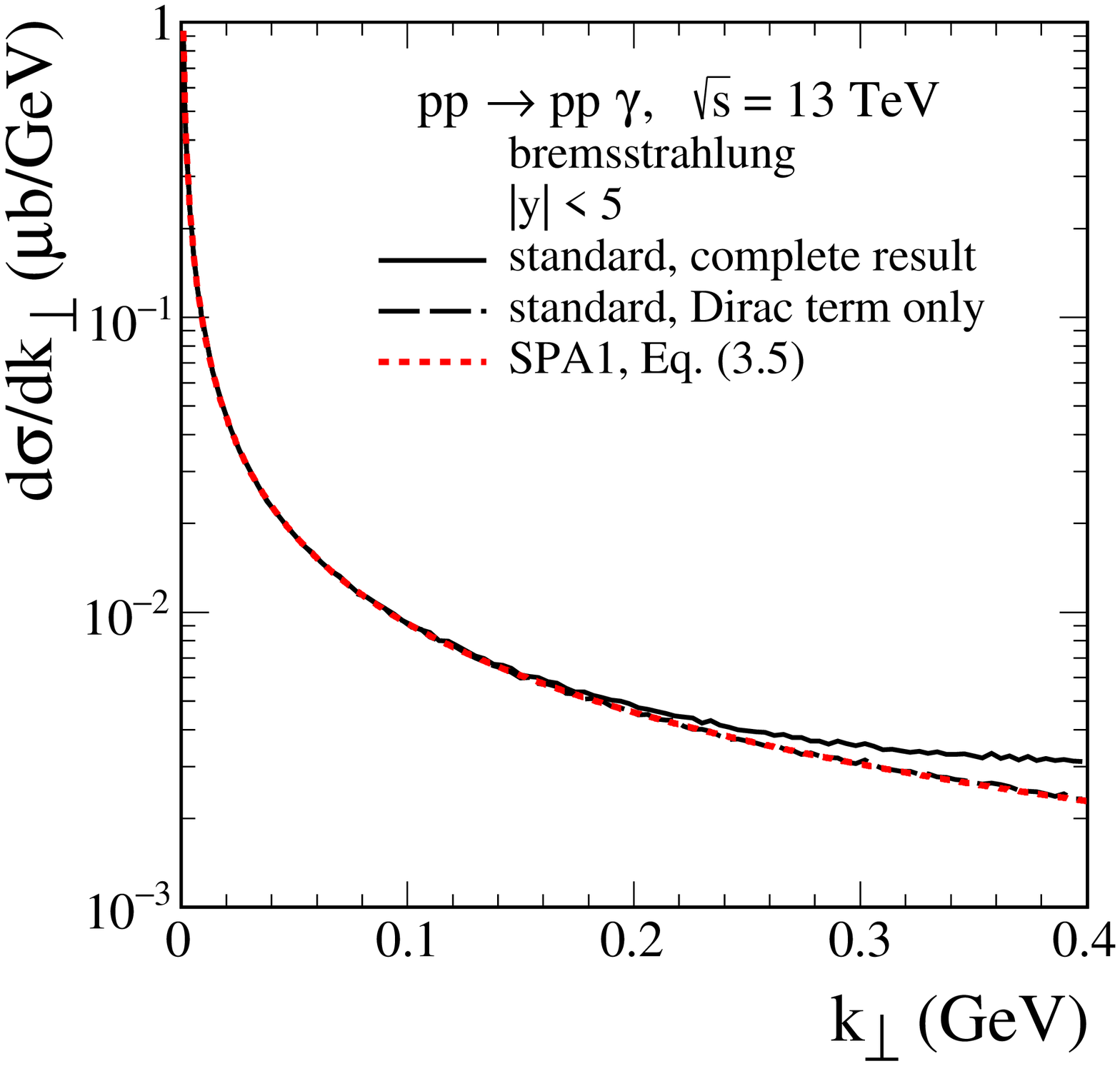}
\includegraphics[width=0.48\textwidth]{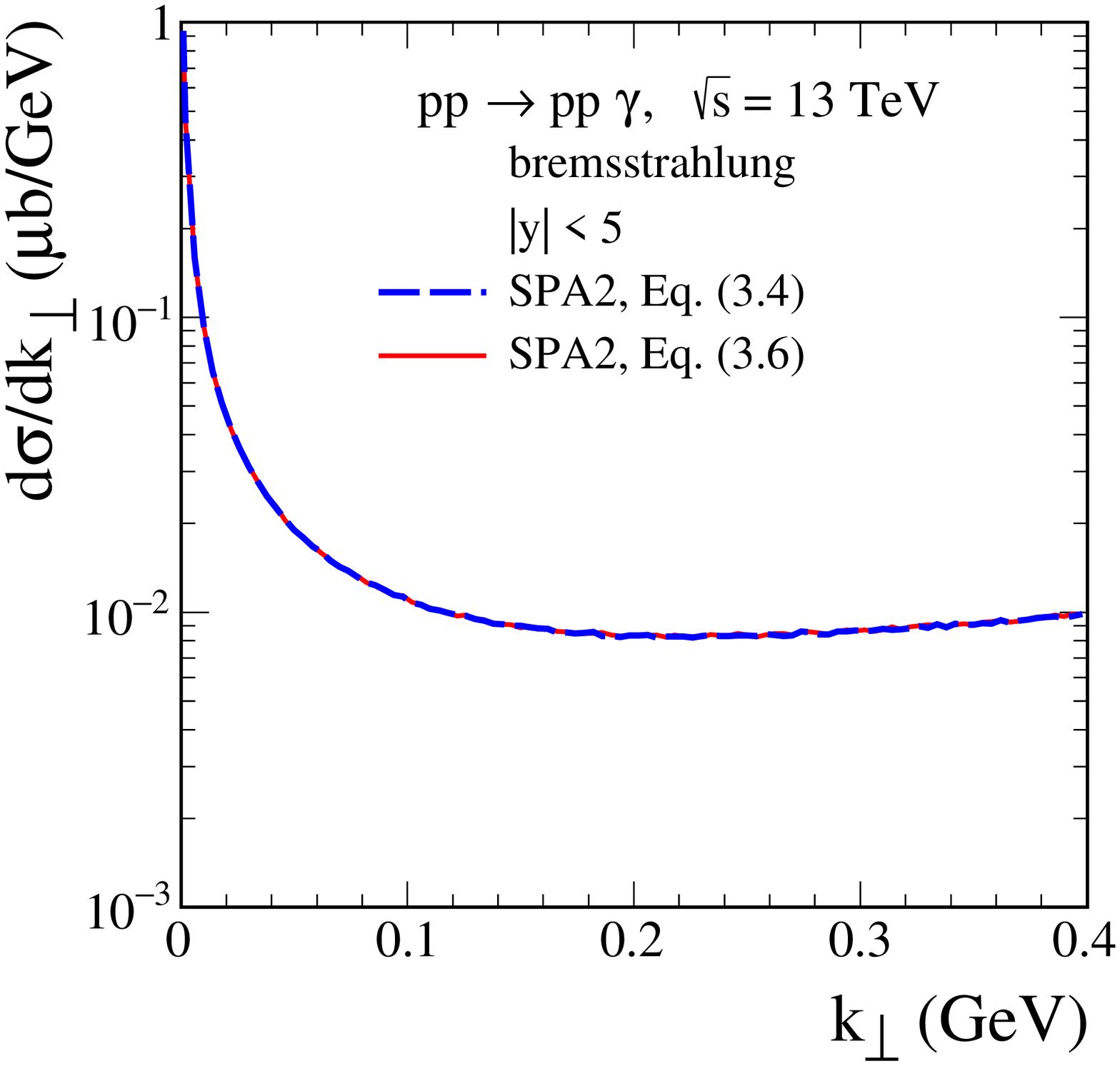}
\caption{\label{fig:SPA2_approx}
\small
Comparison of the cross sections $d\sigma/dk_{\perp}$
for our standard bremsstrahlung model and the SPA1 (left panel) 
and for the two versions of the SPA2 (right panel).
In the left panel, we show the standard complete result
and the result for the Dirac term alone from Fig.~\ref{fig:deco}
together with the SPA1 result (\ref{4.39_aux}).
In the right panel, 
the blue long-dashed line 
corresponds to (\ref{4.42}),
while the red solid line corresponds to (\ref{SPA2_approx}).}
\end{figure}

In Fig.~\ref{fig:SPA2_approx},
we show the results for the SPA1 (\ref{4.39_aux}) and 
SPA2 (\ref{SPA2_approx}) Ans{\"a}tze
using the high-energy small-angle approximation.
The calculations were done for $\sqrt{s} = 13$~TeV,
$1 \, {\rm MeV} < k_{\perp} < 400 \, {\rm MeV}$, and
$|\rm y| < 5$.
From the left panel, we see that
the SPA1 is in very good agreement with our 
standard result if we include there only the Dirac terms.
We have checked numerically that
the results from (\ref{4.39}) and (\ref{4.39_aux}) overlap.
From the right panel, we see that
the SPA2 results (\ref{4.42}) and (\ref{SPA2_approx})
are very close to each other.
%-------------------------------------------------------
%--------------------------------------------------------

%-----------------------------------
\subsection{Comments on the photon radiation in connection with diffractive excitation of the proton}
\label{sec:3C}
%-----------------------------------

\begin{figure}[!h]
\includegraphics[width=5.5cm]{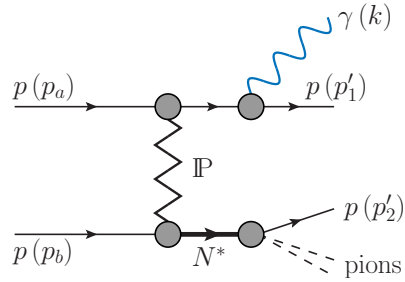}
\caption{A diagram for the reaction (\ref{3C_4.40})
with photon emission from the proton line and hadronic
decay of the $N^{*}$.
The diagrams where the photon is attached 
in all other possible ways to the charged particle lines
are not shown.}
\label{fig:100}
\end{figure}

\begin{figure}[!h]
(a)\includegraphics[width=5.5cm]{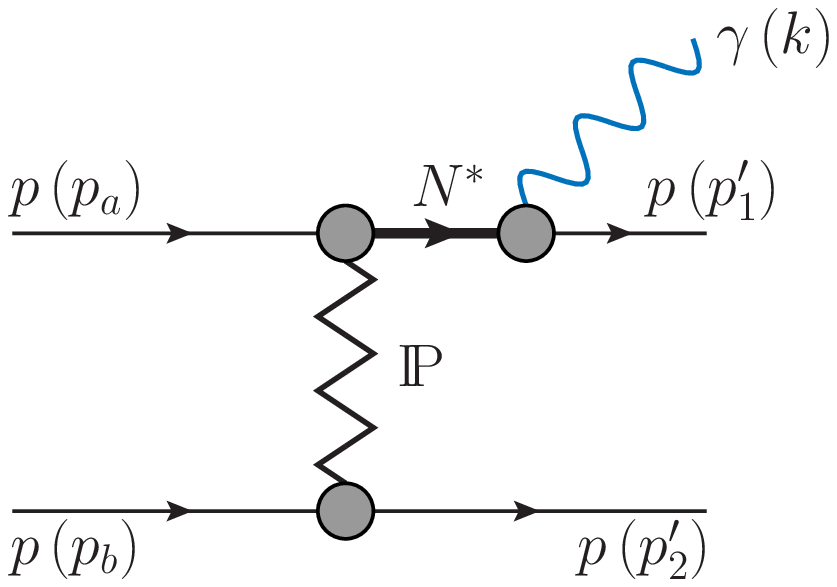} \quad
(b)\includegraphics[width=5.5cm]{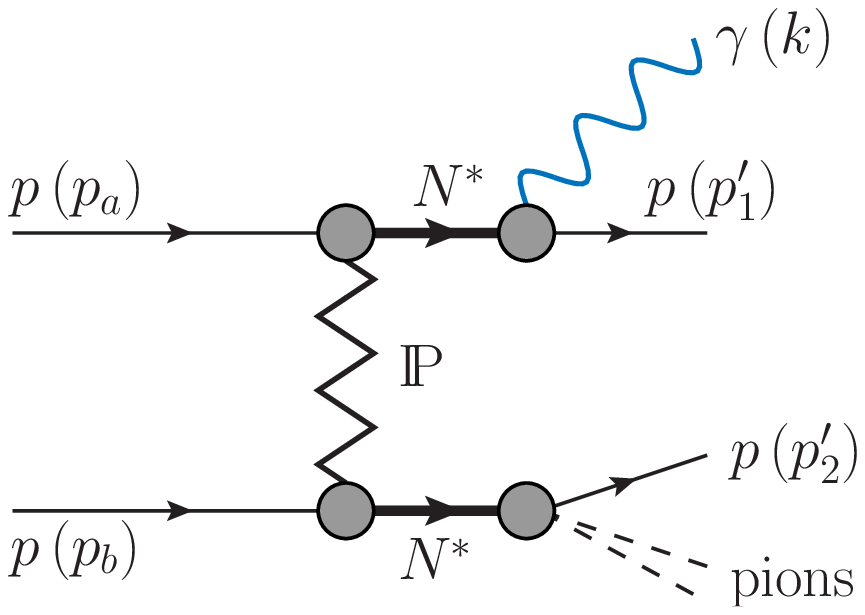}
\caption{Examples of diagrams: (a) for the reaction (\ref{3C_4.41})
and (b) for the reaction (\ref{3C_4.42}).}
\label{fig:101}
\end{figure}

One may ask about the role of diffractive excitations
of the protons in connection with photon radiation.
There are two types of exclusive processes to be considered,
\begin{eqnarray}
p + p \to p  + \gamma + N^{*}\,,
\label{3C_4.40} 
\end{eqnarray}
with the $N^{*}$ charged nucleon resonance 
decaying hadronically,
typically $N^{*} \to p \pi^{0}$, $p \pi^{0}\pi^{0}$, 
$p \pi^{+}\pi^{-}$, and
\begin{eqnarray}
&&p + p \to (N^{*} \to p  + \gamma) + p\,,
\label{3C_4.41} \\
&&p + p \to (N^{*} \to p  + \gamma) 
+ (N^{*} \to {\rm hadrons})\,.
\label{3C_4.42}
\end{eqnarray}
Relevant examples of diagrams for these processes
with pomeron exchange are shown in Fig.~\ref{fig:100}
for the reaction (\ref{3C_4.40})
and in Fig.~\ref{fig:101} 
for (\ref{3C_4.41}) and (\ref{3C_4.42}).
%-------------------------------------------------------------
%-------------------------------------------------------------
%-------------------------------------------------------------
%-------------------------------------------------------------

The reactions (\ref{3C_4.40}) and (\ref{3C_4.42}) lead to
a final state $p + \gamma + p + {\rm pion(s)}$.
Thus, there is no interference with the process
$p p \to p p \gamma$ which we study in this paper.
Of course, the question of whether experimentally the processes
(\ref{3C_4.40}) and (\ref{3C_4.42}) can be separated
from the elastic process (\ref{2.2})
is a very relevant one.
However, this will depend on the experiment
and must be studied by the experimentalists themselves.
The reaction (\ref{3C_4.41}), on the other hand, leads to the final state
$p p \gamma$ and, thus, should be discussed here.
Let us first note that the diagram of Fig.~\ref{fig:101}~(a)
will \underline{not} lead in the amplitude to a bremsstrahlung term
$\propto \omega^{-1}$.
Indeed, this diagram will lead in the amplitude to a factor,
for real photons where $k^{2} = 0$,
\begin{eqnarray}
\left[ (p_{1}' + k)^{2} - m_{N^{*}}^{2} 
+ i m_{N^{*}} \Gamma_{N^{*}}\right]^{-1} =
\left[ 2 (p_{1}' \cdot k) - (m_{N^{*}}^{2} - m_{p}^{2})
+ i m_{N^{*}} \Gamma_{N^{*}}\right]^{-1}\,.
\label{3C_4.43} 
\end{eqnarray}
Here, $m_{N^{*}}$ and $\Gamma_{N^{*}}$ are the mass and width
of the $N^{*}$ resonance considered.
Clearly, Eq.~(\ref{3C_4.43}) will \underline{not} lead
to a singular contribution as $k \to 0$.
Thus, in our terminology, it is part of the structure term (g) 
of Fig.~\ref{fig:pp_pp_gam}.
In the following, we give a short discussion of $N^{*}$ resonances
which can contribute to the photon flux via the diagrams
of the type of Fig.~\ref{fig:101}~(a).

%In the present paper we discuss soft-photon radiation only by protons.
%Photons may also be produced from the radiative decays of
%diffractively excited nucleon resonances.
The $N(1440)$ with $J^{P} = 1/2^{+}$, 
$N(1520)$ with $J^{P} = 3/2^{-}$, and
$N(1680)$ with $J^{P} = 5/2^{+}$ states 
are potential candidates
for the role of $N^{*}$ in (\ref{3C_4.41}).
These resonances satisfy the Gribov-Morrison rule;
see, e.g., Chap.~3.9 of Ref.~\cite{Donnachie:2002en}.
The Roper resonance $N(1440)$ is doubtful.
This poorly known state was considered, however,
in Refs.~\cite{Jenkovszky:2012hf,Jenkovszky:2018itd} 
as an important contribution to
forward low-mass single diffraction dissociation.
In Ref.~\cite{Lebiedowicz:2013vya}, it was shown
for the $pp \to pp \pi^{0}$ reaction that
the nonresonant diffractive processes 
(Drell-Hiida-Deck type model)
lead to an enhancement in the invariant mass 
$M_{\pi p} \approx 1.4$~GeV.
Thus, one cannot be sure whether the Roper resonance
plays an important role there.
%if contribute, could be isolated.
%so that there is difficult to isolate.
%but remember that the contributions
%discussed in \cite{Lebiedowicz:2013vya}
%were not taken into account there. 
Coming back to the $pp \to (N^{*} \to p \gamma) p$ reaction,
the Roper resonance has much smaller branching ratio \cite{Zyla:2020zbs}
${\cal BR}(N(1440) \to p \gamma) = 0.035-0.048 \,\%$
than the $N(1520)$,
${\cal BR}(N(1520) \to p \gamma) = 0.31-0.52 \,\%$,
and the $N(1680)$,
${\cal BR}(N(1680) \to p \gamma) = 0.21-0.32 \,\%$.
For the $N(1680)$ resonance, a sizeable cross section 
$\sigma(pp \to p N(1680)) = 170 \pm 60$~$\mu$b 
was estimated at the CERN ISR energy
of $\sqrt{s} = 45$~GeV \cite{deKerret:1976ze}.
One should, however, bear in mind that at ISR energies
secondary reggeon exchanges are still important.
For the bremsstrahlung-type processes at the LHC energies,
the reggeon-exchange contributions 
are very small; thus, diffractively 
excited $N^{*}$ resonances are there due to pomeron exchange 
and should be produced preferentially
in forward/backward rapidity region.
%As example, for $pp \to pp \pi^{0}$ this is 
%at $|{\rm y}| \sim 9-10$;
%see Fig.~7 of \cite{Lebiedowicz:2013vya}.
%The $pp \to pp \gamma$ reaction gives also
%a sizeable contribution to the low mass 
%single diffractive cross section \cite{Lebiedowicz:2013xlb}.
All these processes require dedicated studies
if the low $\gamma p$ invariant-mass region
can be measured in the forward rapidity range.

\begin{figure}[!ht]
\includegraphics[width=0.48\textwidth]{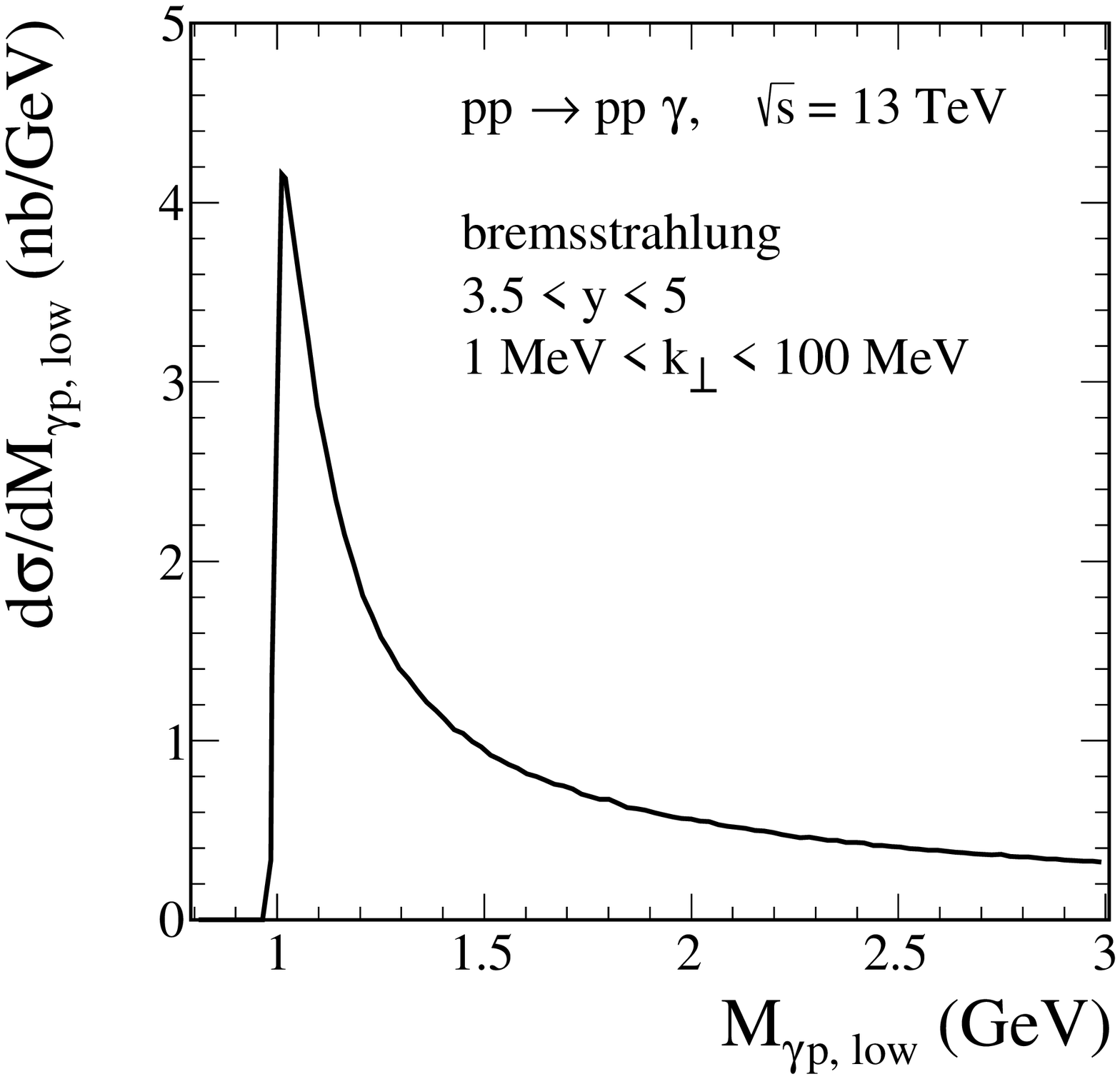}
\includegraphics[width=0.48\textwidth]{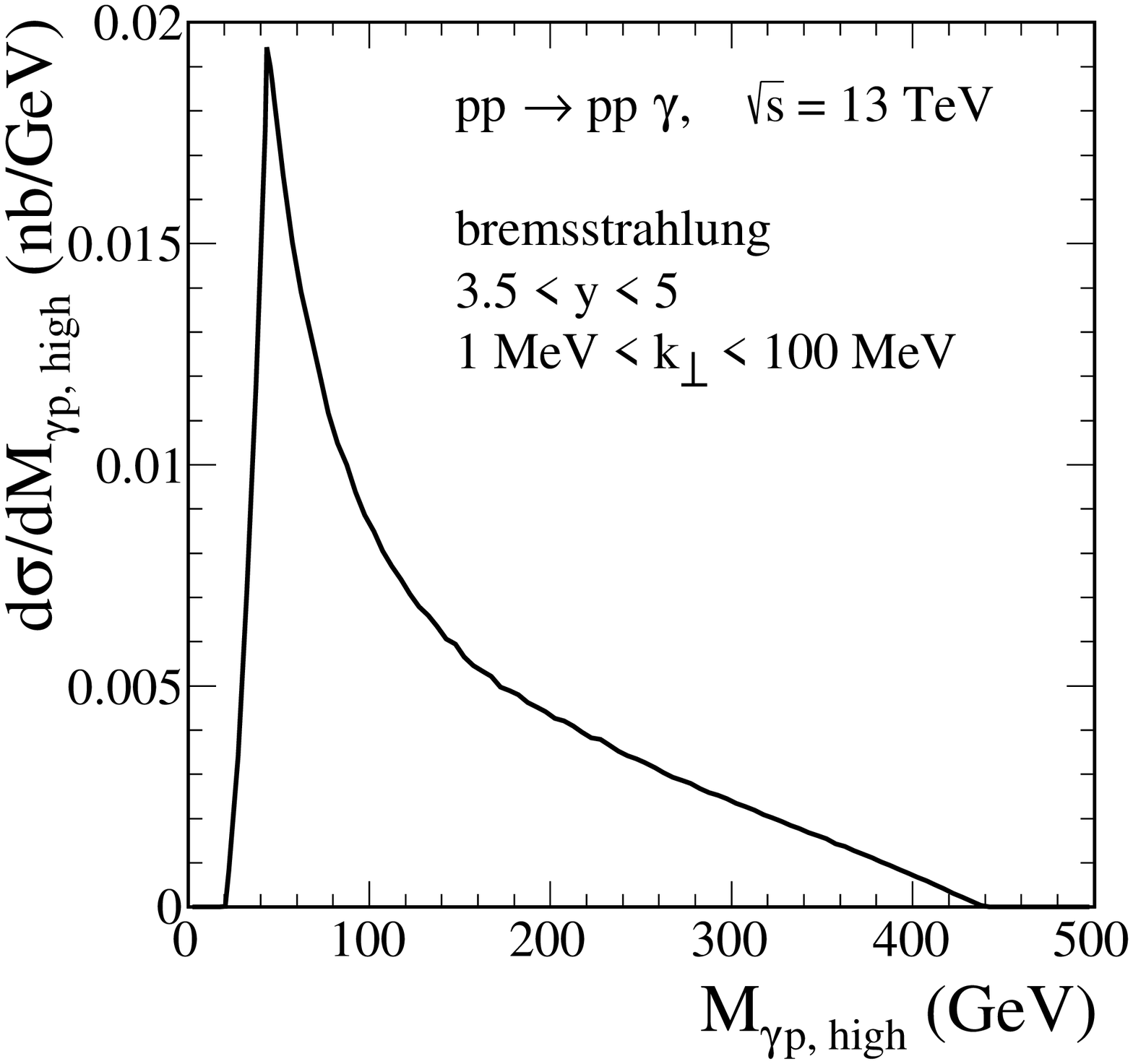}
\caption{\label{fig:W1}
\small
The distributions in $M_{\gamma p, \,{\rm low}}$ (left panel)
and in $M_{\gamma p, \,{\rm high}}$ (right panel)
for our standard bremsstrahlung (nonresonant) model.}
\end{figure}

\begin{figure}[!ht]
\includegraphics[width=0.48\textwidth]{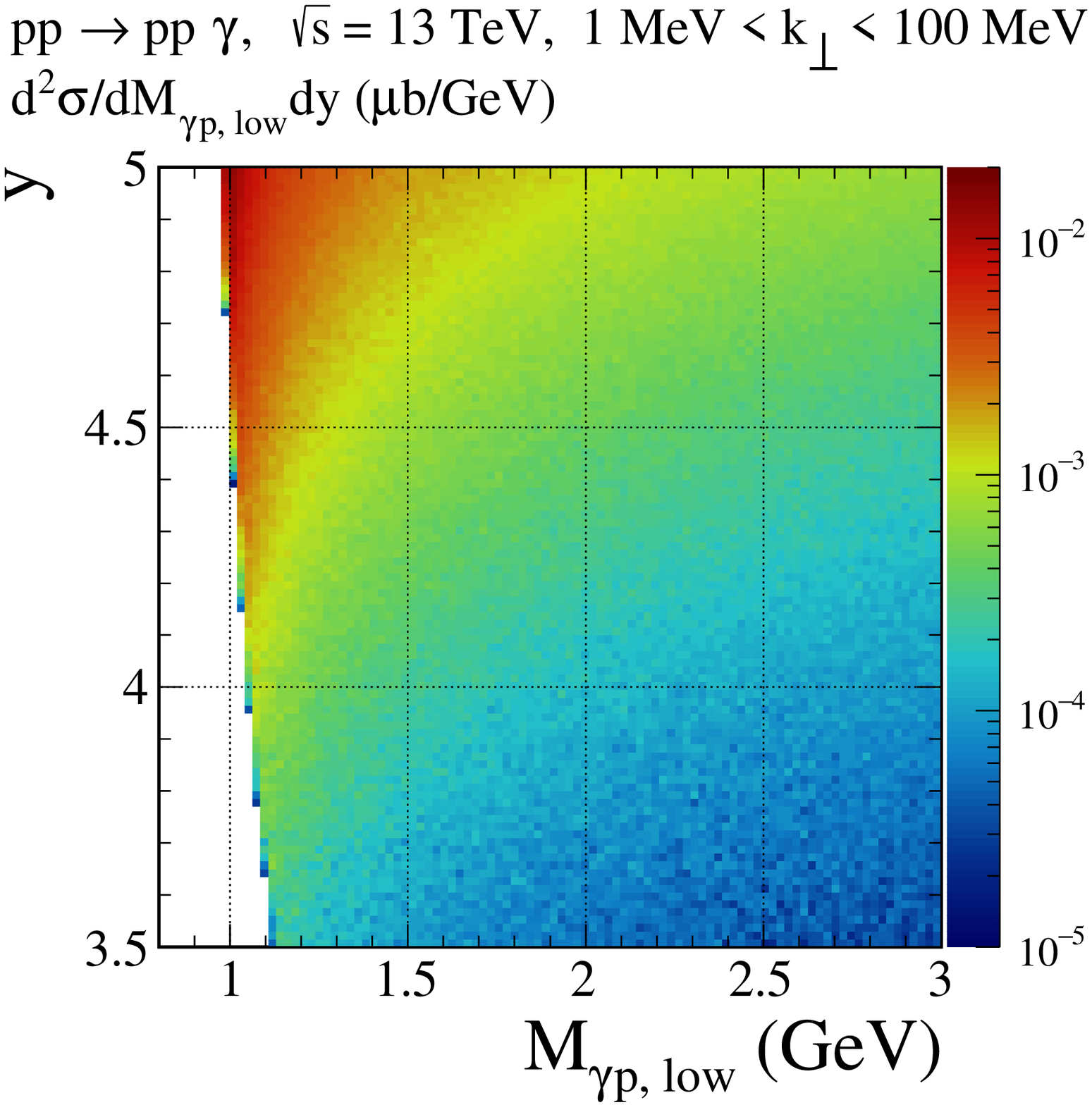} \quad
\includegraphics[width=0.48\textwidth]{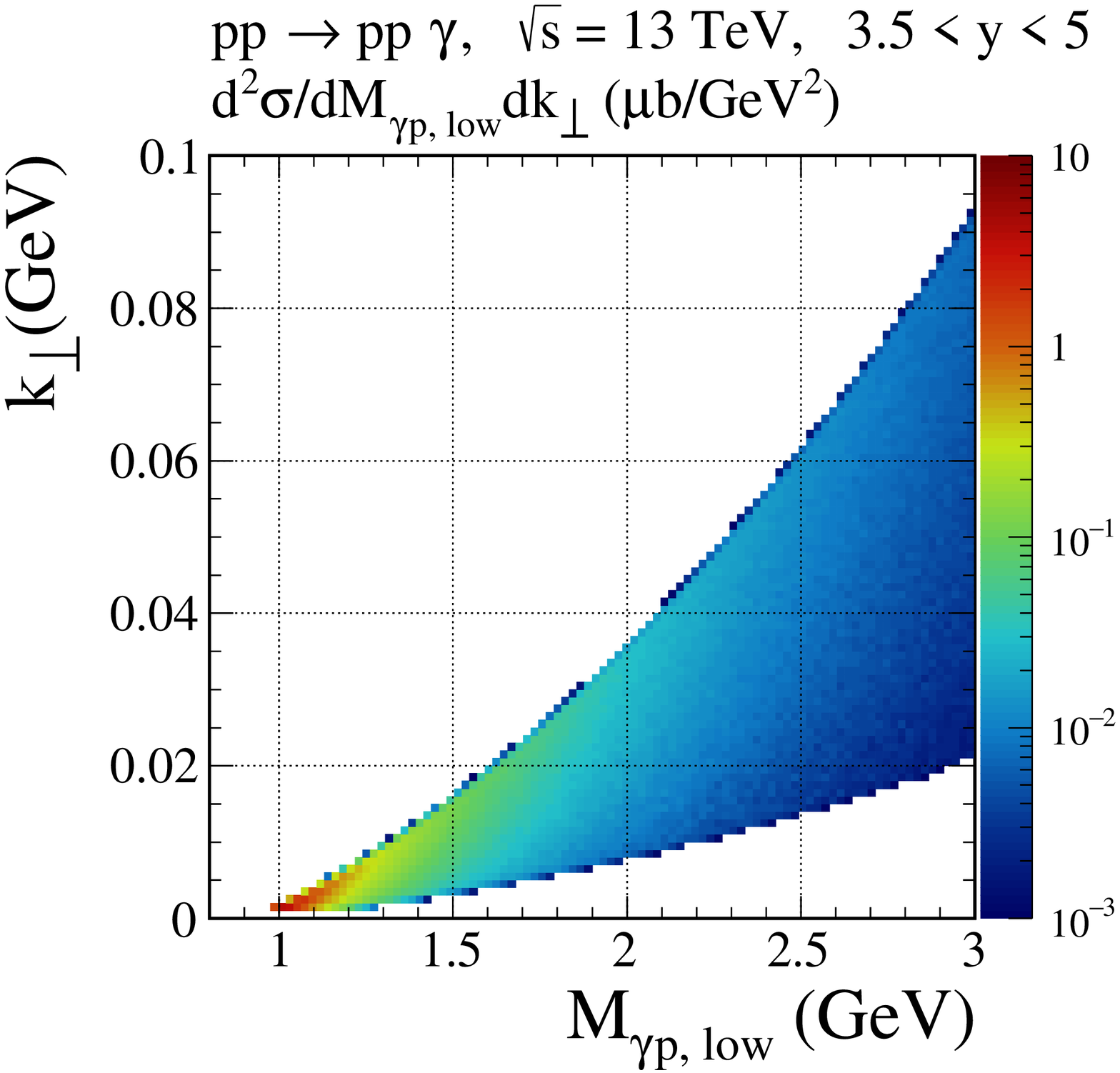}
\caption{\label{fig:2dim_W1}
\small
The two-dimensional distributions in
($M_{\gamma p, \,{\rm low}}, {\rm y}$) and 
in ($M_{\gamma p, \,{\rm low}}, k_{\perp}$)
for the $pp \to pp \gamma$ reaction
for our standard bremsstrahlung (nonresonant) model.}
\end{figure}

In Fig.~\ref{fig:W1}, we show
the differential cross sections 
$d\sigma/dM_{\gamma p, \,{\rm low}}$
(the same sign rapidity of $\gamma$ and $p$)
and $d\sigma/dM_{\gamma p, \,{\rm high}}$
(the opposite sign rapidity of $\gamma$ and $p$)
for our standard bremsstrahlung (nonresonant) model.
%Here $M_{\gamma p, \,{\rm low}}$ and $M_{\gamma p, \,{\rm high}}$ 
%means the lower and higher invariant mass
%of the $\gamma p$ system.
The calculations were done for $\sqrt{s} = 13$~TeV,
$3.5 < {\rm y} < 5$, 
and $1 \; {\rm MeV} < k_{\perp} < 100 \; {\rm MeV}$.
Figure~\ref{fig:2dim_W1} shows the two-dimensional distributions
in ($M_{\gamma p, \,{\rm low}}, {\rm y}$) and 
in ($M_{\gamma p, \,{\rm low}}, k_{\perp}$).
In the present paper we consider soft-photon emission 
only by protons.
As already mentioned, 
photons may also be produced from the radiative decays of
diffractively excited nucleon resonances.
Candidates are $N(1440)$, $N(1520)$, and $N(1680)$.
If these processes contribute significantly 
to our $pp \to pp \gamma$ reaction, then we should see them
in the $M_{\gamma p, \,{\rm low}}$ distribution
(possibly distorted by interference effects)
as a resonance enhancement at $M_{\gamma p} = m_{N^{*}}$
with a width $\Gamma_{N^{*}}$
over our nonresonant term.
However, we expect that the decay photons will be emitted at rapidities larger than covered by the dedicated ALICE~3 detector.
Other background contributions, for instance,
soft photons from central-exclusive production
via the fusion processes $\gamma \Pom \to \gamma$,
$\Ode \Pom \to \gamma$, etc., should be taken into account.
Possible interference effects between various mechanisms 
should be also included. 
This goes beyond the scope of the present paper. 
We leave a detailed analysis of other contributions 
for future studies.
Finally, we conclude that the measurement of 
forward/backward protons 
would be crucial for a better understanding of the mechanisms
of the $pp \to pp \gamma$ reaction.
%-------------------------------------------------------
%--------------------------------------------------------
%-------------------------------------------------------
%--------------------------------------------------------

%--------------------------
\section{Conclusions}
\label{sec:4}
%--------------------------
In this paper, we have presented a detailed discussion of
the reactions $pp \to pp$, $p \bar{p} \to p \bar{p}$,
and $pp \to pp \gamma$ at high c.m. energies using
the framework of the tensor-pomeron model \cite{Ewerz:2013kda}.
To calculate the amplitudes for these reactions we have
considered pomeron ($\Pom$), odderon ($\Ode$),
and reggeon ($f_{2 \Reg}$, $a_{2 \Reg}$,
$\omega_{\Reg}$, $\rho_{\Reg}$) exchanges.

Our theoretical results for elastic $pp$ and $p \bar{p}$
scattering are given in Secs.~\ref{sec:2A} and \ref{sec:2B}
and compared to experimental data 
in Secs.~\ref{sec:3A} and \ref{sec:3B}.
With our model, adjusting the pomeron and odderon parameters,
we get a reasonable fit to the TOTEM data
on $pp$ elastic scattering for $\sqrt{s} = 13$~TeV and
$|t| \lesssim 0.3$~GeV$^{2}$; see Fig.~\ref{fig:dsig_dt}.
We emphasize that we are not  out to produce a precision fit
to all $pp$ and $p \bar{p}$ elastic scattering data.
We only need a reasonable description of the data for
$\sqrt{s} = 13$~TeV as a prerequisite for the calculation
of photon radiation in $pp$ collisions at this energy.
Nevertheless, we also had a look at the data for $\rho$,
the ratio of real and imaginary parts of the forward-scattering
amplitudes.
We found that, taking the TOTEM data for granted,
we need an odderon contribution in the framework of our model,
and a double Regge pole structure of this odderon
with intercept slightly above 1 seems to be preferred; 
see Fig.~\ref{fig:rho}.
Clearly, further experimental studies of reactions where
odderon effects could be present would be very welcome.
Examples of such reactions
are photoproduction of $f_{2}$ mesons, $\gamma p \to f_{2} p$,
and central exclusive production of single and double
$\phi$ mesons, $pp \to p \phi p$ and $pp \to p \phi \phi p$;
see, for instance, 
Refs.~\cite{Bolz:2014mya,Lebiedowicz:2019boz,Lebiedowicz:2019jru}.
All these reactions can be studied at ALICE3.

Turning now to the main topic of our paper,
the calculation of soft-photon radiation in $pp$ collisions,
we have given the explicit formulas for this process
in the framework of our tensor-pomeron model.
The amplitudes corresponding to photon emission
from the external proton lines
[see the diagrams of Fig.~\ref{fig:pp_pp_gam}~(a), (b), (d), and (e)]
are determined by the off-shell 
$pp \to pp$ scattering amplitude.
By construction, the amplitudes (c) and (f),
the contact terms, have to satisfy gauge-invariance constraints
involving the previous amplitudes.
In this way, we obtained our standard results 
for the $pp \to pp \gamma$ reaction.
%It should be emphasized, however, that this solution
%is not unique and there ``anomalous'' terms
%[Fig.~\ref{fig:pp_pp_gam}~(g)],
%not directly related to the $pp \to pp$ amplitude, could come up.
%We considered then as ``standard'' model our amplitudes
%to the $pp \to pp \gamma$ reaction without such anomalous terms.
The relevant distributions are shown 
in Figs.~\ref{fig:ff}--\ref{fig:ratios_y24}.

We have compared our standard results to two soft-photon approximations SPA1 and SPA2; see Sec.~\ref{sec:SPA}. 
In the SPA1, we considered
only the pole terms $\propto \omega^{-1}$ 
in the radiative amplitudes.
%In the SPA1 the photon momentum $k$ was, on purpose,
%omitted in the energy-momentum conserving $\delta$ function
%in the evaluation of the cross section.
This SPA1 agrees rather well, 
at the percent level, with our exact model (or standard result)
in the kinematic range considered;
see Figs.~\ref{fig:y24} and \ref{fig:ratios_y24}.
For $1 \; {\rm MeV} < k_{\perp} < 100 \; {\rm MeV}$
and $3.5  < |\rm y| < 5.0$, for instance,
we find agreement of SPA1 with our standard result at
the percent level up to $\omega \cong 2$~GeV.
The SPA2 is a good approximation 
to the standard result, within 1$\, \%$ accuracy,
for $k_{\perp} \lesssim 22$~MeV
and $\omega \lesssim 0.35$~GeV 
considering $|\rm y| < 3.5$
and up to $\omega \cong 1.7$~GeV 
for $3.5 < |\rm y| < 5.0$; see Fig.~\ref{fig:map_y24}.

In Sec.~\ref{sec:2C}, we have written the amplitude
for $pp \to pp \gamma$ in the standard,
straightforward, way with the photon coupling
to the protons via the Dirac and Pauli terms.
It turned out that the corresponding formulas
(\ref{2.26}), (\ref{2.27}), (\ref{2.43}),
and (\ref{2.46d})--(\ref{2.46f}) were
not convenient for numerical work.
Therefore, we have rewritten these amplitudes in
Appendix~\ref{sec:appendixB} as a sum
of seven terms, labeled by $j = 1, \ldots, 7$,
for the tensor exchanges and
of four terms, labeled by $j' = 1, \ldots, 4$, 
for the vector exchanges
[see (\ref{B3})--(\ref{B16})].
Each of these 11 terms is separately gauge invariant.
The pole terms are only contained 
in the $j=1$ tensor term and $j'=1$ vector term.
The other terms have no singularity for $\omega \to 0$.
Thus, one expects that for soft-photon emission
the terms with $j = 1$ and $j' = 1$ will be dominant.
We found that indeed this is correct but there is a very
interesting effect guaranteeing this.
Considering for pomeron exchange the gauge-invariant
and nonsingular terms with $j = 2$ and $j = 4$ alone,
we find that the pole term only dominates over
these $j = 2$ and $j = 4$ terms individually for very small
$k_{\perp} \approx \omega \lesssim 2 m_{p}^{2}/\sqrt{s}
\cong 0.15$~MeV
[see (\ref{B20}), (\ref{B21}), and Fig.~\ref{fig:B2}].
This effect is, in essence, due to the Pauli coupling
of the photon.
We note that in the literature such small values for $\omega$
as a limit for the dominance of the $\omega^{-1}$ term
are mentioned.
For instance, in Ref.~\cite{DelDuca:1990gz},
it is argued that for hard high-energy elastic processes
Low's original result 
gives a reliable representation of the radiative amplitude
only in the vanishingly small region
$\omega \lesssim m^{2}/Q$ in the limit $Q \to \infty$.
Here, $Q$ is the scale of the hard process, and $m$ 
is the charged particle mass.
But since in this work
only \underline{hard} processes with photon emission 
are considered, these arguments do not apply to our case.
We consider the exclusive \underline{soft} process 
$pp \to pp \gamma$ with soft-photon emission.
We have, of course, to take \underline{all}
contributions with different labels $j$ into account
and add them coherently; see (\ref{B3}).
We find then large cancellations between 
the $j = 2$ and $j = 4$
terms; see Fig.~\ref{fig:B1}.
This leads to a much larger region in $k_{\perp}$
and $\omega$ where the pole terms alone
give a good representation of the radiative amplitude;
see Fig.~\ref{fig:map_y24}.
Our conclusion is that simple order of magnitude estimates
for the $\omega$ regions where the pole term dominates,
using only parts of the amplitudes,
may give completely wrong results.
It is essential to add coherently all the various parts
of the amplitude for soft-photon emission
in order not to miss important interference effects.

In this article, we have only discussed
the bremsstrahlung-type emission of soft photons
in $pp$ collisions.
Anomalous emission terms have been subsumed
in the amplitude ${\cal M}_{\mu}^{(g)}$,
which must satisfy (\ref{2.45})
and can have no singularity for $k \to 0$.
There are, however, quite conventional contributions
to ${\cal M}_{\mu}^{(g)}$, for instance,
soft photons from central-exclusive production
via the fusion processes $\gamma \Pom \to \gamma$,
$\Ode \Pom \to \gamma$.
The contributions of such processes are expected
to be important in the midrapidity region, around ${\rm y} = 0$.
In a future paper, we plan to study 
central-exclusive production of photons
within the tensor-pomeron approach.

Finally, we emphasize that we have taken care
to write the formulas for the $pp \to pp \gamma$ amplitude
in such a way that they also apply to soft virtual photon production,
for instance, $pp \to pp (\gamma^{*} \to e^{+} e^{-})$.
Thus, our Eqs.~(\ref{2.26}), (\ref{2.27}), (\ref{2.43}),
and (\ref{2.46d})--(\ref{2.46f}),
as well as (\ref{B1})--(\ref{B16}), 
can be directly used for
soft virtual photon production.
But further investigations of this interesting topic go beyond
the scope of our present article.

We hope that our theoretical studies
of the exclusive $pp \to pp \gamma$ reaction 
will find experimental counterparts with measurements
of soft photons at the Relativistic Heavy Ion Collider
and at the LHC, for instance,
with the planned ALICE3 detector
\cite{Adamova:2019vkf,QM2022_PBM,EMMI_RRTF}.

%--------------------
\acknowledgments
%--------------------
We thank
Johanna Stachel, Peter Braun-Munzinger, and Carlo Ewerz
for very useful discussions.
We are grateful to Carlo Ewerz for
reading the manuscript.
This work and the stay of Piotr Lebiedowicz 
in Heidelberg were supported by the Bekker Program 
of the Polish National Agency for Academic Exchange
under Contract No. BPN/BEK/2021/2/00009/U/00001.
This study was also partially supported by
the Polish National Science Centre under Grant
No. 2018/31/B/ST2/03537.

\newpage
%--------------------
\appendix
%--------------------
\section{List of quantities and their values used in the paper}
\label{sec:appendixA}

The quantities listed in the following are 
in essence taken from Chap.~3 of Ref.~\cite{Ewerz:2013kda}.
Here, the propagators and vertices involving the pomeron, 
the odderon, and the reggeons are to be understood 
as effective propagators and vertices.

The effective propagators for $\Pom$, reggeons, and $\Ode$ are as follows:
%%%%%%%%%%%%%%%%%%%%%%%%%%%%%%%%%%%%%%%
\begin{itemize}
%%%%%%%%%%%%%%%%%%%%%%%%%%%%%%%%%%%%%%%
\item pomeron $\Pom$ (see (3.10), (3.11),
and Sec.~6.1 of Ref.~\cite{Ewerz:2013kda}):\\
%%%%%%%%%%%%%%%%%%%%%%%%%%%%%%%%%%%%%%%
\includegraphics[width=150pt]{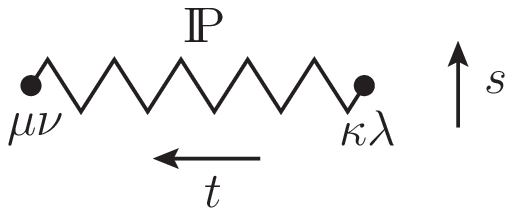} 
\vspace*{-0.45cm}
\begin{eqnarray}
&&i\Delta^{(\Pom)}_{\mu \nu, \kappa \lambda}(s,t) = 
\frac{1}{4s} \left( g_{\mu \kappa} g_{\nu \lambda} 
                  + g_{\mu \lambda} g_{\nu \kappa}
                  - \frac{1}{2} g_{\mu \nu} g_{\kappa \lambda} \right)
(-i s \alpha'_{\Pom})^{\alpha_{\Pom}(t)-1}\,,
\label{A1}\\
&&\alpha_{\Pom}(t) = \alpha_{\Pom}(0)+\alpha'_{\Pom}\,t\,, 
\quad \alpha_{\Pom}(0) = 1 + \epsilon_{\Pom}\,, \nonumber\\
&&\alpha'_{\Pom} = 0.25 \; \mathrm{GeV}^{-2}\,.
\label{A2}
\end{eqnarray}
The default value for $\epsilon_{\Pom}$ 
from Ref.~\cite{Ewerz:2013kda} is $\epsilon_{\Pom} = 0.0808$.
In our present paper we find from a comparison to the data 
from TOTEM measurements at $\sqrt{s} = 13$~TeV
\cite{TOTEM:2017sdy,TOTEM:2018hki} 
a slightly higher value: $\epsilon_{\Pom} = 0.0865$;
see Fig.~\ref{fig:dsig_dt}.
This value for the soft pomeron intercept
is in agreement with those obtained 
in Refs.~\cite{Cudell:1996sh,Luna:2003kw,Luna:2004gr}.
%for the simple Regge pole exchange.
Our value above also agrees,
within practically 1 standard deviation, with that obtained 
in Ref.~\cite{Britzger:2019lvc} from a fit to photoproduction
and low-x deep inelastic scattering which
gave $\epsilon_{\Pom} = 0.0935(^{+76}_{-64})$.

%%%%%%%%%%%%%%%%%%%%%%%%%%%%%%%%%%%%%%%
\item reggeons $\Reg_{+} = f_{2 \Reg}, a_{2 \Reg}$ (see (3.12), (3.13), and Sec.~6.3 of Ref.~\cite{Ewerz:2013kda}):\\
%%%%%%%%%%%%%%%%%%%%%%%%%%%%%%%%%%%%%%%
\includegraphics[width=150pt]{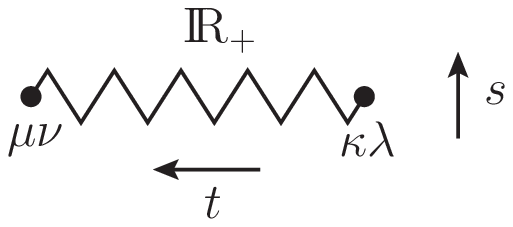} 
\vspace*{-0.45cm}
\begin{eqnarray}
&&i\Delta^{(\Reg_{+})}_{\mu \nu, \kappa \lambda}(s,t) = 
\frac{1}{4s} \left( g_{\mu \kappa} g_{\nu \lambda} 
                  + g_{\mu \lambda} g_{\nu \kappa}
                  - \frac{1}{2} g_{\mu \nu} g_{\kappa \lambda} \right)
(-i s \alpha'_{\Reg_{+}})^{\alpha_{\Reg_{+}}(t)-1}\,,
\label{A3}\\
&&\alpha_{\Reg_{+}}(t) = \alpha_{\Reg_{+}}(0)+\alpha'_{\Reg_{+}}\,t\,, \nonumber\\
&&\alpha_{\Reg_{+}}(0) = 0.5475\,, \nonumber\\
&&\alpha'_{\Reg_{+}} = 0.9 \; \mathrm{GeV}^{-2}\,.
\label{A4}
\end{eqnarray}
\newpage
%%%%%%%%%%%%%%%%%%%%%%%%%%%%%%%%%%%%%%%
\item reggeons $\Reg_{-} = \omega_{\Reg}, \rho_{\Reg}$ (see (3.14), (3.15), and Sec.~6.3 of Ref.~\cite{Ewerz:2013kda}):\\
%%%%%%%%%%%%%%%%%%%%%%%%%%%%%%%%%%%%%%%
\includegraphics[width=150pt]{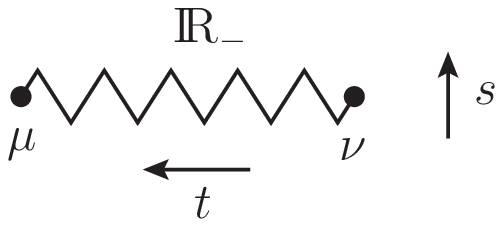} 
\vspace*{-0.45cm}
\begin{eqnarray}
&&i\Delta^{(\Reg_{-})}_{\mu \nu}(s,t) = 
i g_{\mu \nu} \frac{1}{M_{-}^{2}}
(-i s \alpha'_{\Reg_{-}})^{\alpha_{\Reg_{-}}(t)-1}\,,
\label{A5}\\
&&\alpha_{\Reg_{-}}(t) = \alpha_{\Reg_{-}}(0)+\alpha'_{\Reg_{-}}\,t\,, \nonumber\\
&&\alpha_{\Reg_{-}}(0) = 0.5475\,, \nonumber\\
&&\alpha'_{\Reg_{-}} = 0.9 \; \mathrm{GeV}^{-2}\,, \nonumber\\
&&M_{-} = 1.41 \; \mathrm{GeV}\,.
\label{A6}
\end{eqnarray}
The numbers for the reggeon parameters 
in (\ref{A4}) and (\ref{A6})
are taken from Ref.~\cite{Donnachie:1992ny}
and Figs.~3.1 and 3.2 of Ref.~\cite{Donnachie:2002en},
except for $M_{-}$,
which is discussed in Sec.~6.3 of Ref.~\cite{Ewerz:2013kda}.
%%%%%%%%%%%%%%%%%%%%%%%%%%%%%%%%%%%%%%%
\item odderon $\Ode$\\
%%%%%%%%%%%%%%%%%%%%%%%%%%%%%%%%%%%%%%%
\includegraphics[width=150pt]{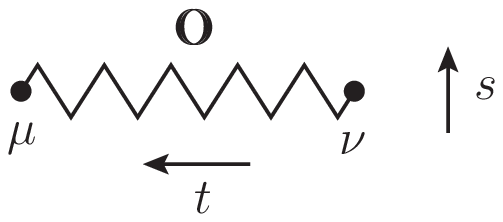} 
%\vspace*{-0.45cm}
\newline
Our Ansatz for a single-pole odderon is as in Ref.~\cite{Ewerz:2013kda}
[see (3.16), (3.17), and Sec.~6.2 therein]: 
\begin{eqnarray}
&&i\Delta^{(\Ode)}_{\mu \nu}(s,t) = 
-i g_{\mu \nu} \frac{\eta_{\Ode}}{M_{0}^{2}}
(-i s \alpha'_{\Ode})^{\alpha_{\Ode}(t)-1}\,,
\label{A7}\\
&&\alpha_{\Ode}(t) = \alpha_{\Ode}(0)+\alpha'_{\Ode}\,t\,, 
\quad \alpha_{\Ode}(0) = 1 + \epsilon_{\Ode}\,, \nonumber\\
&&M_{0} = 1 \; \mathrm{GeV}\,, \nonumber\\
&&\eta_{\Ode} = \pm 1\,, \nonumber\\
&&\alpha'_{\Ode} = 0.25 \; \mathrm{GeV}^{-2}\,.
\label{A8}
\end{eqnarray}
Here, $M_{0} = 1$~GeV is introduced for dimensional reasons,
and $\alpha'_{\Ode} = \alpha'_{\Pom}$ (\ref{A2}) is set as default.

For a double-pole odderon, we set
\begin{equation}
i\widetilde{\Delta}^{(\Ode)}_{\mu \nu}(s,t) = i\Delta^{(\Ode)}_{\mu \nu}(s,t)
\left[
C_{1} + C_{2} \ln \left( -i s \alpha_{\Ode}' \right)
\right] \,.
\label{A9}
\end{equation}
Here, $C_{1}$ and $C_{2}$ are real constants.

From the comparison with the $\rho$ values of $pp$ and $p \bar{p}$
scattering, we find that the following values
give a reasonable description of the TOTEM data 
(see Fig.~\ref{fig:rho}):
\begin{eqnarray}
&&\eta_{\Ode} = -1\,, \quad 
\alpha'_{\Ode} = 0.25 \; \mathrm{GeV}^{-2} \,, \quad 
\epsilon_{\Ode} = 0.0800 \,, \nonumber \\
&&(C_{1},C_{2}) = (-1.0,0.1), (-1.5,0.2), (-2.0,0.3)\,.
\label{A10}
\end{eqnarray}
%
%%%%%%%%%%%%%%%%%%%%%%%%%%%%%%%%%%%%%%%
\end{itemize}
%%%%%%%%%%%%%%%%%%%%%%%%%%%%%%%%%%%%%%%

The effective proton-pomeron,
proton-reggeon, and proton-odderon vertices are taken 
as in Ref.~\cite{Ewerz:2013kda}, but
at least for the $\Pom pp$ vertex,
we use a different form factor 
in order to fit the TOTEM data 
\cite{TOTEM:2017sdy,TOTEM:2018hki}
at $\sqrt{s} = 13$~TeV and in the low $|t|$ region.

Now here is the list of effective vertices.
%%%%%%%%%%%%%%%%%%%%%%%%%%%%%%%%%%%%%%%
\begin{itemize}
%%%%%%%%%%%%%%%%%%%%%%%%%%%%%%%%%%%%%%%
\item $\Pom p p$ vertex (see (3.43), (3.44),
and Sec.~6.1 of Ref.~\cite{Ewerz:2013kda}):\\
%%%%%%%%%%%%%%%%%%%%%%%%%%%%%%%%%%%%%%%
\includegraphics[width=140pt]{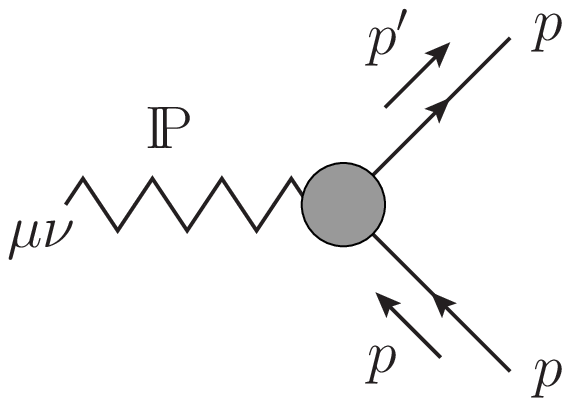} 
\includegraphics[width=140pt]{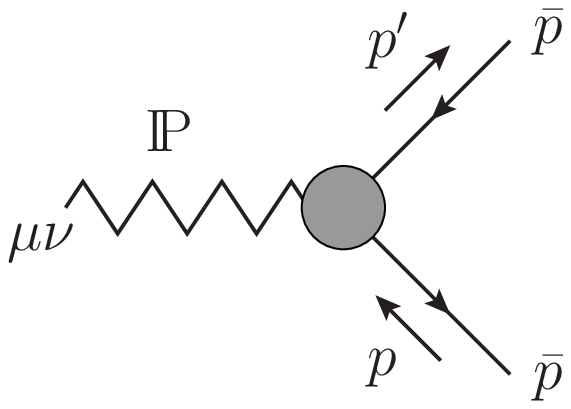} 
%\newline
\vspace*{-0.45cm}
\begin{eqnarray}
i\Gamma_{\mu \nu}^{(\Pom pp)}(p',p)
&=& i\Gamma_{\mu \nu}^{(\Pom \bar{p} \bar{p})}(p',p)\nonumber\\
&=&-i 3 \beta_{\Pom pp} F_{1}[(p'-p)^{2}]\nonumber\\
&&\times \left\lbrace 
\frac{1}{2} 
\left[ \gamma_{\mu}(p'+p)_{\nu} 
     + \gamma_{\nu}(p'+p)_{\mu} \right]
- \frac{1}{4} g_{\mu \nu} (\slash{p}' + \slash{p})
\right\rbrace \,, \quad
\label{A11}\\
\beta_{\Pom pp} &=& 1.87 \; \mathrm{GeV}^{-1}\,.
\label{A12}
\end{eqnarray}

In Ref.~\cite{Ewerz:2013kda}, the form factor $F_{1}(t)$
was taken as the electromagnetic form factor of the proton,
\begin{eqnarray} 
&&F_{1}(t) = \big(1 - \frac{t}{4 m_{p}^{2}} \frac{\mu_{p}}{\mu_{N}}\big)
\big(1 - \frac{t}{4 m_{p}^{2}} \big)^{-1}\, G_{D}(t)\,,
\nonumber \\
&&
\mu_{N} = \frac{e}{2 m_{p}}, \quad
\frac{\mu_{p}}{\mu_{N}} = 2.7928\,,
\nonumber \\
&&
G_{D}(t) = \big(1 - \frac{t}{m_{D}^{2}} \big)^{-2}, \quad
m_{D}^{2} = 0.71\;{\rm GeV}^{2}\,.
\label{A12_aux}
\end{eqnarray}

In the present paper, we use
$F_{1}(t) \to F(t) = \exp(-b\,|t|)$ with $b = 2.95$~GeV$^{-2}$ adjusted to the TOTEM data
(see Fig.~\ref{fig:dsig_dt}).

%%%%%%%%%%%%%%%%%%%%%%%%%%%%%%%%%%%%%%%
\item $\Reg_{+} p p$ vertex, 
where $\Reg_{+} = f_{2\Reg}, a_{2\Reg}$
(see (3.49)--(3.52) 
and Sec.~6.3 of Ref.~\cite{Ewerz:2013kda}):\\
%%%%%%%%%%%%%%%%%%%%%%%%%%%%%%%%%%%%%%%
\includegraphics[width=140pt]{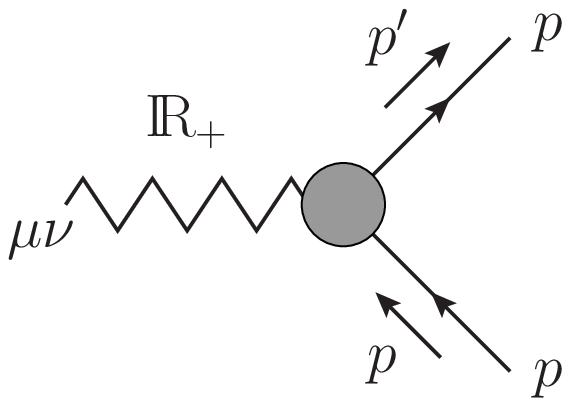} 
\includegraphics[width=140pt]{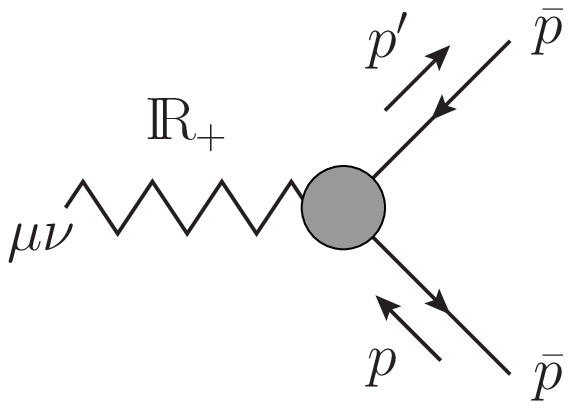} 
%\newline
\vspace*{-0.45cm}
\begin{eqnarray}
i\Gamma_{\mu \nu}^{(f_{2 \Reg} pp)}(p',p)
&=& i\Gamma_{\mu \nu}^{(f_{2 \Reg} \bar{p} \bar{p})}(p',p)\nonumber\\
&=&-i g_{f_{2 \Reg} pp} \frac{1}{M_{0}}F_{1}[(p'-p)^{2}]\nonumber\\
&&\times \left\lbrace 
\frac{1}{2} 
\left[ \gamma_{\mu}(p'+p)_{\nu} 
     + \gamma_{\nu}(p'+p)_{\mu} \right]
- \frac{1}{4} g_{\mu \nu} (\slash{p}' + \slash{p})
\right\rbrace \,, \quad
\label{A13}\\
g_{f_{2 \Reg} pp} &=& 11.04\,, \quad M_{0} = 1 \; \mathrm{GeV}\,;
\label{A14}
\end{eqnarray}
\begin{eqnarray}
i\Gamma_{\mu \nu}^{(a_{2 \Reg} pp)}(p',p)
&=& i\Gamma_{\mu \nu}^{(a_{2 \Reg} \bar{p} \bar{p})}(p',p)\nonumber\\
&=&-i g_{a_{2 \Reg} pp} \frac{1}{M_{0}}F_{1}[(p'-p)^{2}]\nonumber\\
&&\times \left\lbrace 
\frac{1}{2} 
\left[ \gamma_{\mu}(p'+p)_{\nu} 
     + \gamma_{\nu}(p'+p)_{\mu} \right]
- \frac{1}{4} g_{\mu \nu} (\slash{p}' + \slash{p})
\right\rbrace \,, \quad
\label{A15}\\
g_{a_{2 \Reg} pp} &=& 1.68\,, \quad M_{0} = 1 \; \mathrm{GeV}\,.
\label{A16}
\end{eqnarray}
%

%%%%%%%%%%%%%%%%%%%%%%%%%%%%%%%%%%%%%%%
\item $\Reg_{-} p p$ vertex,
where $\Reg_{-} = \omega_{\Reg}, \rho_{\Reg}$
(see (3.59)--(3.62) 
and Sec.~6.3 of \cite{Ewerz:2013kda}):\\
%%%%%%%%%%%%%%%%%%%%%%%%%%%%%%%%%%%%%%%
\includegraphics[width=140pt]{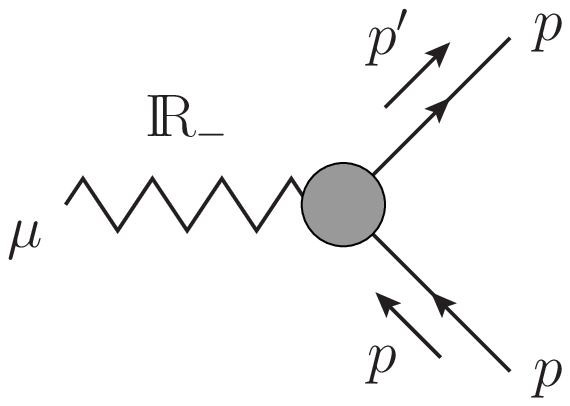} 
\includegraphics[width=140pt]{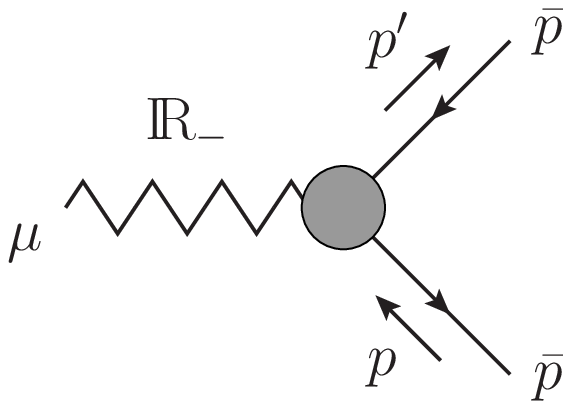} 
%\newline
\vspace*{-0.3cm}
\begin{eqnarray}
i\Gamma_{\mu}^{(\omega_{\Reg} pp)}(p',p)
&=& -i\Gamma_{\mu}^{(\omega_{\Reg} \bar{p} \bar{p})}(p',p)\nonumber\\
&=&-i g_{\omega_{\Reg} pp} F_{1}[(p'-p)^{2}] \gamma_{\mu}\,,
\label{A17}\\
g_{\omega_{\Reg} pp} &=& 8.65\,;
\label{A18}
\end{eqnarray}
\begin{eqnarray}
i\Gamma_{\mu}^{(\rho_{\Reg} pp)}(p',p)
&=& -i\Gamma_{\mu}^{(\rho_{\Reg} \bar{p} \bar{p})}(p',p)\nonumber\\
&=&-i g_{\rho_{\Reg} pp} F_{1}[(p'-p)^{2}] \gamma_{\mu}\,,
\label{A19}\\
g_{\rho_{\Reg} pp} &=& 2.02\,.
\label{A20}
\end{eqnarray}
%

%%%%%%%%%%%%%%%%%%%%%%%%%%%%%%%%%%%%%%%
\item $\Ode p p$ vertex (see (3.68), (3.69),
and Sec.~6.2 of Ref.~\cite{Ewerz:2013kda}):\\
%%%%%%%%%%%%%%%%%%%%%%%%%%%%%%%%%%%%%%%
\includegraphics[width=140pt]{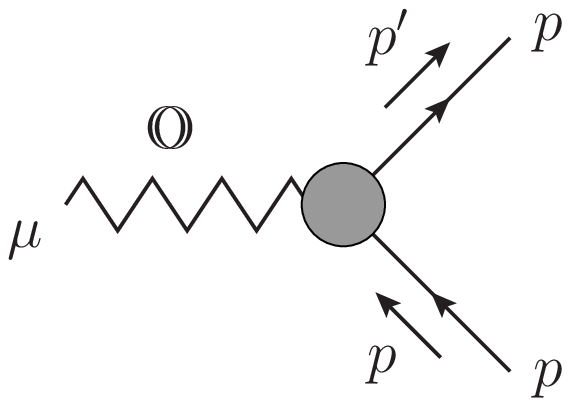} 
\includegraphics[width=140pt]{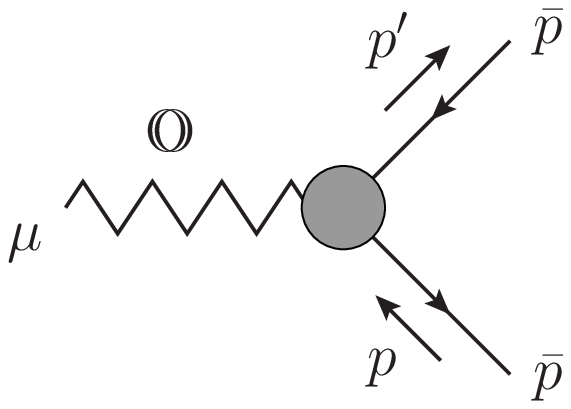} 
%\newline
\vspace*{-0.3cm}
\begin{eqnarray}
i\Gamma_{\mu}^{(\Ode pp)}(p',p)
&=& -i\Gamma_{\mu}^{(\Ode \bar{p} \bar{p})}(p',p)\nonumber\\
&=&-i 3 \beta_{\Ode pp} M_{0} \,F_{1}[(p'-p)^{2}] \gamma_{\mu}\,,
\label{A21}\\
\beta_{\Ode pp} &=& 0.2 \; \mathrm{GeV}^{-1}\,,
\quad M_{0} = 1 \; \mathrm{GeV}\,.
\label{A22}
\end{eqnarray}
The coupling constant $\beta_{\Ode pp}$ 
is taken as free parameter to be determined by experiment.
In the present study, we assume (\ref{A22}):
$\beta_{\Ode pp} =
0.1 \times \beta_{\Pom pp} \simeq 0.2\;{\rm GeV}^{-1}$.
%%%%%%%%%%%%%%%%%%%%%%%%%%%%%%%%%%%%%%%
\end{itemize}
%%%%%%%%%%%%%%%%%%%%%%%%%%%%%%%%%%%%%%%

At the end of this Appendix, we give the formulas
corresponding to (\ref{2.36}), (\ref{2.37}), and (\ref{2.41})
for the case of a double-pole odderon (\ref{A9}).
We have here to replace
${\cal F}_{\Ode pp}(s,t)$ by 
$\widetilde{\cal F}_{\Ode pp}(s,t)$
according to (\ref{2.10_new}).
In this way, we get instead of (\ref{2.36})
\begin{eqnarray}
\widetilde{\cal F}_{\Ode pp}(s',t_{2}) = 
\widetilde{\cal F}_{\Ode pp}(s,t_{2})
+ \varkappa \, \widetilde{\Delta \cal F}_{\Ode pp}(s,t_{2},\varkappa)\,,
\label{A24}
\end{eqnarray}
where
\begin{eqnarray}
\widetilde{\Delta \cal F}_{\Ode pp}(s,t_{2},\varkappa) 
&=&
{\cal F}_{\Ode pp}(s,t_{2})
\big\lbrace
  C_{1} \big(1 - \alpha_{\Ode}(t_{2}) \big) 
g_{\Ode}(\varkappa, t_{2}) \nonumber \\
&&+ C_{2} \big[ \big(1 - \alpha_{\Ode}(t_{2}) \big) 
g_{\Ode}(\varkappa, t_{2}) 
\ln\big(-is(1-\varkappa) \alpha_{\Ode}'\big)
+\frac{1}{\varkappa} \ln(1 - \varkappa) \big]
\big\rbrace
\,. \nonumber \\
\label{A25}
\end{eqnarray}
For $\Delta{\cal F}_{V}(s,t,\varkappa)$
in (\ref{2.41}), this gives the replacement
\begin{eqnarray}
\widetilde{\Delta \cal F}_{V}(s,t,\varkappa) &=&
\widetilde{\Delta \cal F}_{\Ode pp}(s,t,\varkappa)
+ (1 - \alpha_{\omega_{\Reg}}(t)) \,g_{\omega_{\Reg}}(\varkappa ,t)\,
{\cal F}_{\omega_{\Reg} pp}(s,t)\nonumber\\ 
&&+ (1 - \alpha_{\rho_{\Reg}}(t)) \,g_{\rho_{\Reg}}(\varkappa ,t)\,
{\cal F}_{\rho_{\Reg} pp}(s,t)\,.
\label{A26}
\end{eqnarray}

\section{The amplitudes with photon emission}
\label{sec:appendixB}

Here, we rewrite the amplitude 
${\cal M}_{\mu}^{(a + b + c)}
=
{\cal M}_{\mu}^{(a)} + {\cal M}_{\mu}^{(b)}
+ {\cal M}_{\mu}^{(c)}$,
see (\ref{2.26}), (\ref{2.27}), and (\ref{2.43}),
in a way that is more suitable for numerical computations.

We use the following relations:
\begin{eqnarray}
&&\frac{\slash{p}_{a} - \slash{k} + m_{p}}{(p_{a}-k)^{2}-m_{p}^{2}+i \varepsilon} 
\big(\gamma_{\mu} - \frac{i}{2 m_{p}} \sigma_{\mu \nu} k^{\nu} F_{2}(0) \big) u_{a} 
\nonumber\\
&& \qquad = 
\frac{1}{-2 p_{a} \cdot k + k^{2} + i \varepsilon}
\big\lbrace
2 p_{a \mu} - k_{\mu} + (k_{\mu} - \slash{k} \gamma_{\mu})
+ \frac{F_{2}(0)}{2 m_{p}}
\big[
2 (p_{a \mu} \slash{k} - (p_{a} \cdot k) \gamma_{\mu})
\nonumber\\
&& \qquad \quad
+ 2 m_{p} (k_{\mu} - \slash{k} \gamma_{\mu})
- (\slash{k} k_{\mu} - k^{2} \gamma_{\mu})
\big]
\big\rbrace u_{a}
\,,
\label{B1}
\end{eqnarray}
\begin{eqnarray}
&&
\bar{u}_{1'}
\big(\gamma_{\mu} - \frac{i}{2 m_{p}} \sigma_{\mu \nu} k^{\nu} F_{2}(0) \big) 
\frac{\slash{p}_{1}' + \slash{k} + m_{p}}{(p_{1}'+k)^{2}-m_{p}^{2}+i \varepsilon} 
\nonumber\\
&& \qquad = 
\bar{u}_{1'}
\frac{1}{2 p_{1}' \cdot k + k^{2} + i \varepsilon}
\big\lbrace
2 p_{1 \mu}' + k_{\mu} - (k_{\mu} - \gamma_{\mu} \slash{k})
+ \frac{F_{2}(0)}{2 m_{p}}
\big[
- 2 (p_{1 \mu}' \slash{k} - (p_{1}' \cdot k) \gamma_{\mu})
\nonumber\\
&& \qquad \quad
- 2 m_{p} (k_{\mu} - \gamma_{\mu} \slash{k})
-(k_{\mu} \slash{k} - k^{2} \gamma_{\mu})
\big]
\big\rbrace
\,.
\label{B2}
\end{eqnarray}

Using (\ref{B1}) and (\ref{B2}) and exploiting the properties
of the Dirac spinors,
$\slash{p}_{a} u_{a} = m_{p} u_{a}$,
$\bar{u}_{1'}\slash{p}_{1'}  = \bar{u}_{1'} m_{p}$
etc.,
we can write our ${\cal M}_{\mu}^{(\rm standard)}$
from (\ref{2.50}) as follows:
\begin{eqnarray}
{\cal M}_{\mu}^{(\rm standard)}=
\sum_{j = 1}^{7} 
\big( 
  {\cal M}_{{\rm T},\mu}^{(a + b + c)\,j} 
+ {\cal M}_{{\rm T},\mu}^{(d + e + f)\,j} 
\big)
+ \sum_{j' = 1}^{4} 
\big(
  {\cal M}_{{\rm V},\mu}^{(a + b + c)\,j'}
+ {\cal M}_{{\rm V},\mu}^{(d + e + f)\,j'} 
\big)\,.
\label{B3}
\end{eqnarray}
Here, ${\rm T}$ and ${\rm V}$ stand for 
the tensor- and vector-exchange diagrams,
respectively,
and $j$~and~$j'$ are just labels for the subamplitudes
in the sums on the rhs of (\ref{B3}).
We have
\newpage
\begin{eqnarray}
%%%%%%%%%%%%%%%%%%%%%%%%%%%%%%%%%%%%%%
{\cal M}_{{\rm T},\mu}^{(a + b + c)\,1}&=&
%%%%%%%%%%%%%%%%%%%%%%%%%%%%%%%%%%%%%%
e \bar{u}_{1'} \otimes \bar{u}_{2'}
\big\lbrace
i {\cal F}_{T}(s,t_{2}) 
\big[
\gamma^{\alpha} \otimes \gamma_{\alpha}
(p_{a}+p_{1}',p_{b}+p_{2}')
+
(\slash{p}_{b}+\slash{p}_{2}')\otimes(\slash{p}_{a}+\slash{p}_{1}')
\nonumber\\
&&  
-2 m_{p}^{2} \mathbb{1} \otimes \mathbb{1} \big] 
\big[
\frac{2p_{a \mu} - k_{\mu}}{-2p_{a} \cdot k + k^{2}+i \varepsilon} 
+
\frac{2p_{1 \mu}' + k_{\mu}}{2p_{1}' \cdot k + k^{2}+i \varepsilon}
\big] 
\big\rbrace
u_{a} \otimes u_{b}\,,
\label{B4}\\
%%%%%%%%%%%%%%%%%%%%%%%%%%%%%%%%%%%%%%
{\cal M}_{{\rm T},\mu}^{(a + b + c)\,2}&=&
%%%%%%%%%%%%%%%%%%%%%%%%%%%%%%%%%%%%%%
e \bar{u}_{1'} \otimes \bar{u}_{2'}
\big\lbrace
i {\cal F}_{T}(s',t_{2})
\frac{1}{-2p_{a} \cdot k + k^{2}+i \varepsilon}
\big[
\gamma^{\alpha} \otimes \gamma_{\alpha}
(p_{a}+p_{1}'-k,p_{b}+p_{2}')\nonumber\\
&&
+
(\slash{p}_{b}+\slash{p}_{2}')\otimes(\slash{p}_{a}+\slash{p}_{1}'-\slash{k})
\big]
\big[
k_{\mu} - \slash{k} \gamma_{\mu}
+ \frac{F_{2}(0)}{2 m_{p}}
\big(
2 p_{a \mu} \slash{k} - 2 (p_{a} \cdot k) \gamma_{\mu}
\nonumber\\
&& 
+ 2 m_{p} (k_{\mu} - \slash{k} \gamma_{\mu})
- (\slash{k} k_{\mu} - k^{2} \gamma_{\mu})
\big)
\big] \otimes \mathbb{1}
\big\rbrace
u_{a} \otimes u_{b}\,, 
\label{B5}\\
%%%%%%%%%%%%%%%%%%%%%%%%%%%%%%%%%%%%%%
{\cal M}_{{\rm T},\mu}^{(a + b + c)\,3}&=&
%%%%%%%%%%%%%%%%%%%%%%%%%%%%%%%%%%%%%%
-e \bar{u}_{1'} \otimes \bar{u}_{2'}
\big\lbrace
i{\cal F}_{T}(s',t_{2})
\frac{m_{p}}{-2p_{a} \cdot k + k^{2}+i \varepsilon}
\big[
2 p_{a \mu} \slash{k} - 2 (p_{a} \cdot k) \gamma_{\mu}
+ 2 m_{p} (k_{\mu} - \slash{k} \gamma_{\mu})
\nonumber\\
&& 
- (\slash{k} k_{\mu} - k^{2} \gamma_{\mu})
+ \frac{F_{2}(0)}{2 m_{p}}
\big(
(- 2 (p_{a} \cdot k) + k^{2} + 4m_{p}^{2})(k_{\mu} - \slash{k} \gamma_{\mu})
\nonumber\\
&& 
+ 2 m_{p} (2 p_{a \mu} \slash{k} 
- 2 (p_{a} \cdot k)\gamma_{\mu})
- 2 m_{p} (\slash{k} k_{\mu} - k^{2} \gamma_{\mu})
\big)
\big] \otimes \mathbb{1}
\big\rbrace
u_{a} \otimes u_{b}\,,
\label{B6}\\
%%%%%%%%%%%%%%%%%%%%%%%%%%%%%%%%%%%%%%
{\cal M}_{{\rm T},\mu}^{(a + b + c)\,4}&=&
%%%%%%%%%%%%%%%%%%%%%%%%%%%%%%%%%%%%%%
e \bar{u}_{1'} \otimes \bar{u}_{2'}
\big\lbrace
i {\cal F}_{T}(s,t_{2})
\frac{1}{2p_{1}' \cdot k + k^{2}+i \varepsilon}
\big[-(k_{\mu} - \gamma_{\mu} \slash{k})
\nonumber\\
&& + \frac{F_{2}(0)}{2 m_{p}}
\big(
-2 p_{1 \mu}' \slash{k} + 2 (p_{1}' \cdot k) \gamma_{\mu}
- 2 m_{p} (k_{\mu} - \gamma_{\mu} \slash{k})
- (k_{\mu} \slash{k} - k^{2} \gamma_{\mu})
\big)
\big] \otimes \mathbb{1}\nonumber\\
&& 
\times
\big[
\gamma^{\alpha} \otimes \gamma_{\alpha}
(p_{a}+p_{1}'+k,p_{b}+p_{2}')
+
(\slash{p}_{b}+\slash{p}_{2}')\otimes(\slash{p}_{a}+\slash{p}_{1}'+\slash{k})
\big]
\big\rbrace
u_{a} \otimes u_{b}\,,
\label{B7}\\
%%%%%%%%%%%%%%%%%%%%%%%%%%%%%%%%%%%%%%
{\cal M}_{{\rm T},\mu}^{(a + b + c)\,5}&=&
%%%%%%%%%%%%%%%%%%%%%%%%%%%%%%%%%%%%%%
-e \bar{u}_{1'} \otimes \bar{u}_{2'}
\big\lbrace
i{\cal F}_{T}(s,t_{2})
\frac{m_{p}}{2p_{1}' \cdot k + k^{2}+i \varepsilon}
\big[
-2 p_{1 \mu}' \slash{k} + 2 (p_{1}' \cdot k) \gamma_{\mu}
- 2 m_{p} (k_{\mu} - \gamma_{\mu} \slash{k})
\nonumber\\
&& 
- (\slash{k} k_{\mu} - k^{2} \gamma_{\mu})
- \frac{F_{2}(0)}{2 m_{p}}
\big(
(2 (p_{1}' \cdot k) + k^{2} + 4m_{p}^{2})(k_{\mu} - \gamma_{\mu} \slash{k})
\nonumber\\
&&  
+ 2 m_{p} (2 p_{1 \mu}' \slash{k}
- 2 (p_{1}' \cdot k)\gamma_{\mu})
+ 2 m_{p} (\slash{k} k_{\mu} - k^{2} \gamma_{\mu})
\big)
\big] \otimes \mathbb{1}
\big\rbrace
u_{a} \otimes u_{b}\,,
\label{B8}\\
%%%%%%%%%%%%%%%%%%%%%%%%%%%%%%%%%%%%%%
{\cal M}_{{\rm T},\mu}^{(a + b + c)\,6}&=&
%%%%%%%%%%%%%%%%%%%%%%%%%%%%%%%%%%%%%%
e \bar{u}_{1'} \otimes \bar{u}_{2'}
\big\lbrace
i \Delta{\cal F}_{T}(s,t_{2},\varkappa)
\big[
\gamma^{\alpha} \otimes \gamma_{\alpha}
(p_{a}+p_{1}'-k,p_{b}+p_{2}')
\nonumber\\
&& +
(\slash{p}_{b}+\slash{p}_{2}')\otimes(\slash{p}_{a}+\slash{p}_{1}'-\slash{k})
- m_{p} (2 m_{p} - \slash{k}) \otimes \mathbb{1}
\big]
\nonumber\\
&& 
\times
\big[
\frac{(2 p_{a} + 2 p_{b} -k, k)}{s} 
\frac{2p_{a \mu} - k_{\mu}}{-2p_{a} \cdot k + k^{2}+i \varepsilon}
+ \frac{(2 p_{a} + 2 p_{b} -k)_{\mu}}{s}
\big]
\big\rbrace
u_{a} \otimes u_{b}\,,
\label{B9}\\
%%%%%%%%%%%%%%%%%%%%%%%%%%%%%%%%%%%%%%
{\cal M}_{{\rm T},\mu}^{(a + b + c)\,7}&=&
%%%%%%%%%%%%%%%%%%%%%%%%%%%%%%%%%%%%%%
e \bar{u}_{1'} \otimes \bar{u}_{2'}
\big\lbrace
i {\cal F}_{T}(s,t_{2})
\big[
\big(
  \gamma^{\alpha} \otimes \gamma_{\alpha} (p_{b}+p_{2}',k)
+ (\slash{p}_{b}+\slash{p}_{2}') \otimes \slash{k}
- m_{p} \slash{k} \otimes \mathbb{1}
\big)
\nonumber\\
&& \times
\big(
\frac{2p_{1 \mu}' + k_{\mu}}{2p_{1}' \cdot k + k^{2}+i \varepsilon}
-
\frac{2p_{a \mu} - k_{\mu}}{-2p_{a} \cdot k + k^{2}+i \varepsilon}
\big)
\nonumber\\
&& 
- 2 \gamma^{\alpha} \otimes \gamma_{\alpha} (p_{b}+p_{2}')_{\mu}
- 2 (\slash{p}_{b}+\slash{p}_{2}') \otimes \gamma_{\mu}
+ 2 m_{p} \gamma_{\mu} \otimes \mathbb{1}
\big]
\big\rbrace
u_{a} \otimes u_{b}\,.
\label{B10}
\end{eqnarray}

For the subamplitudes with the vector exchanges we get
\begin{eqnarray}
%%%%%%%%%%%%%%%%%%%%%%%%%%%%%%%%%%%%%%
{\cal M}_{{\rm V},\mu}^{(a + b + c)\,1}&=&
%%%%%%%%%%%%%%%%%%%%%%%%%%%%%%%%%%%%%%
-e \bar{u}_{1'} \otimes \bar{u}_{2'}
\big\lbrace
{\cal F}_{V}(s,t_{2}) 
\gamma^{\alpha} \otimes \gamma_{\alpha}
\big[
\frac{2p_{a \mu} - k_{\mu}}{-2p_{a} \cdot k + k^{2}+i \varepsilon} 
+
\frac{2p_{1 \mu}' + k_{\mu}}{2p_{1}' \cdot k + k^{2}+i \varepsilon}
\big] 
\big\rbrace
\nonumber\\
&& \times
u_{a} \otimes u_{b}\,,
\label{B11}\\
%%%%%%%%%%%%%%%%%%%%%%%%%%%%%%%%%%%%%%
{\cal M}_{{\rm V},\mu}^{(a + b + c)\,2}&=&
%%%%%%%%%%%%%%%%%%%%%%%%%%%%%%%%%%%%%%
-e \bar{u}_{1'} \otimes \bar{u}_{2'}
\big\lbrace
\Delta{\cal F}_{V}(s,t_{2},\varkappa)
\gamma^{\alpha} \otimes \gamma_{\alpha}
\big[ \frac{(2 p_{a} + 2 p_{b} -k, k)}{s} 
\frac{2p_{a \mu} - k_{\mu}}{-2p_{a} \cdot k + k^{2}+i \varepsilon}
\nonumber\\
&& + \frac{(2 p_{a} + 2 p_{b} -k)_{\mu}}{s}
\big]
\big\rbrace
u_{a} \otimes u_{b}\,,
\label{B12}\\
%%%%%%%%%%%%%%%%%%%%%%%%%%%%%%%%%%%%%%
{\cal M}_{{\rm V},\mu}^{(a + b + c)\,3}&=&
%%%%%%%%%%%%%%%%%%%%%%%%%%%%%%%%%%%%%%
-e \bar{u}_{1'} \otimes \bar{u}_{2'}
\big\lbrace
{\cal F}_{V}(s',t_{2})
\frac{1}{-2p_{a} \cdot k + k^{2}+i \varepsilon}
\gamma^{\alpha} 
\big[
k_{\mu} - \slash{k} \gamma_{\mu}
\nonumber\\
&& + \frac{F_{2}(0)}{2 m_{p}}
\big(
2 p_{a \mu} \slash{k} - 2 (p_{a} \cdot k) \gamma_{\mu}
+ 2 m_{p} (k_{\mu} - \slash{k} \gamma_{\mu})
- (\slash{k} k_{\mu} - k^{2} \gamma_{\mu})
\big)
\big] \otimes \gamma_{\alpha}
\big\rbrace
\nonumber\\
&& \times
u_{a} \otimes u_{b}\,,
\label{B13}\\
%%%%%%%%%%%%%%%%%%%%%%%%%%%%%%%%%%%%%%
{\cal M}_{{\rm V},\mu}^{(a + b + c)\,4}&=&
%%%%%%%%%%%%%%%%%%%%%%%%%%%%%%%%%%%%%%
-e \bar{u}_{1'} \otimes \bar{u}_{2'}
\big\lbrace
{\cal F}_{V}(s,t_{2})
\frac{1}{2p_{1}' \cdot k + k^{2}+i \varepsilon}
\big[-(k_{\mu} - \gamma_{\mu} \slash{k}) \nonumber\\
&& + \frac{F_{2}(0)}{2 m_{p}}
\big(
-2 p_{1 \mu}' \slash{k} + 2 (p_{1}' \cdot k) \gamma_{\mu}
- 2 m_{p} (k_{\mu} - \gamma_{\mu} \slash{k})
- (k_{\mu} \slash{k} - k^{2} \gamma_{\mu})
\big)
\big] 
\gamma^{\alpha} \otimes \gamma_{\alpha}
\big\rbrace
\nonumber\\
&& \times
u_{a} \otimes u_{b}\,.
\label{B14}
\end{eqnarray}

According to (\ref{2.46d})--(\ref{2.46f}),
we have
\begin{eqnarray}
{\cal M}_{{\rm T},\mu}^{(d + e + f)\,j} &=& 
\left.{\cal M}_{{\rm T},\mu}^{(a + b + c)\,j}
\right|_{\mathop{^{(p_{a}, \,\lambda_{a}) \leftrightarrow (p_{b},\, \lambda_{b})}_{(p_{1}',\, \lambda_{1}) \leftrightarrow (p_{2}', \,\lambda_{2})}}} \quad {\rm for} \;\; j = 1, \ldots, 7\,,
\label{B15}\\
{\cal M}_{{\rm V},\mu}^{(d + e + f)\,j'} &=& 
\left.{\cal M}_{{\rm V},\mu}^{(a + b + c)\,j'}
\right|_{\mathop{^{(p_{a}, \,\lambda_{a}) \leftrightarrow (p_{b},\, \lambda_{b})}_{(p_{1}',\, \lambda_{1}) \leftrightarrow (p_{2}', \,\lambda_{2})}}} \quad {\rm for} \;\; j' = 1, \ldots, 4\,,
\label{B16}
\end{eqnarray}
where we also exchange the order of the tensor products in
(\ref{B4})--(\ref{B14}).
Note that all subamplitudes
${\cal M}_{{\rm T},\mu}^{(a + b + c)\,j}$
and 
${\cal M}_{{\rm V},\mu}^{(a + b + c)\,j'}$
are separately gauge invariant, as is easy to check,
\begin{eqnarray}
k^{\mu} {\cal M}_{{\rm T},\mu}^{(a + b + c)\,j} &=& 0\,, 
\quad j = 1, \ldots, 7\,, \nonumber \\
k^{\mu} {\cal M}_{{\rm V},\mu}^{(a + b + c)\,j'} &=& 0\,, 
\quad j' = 1, \ldots, 4\,.
\label{B17}
\end{eqnarray}
The same holds for 
${\cal M}_{{\rm T},\mu}^{(d + e + f)\,j}$
and 
${\cal M}_{{\rm V},\mu}^{(d + e + f)\,j'}$.

Note that only the $j = 1$ terms in (\ref{B4}) and (\ref{B15}), 
and the $j' = 1$ terms in (\ref{B11}) and (\ref{B16}),
contain the pole terms proportional to $\omega^{-1}$
for $\omega \to 0$.
The SPA1 and SPA2 results are derived from these terms;
see Sec.~\ref{sec:SPA}.

\begin{figure}[!ht]
\includegraphics[width=0.46\textwidth]{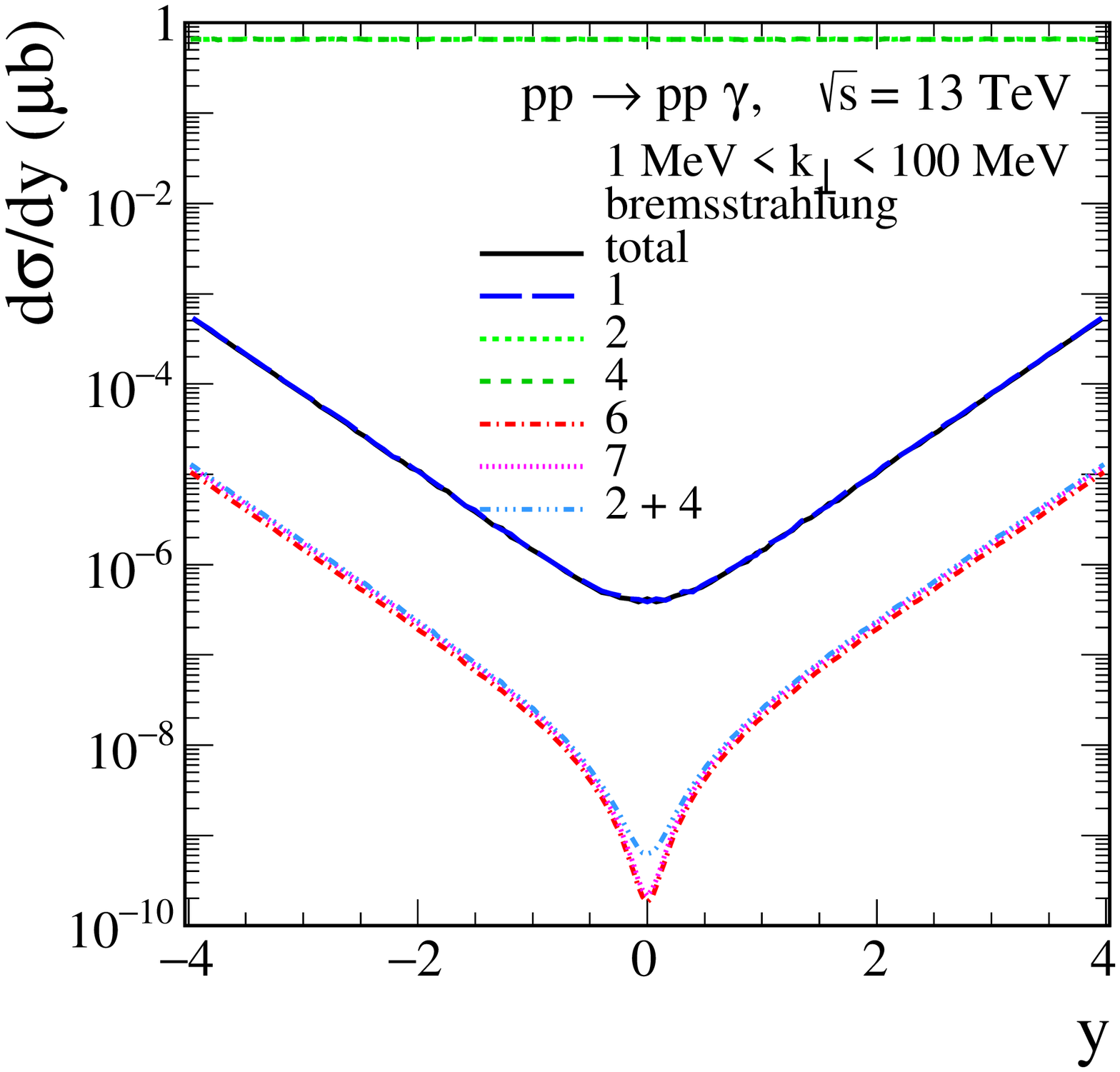}
\includegraphics[width=0.46\textwidth]{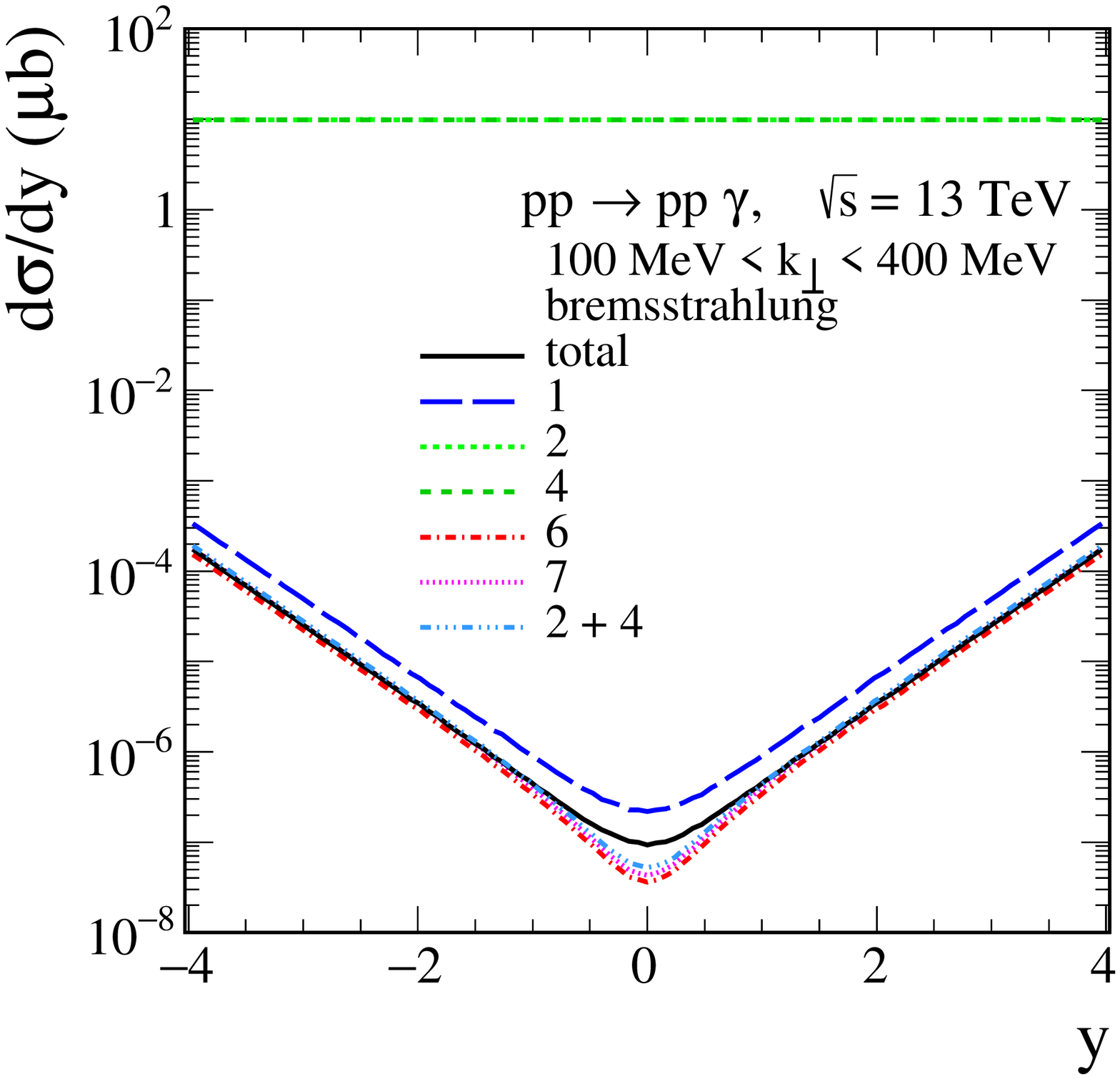}\\
\includegraphics[width=0.46\textwidth]{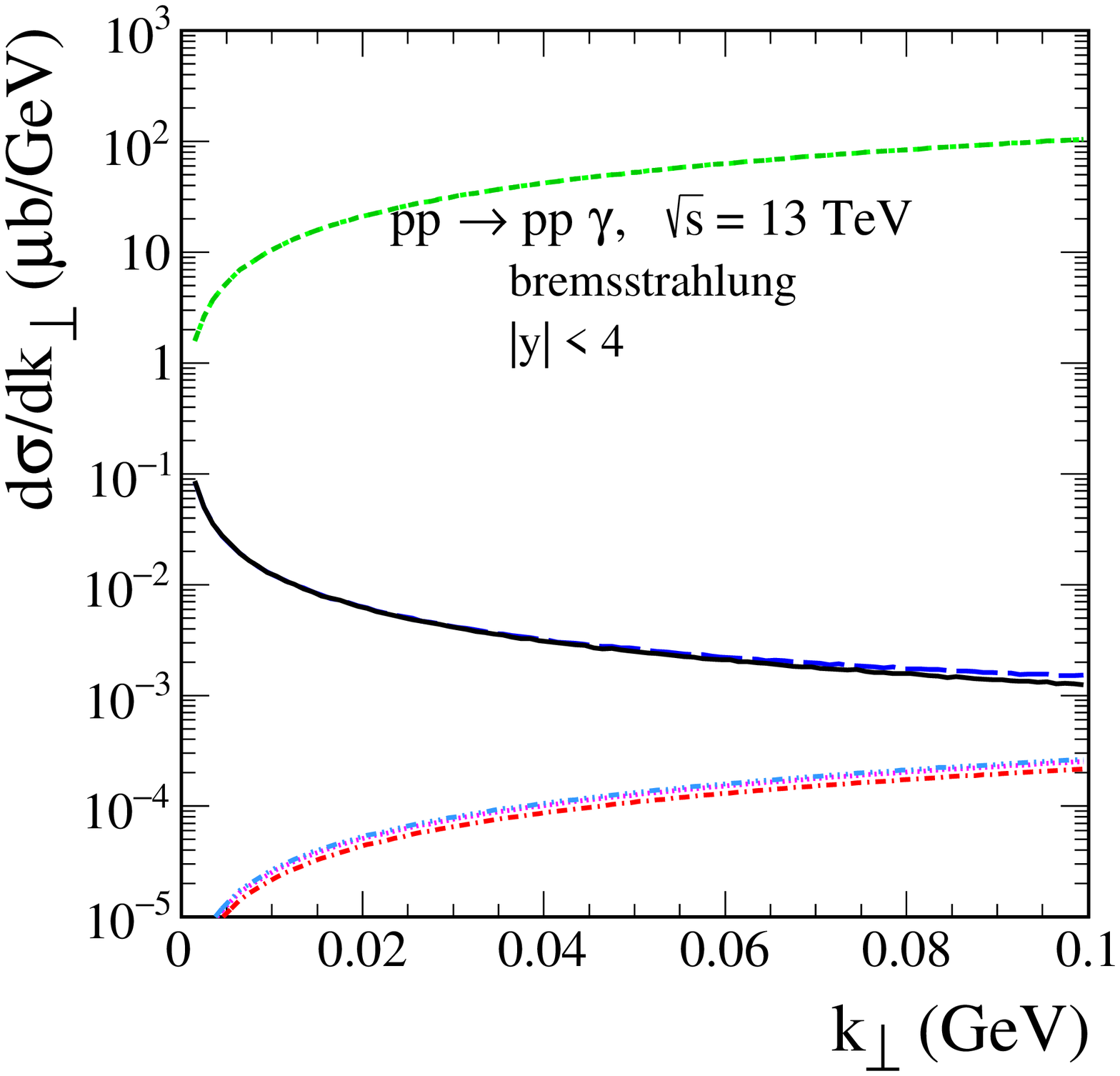}
\includegraphics[width=0.46\textwidth]{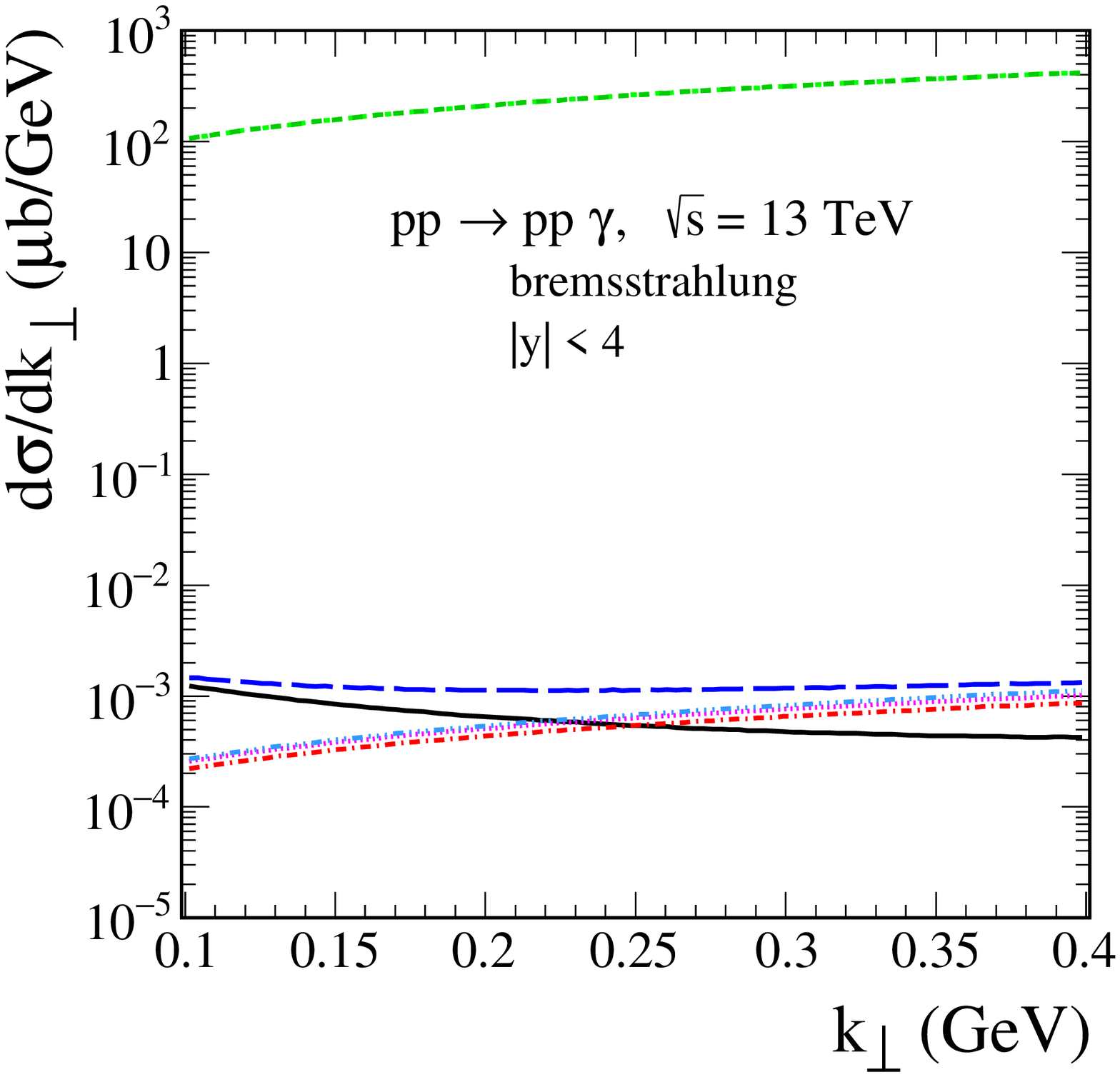}\\
\includegraphics[width=0.46\textwidth]{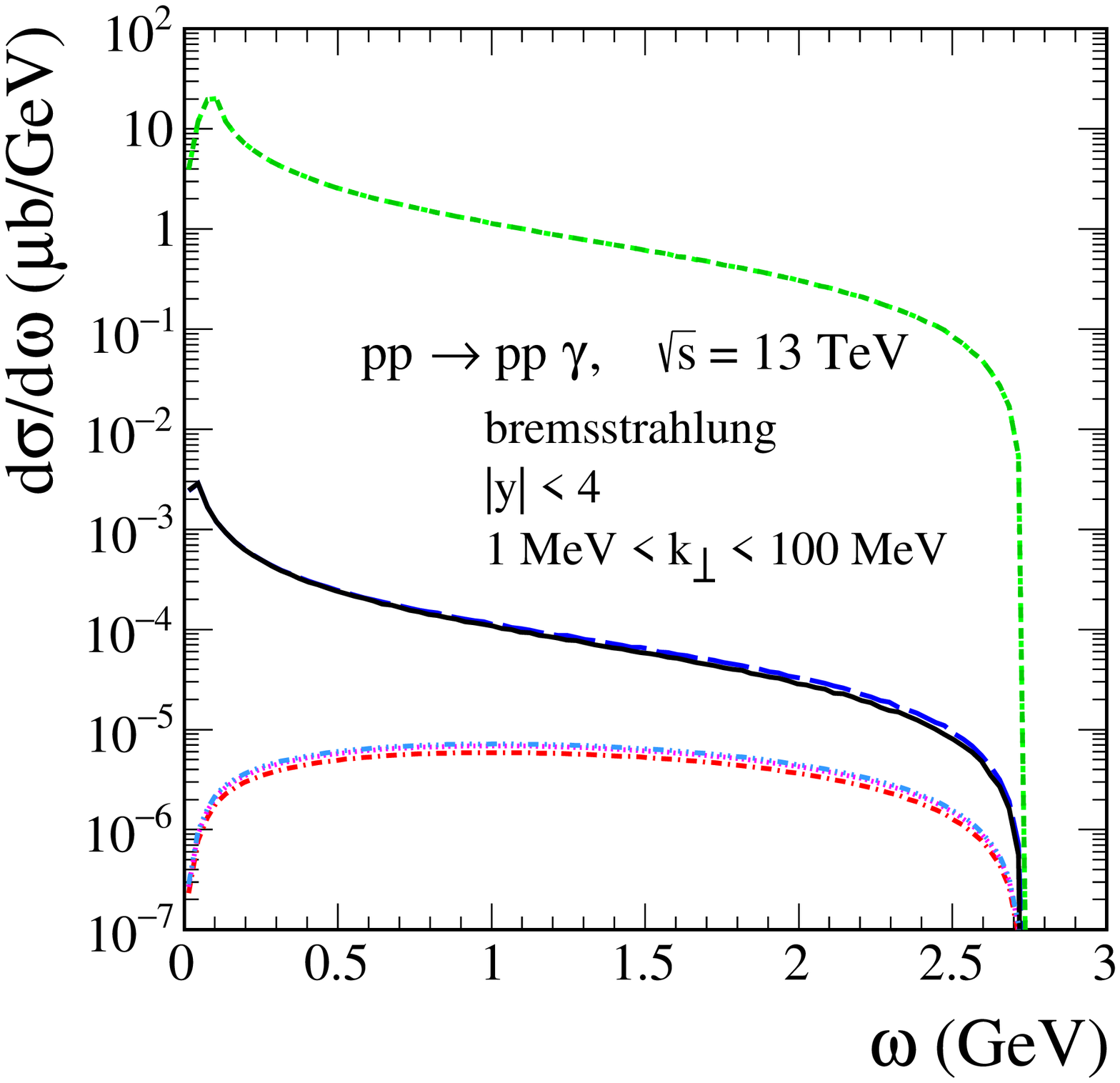}
\includegraphics[width=0.46\textwidth]{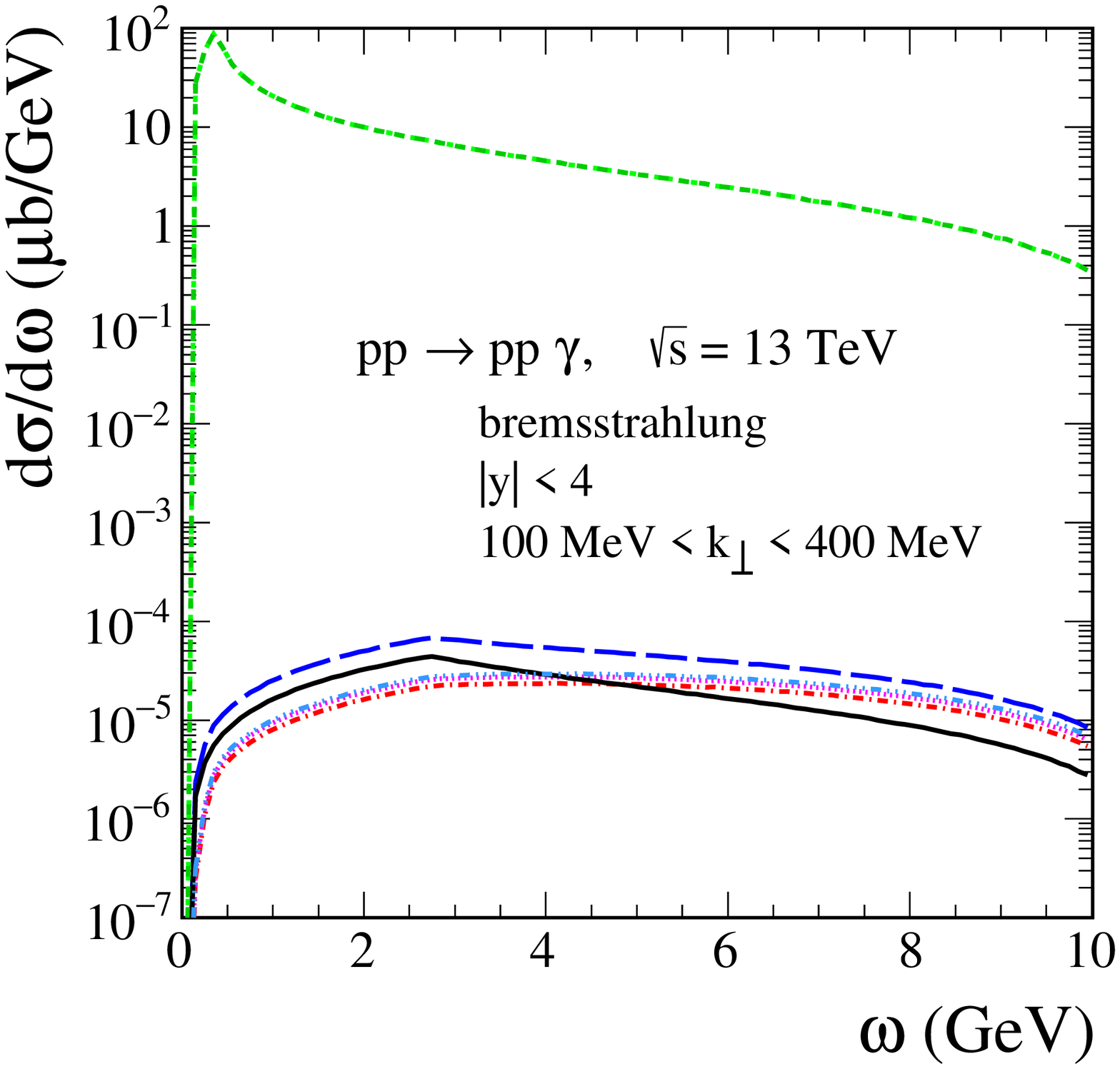}
\caption{\label{fig:B1}
\small
The differential distributions 
for the $pp \to pp \gamma$ reaction
calculated for $\sqrt{s} = 13$~TeV, $|\rm y| < 4$, and
for two $k_{\perp}$ intervals as
specified in the figure legends.
The results for individual $j$ terms
in (\ref{B3}) and their coherent sum (total) are shown.
The coherent sum of the amplitudes $(a + b + c)$ plus $(d + e + f)$
with $j = 2$ and $j = 4$ is denoted by ``$2 + 4$''.}
\end{figure}

In Fig.~\ref{fig:B1}, we show differential distributions
for the $pp \to pp \gamma$ reaction 
for $\sqrt{s} = 13$~TeV, $|\rm y| < 4$,
and for two $k_{\perp}$ intervals:
\mbox{$1\, {\rm MeV} < k_{\perp} < 100\, {\rm MeV}$}
(the left panels),
\mbox{$100\, {\rm MeV} < k_{\perp} < 400\, {\rm MeV}$}
(the right panels).
In these calculations, we limit ourselves 
to the leading pomeron-exchange contribution.
We show the complete result (``total'')
including interference effects
and the results for individual $j$ terms
%${\cal M}_{{\rm T},\mu}^{(a + \ldots + f)\,j}$
(\ref{B4})--(\ref{B10}) plus (\ref{B15}),
except for $j = 3$ and 5, which
are very small and can be safely neglected.
Note that there is significant cancellation 
among the terms $j = 2$ and $4$,
due to destructive interference.
The coherent sum of the amplitudes 
$(a + b + c)$ plus $(d + e + f)$ 
with $j = 2$ and $j = 4$ is denoted by ``$2 + 4$''.

%--------------------------------------------------------
%--------------------------------------------------------

How can we understand these results?
First, we remark that the destructive interference of 
the 2 and 4 terms above is \underline{not} due to
a gauge cancellation.
These terms are separately gauge invariant;
see (\ref{B17}).
Let us have a closer look at the terms 1, 2, and 4
of (\ref{B4}), (\ref{B5}), and (\ref{B7}),
respectively, for real photons,
$k^{2} = 0$, and with transverse momentum only:
\begin{eqnarray}
(k^{\mu}) =
\left( 
\begin{array}{c}
 \omega \\
 \omega \,\bhkperp \\
\end{array} 
\right)
\,, \quad |\bhkperp| = 1\,.
\label{B18}
\end{eqnarray} 
We work in the c.m. system with $|\bpa|$ defining the z axis.
Then, disregarding terms on the rhs of (\ref{B4})
and (\ref{B5}), which are of the same order 
in $|\bpa|$, we get very roughly
\begin{eqnarray}
{\cal M}_{{\rm T},\mu}^{(a + b + c)\,1} & \propto &
\frac{p_{a \mu}}{-p_{a} \cdot k} 
+ \frac{p_{1 \mu}'}{p_{1}' \cdot k} \nonumber \\
& \propto & \frac{\Delta p}{|\bpa| \omega} \,.
\label{B19}
\end{eqnarray}
Here $\Delta p = {\cal O}(m_{p})$ is a measure
of the transverse momentum change from $\bpa$ to $\bpaap$.

\begin{figure}[!ht]
\includegraphics[width=0.48\textwidth]{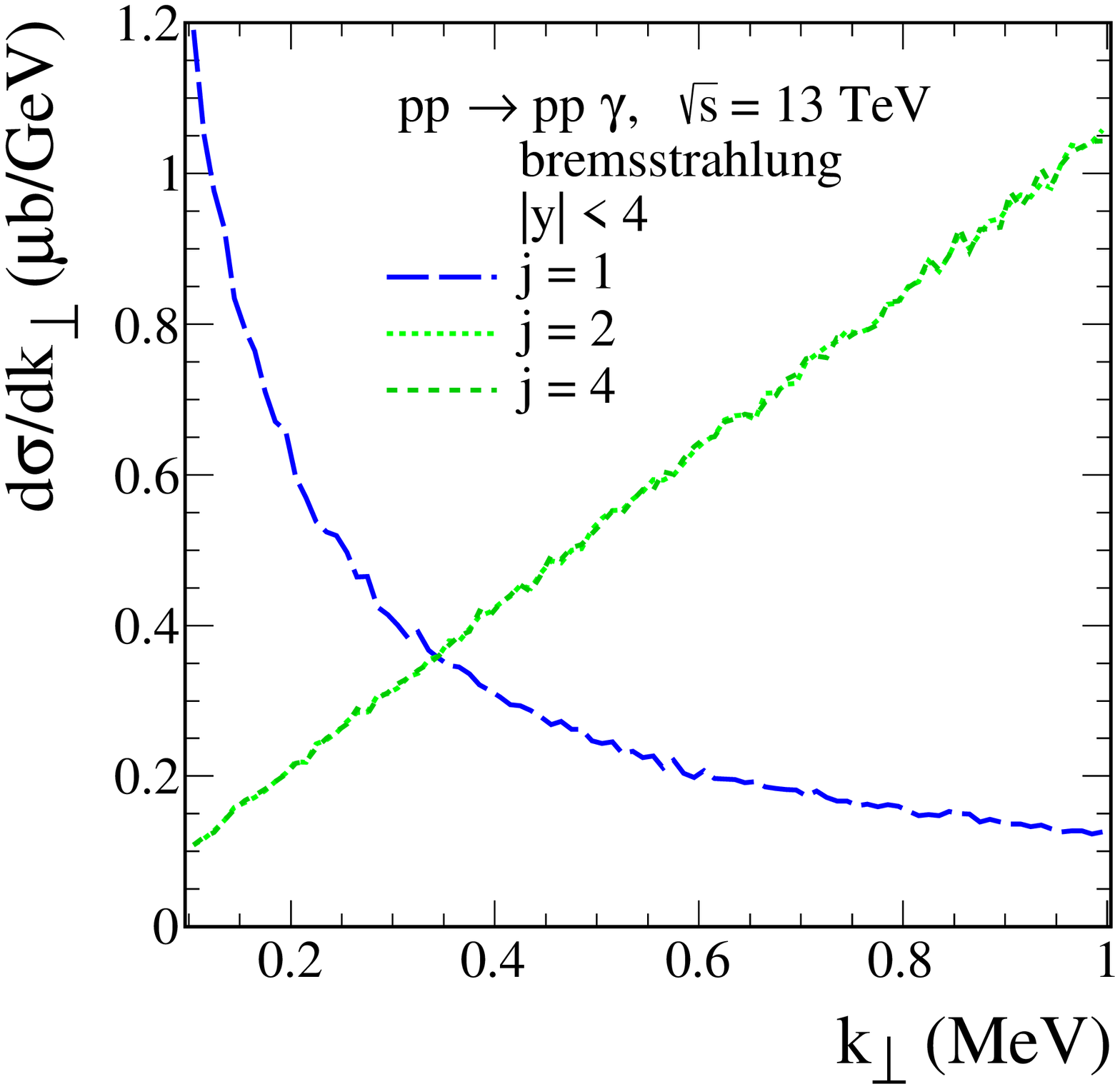}
\includegraphics[width=0.48\textwidth]{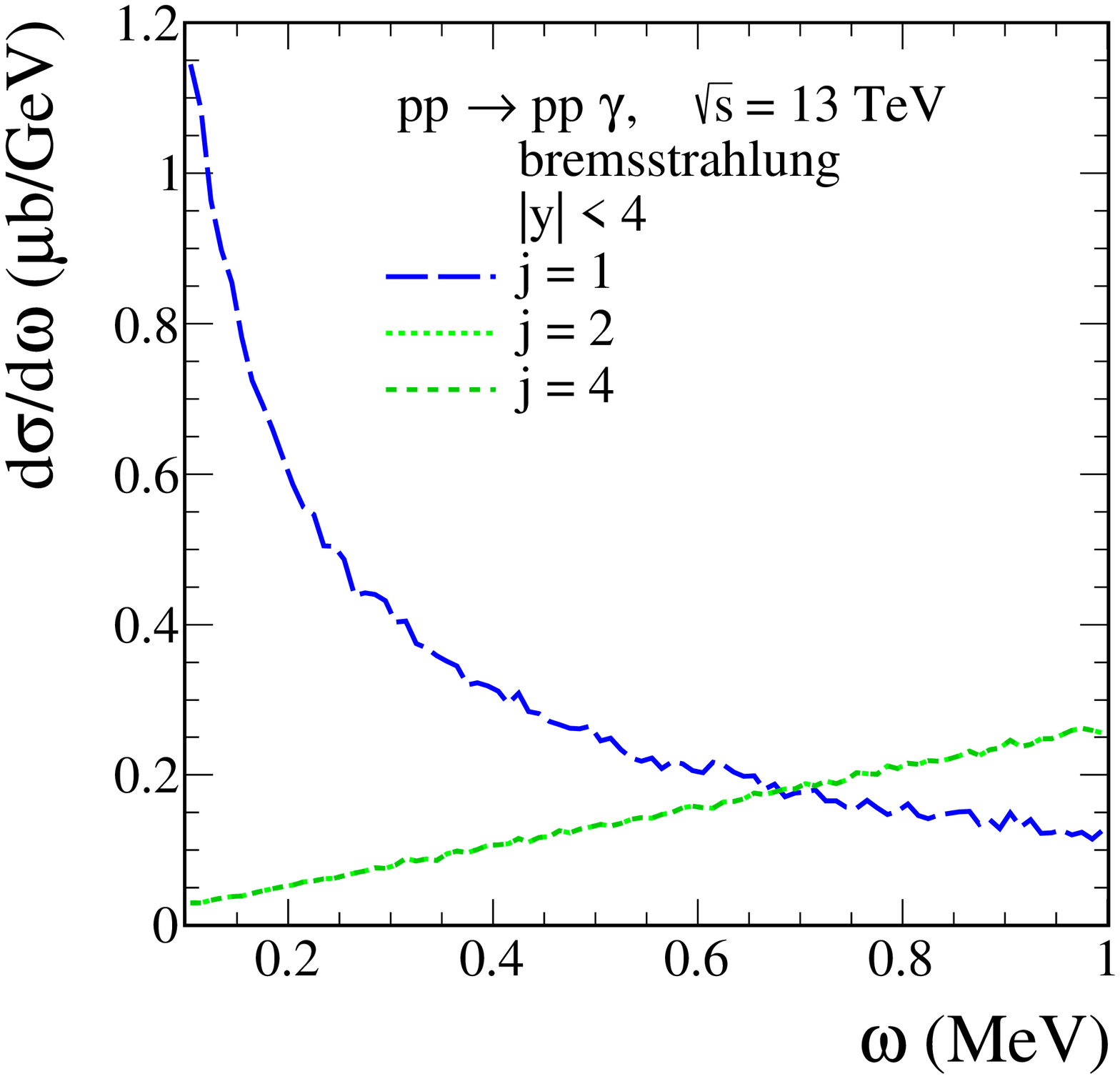}
\caption{\label{fig:B2}
\small
The cross sections $d\sigma/dk_{\perp}$ (left)
and $d\sigma/d\omega$ (right)
for the terms
${\cal M}_{{\rm T},\mu}^{(a + b + c)\,j}
+ {\cal M}_{{\rm T},\mu}^{(d + e + f)\,j}$
for pomeron exchange for $j = 1, 2$, and 4.
See (\ref{B4}), (\ref{B5}), (\ref{B7}), and (\ref{B15}).
The crossing of the term $j = 1$
containing the $\omega^{-1}$ pole 
with the nonpole terms $j = 2$
and 4 occurs for $k_{\perp} \cong 0.35$~MeV
and $\omega \cong 0.7$~MeV.}
\end{figure}

The term ${\cal M}_{{\rm T},\mu}^{(a + b + c)\,2}$
has no singularity for $\omega \to 0$.
The main term here comes from the anomalous magnetic moment
$F_{2}(0)$ and can be estimated as
\begin{eqnarray}
{\cal M}_{{\rm T},\mu}^{(a + b + c)\,2} & \propto &
\frac{1}{|\bpa| \omega} 
\frac{|\bpa| \omega}{m_{p}} = \frac{1}{m_{p}}
\label{B20}
\end{eqnarray}
and similarly for ${\cal M}_{{\rm T},\mu}^{(a + b + c)\,4}$.
Thus, the term ${\cal M}_{{\rm T},\mu}^{(a + b + c)\,1}$
will win over the 2 and 4 terms individually
for $\omega \to 0$.
However, for this to happen, we must require
\begin{eqnarray}
&&\frac{\Delta p}{|\bpa| \omega} \approx
\frac{m_{p}}{|\bpa| \omega} \gtrsim \frac{1}{m_{p}}\,, 
\nonumber \\
&&\omega \lesssim \frac{m_{p}^{2}}{|\bpa|}\,.
\label{B21}
\end{eqnarray}
For our case with $2 |\bpa| = 13$~TeV,
this requires
\begin{eqnarray}
k_{\perp} \approx \omega \lesssim 0.15\; {\rm MeV}\,.
\label{B22}
\end{eqnarray}
Explicit calculations confirm the order of magnitude of this estimate. 
Indeed, from Fig.~\ref{fig:B2}, we see that
the $j = 1$ term exceeds the $j = 2$ term only for
$k_{\perp} \lesssim 0.35$~MeV
and $\omega \lesssim 0.7$~MeV.
The crossing place of these terms
depends on the ${\rm y}$ range since the $j = 1$ term
has a minimum at ${\rm y} = 0$
and grows rapidly with $|{\rm y}|$ increasing,
while the terms 2 and 4 are flat in the midrapidity region
(see the top panels of Fig.~\ref{fig:B1}).
When going with ${\rm y} \to 0$,
the crossing place of the terms 1 and 2 shifts to lower values
of $k_{\perp}$ and $\omega$.
In reality, however, there is destructive interference
between the terms 2 and 4, and their \underline{sum} is harmless,
well below the term 1,
at least for $k_{\perp} < 100$~MeV and $\omega < 2$~GeV,
as we see from the left panels of Fig.~\ref{fig:B1}.
In the right panels of Fig.~\ref{fig:B1} we show the results
for larger $k_{\perp}$ and $\omega$ ranges.
The destructive interference of the terms 2 and 4 is again
a salient feature.

%--------------------------------------------------------
%--------------------------------------------------------

As we mentioned already, the terms with labels
$j = 2$ and 4 contain the Dirac and Pauli couplings,
and the latter one dominates.
In the result ``$2 + 4$'' shown in Fig.~\ref{fig:B1},
we find destructive interference of the Pauli parts
of $j = 2$ and $j = 4$
in the term $(a+b+c)$ and in the term $(d+e+f)$ individually.
But for the Pauli parts, we find numerically that
there is practically no interference between 
the $(a+b+c)$ and the $(d+e+f)$ terms.
In the ``$2 + 4$'' term,
the Dirac part wins over the Pauli part.
The visible dip for $d\sigma/d{\rm y}$ at ${\rm y} \approx 0$
in the ``$2 + 4$'' result (see the left upper panel of Fig.~\ref{fig:B1})
is due to destructive interference of the Dirac parts
between the $(a+b+c)$ and the $(d+e+f)$ terms.
%For the term $j = 1$ at ${\rm y} \approx 0$
%there is a small constructive interference.

For the vector-exchange contributions,
(\ref{B11})--(\ref{B14}) and (\ref{B16}),
the situation is similar to that for the tensor exchanges.
The terms containing the $\omega^{-1}$ pole are
the $j' = 1$ terms. Individually, the vector-exchange terms
with $j' = 3$ and $j' = 4$ are much larger than 
the $j' = 1$ term, 
except for very small $k_{\perp}$ and $\omega$.
But again there is destructive interference 
of the $j' = 3$ and $j' = 4$
terms and the sum $3+4$ is well below the $j' = 1$ term
in the same kinematic regions as shown in Fig.~\ref{fig:B1}.
In this kinematic region, the $j' = 2$ term is very small.

%------------------------------------------------------------------
\bibliography{refs}
%------------------------------------------------------------------

\end{document}